\documentclass[preprintnumbers,prd,singlecolumn,floatfix,preprintnumbers,amsmath,amssymb,nofootinbib,superscriptaddress]{revtex4}
\usepackage{hyperref}

\usepackage{times,graphicx,amssymb, amsmath, natbib}
\usepackage[T1]{fontenc}

\usepackage{epsfig}

\newcommand{\beq}{\begin{equation}}
\newcommand{\eeq}{\end{equation}}

\newcommand{\ber}{\begin{eqnarray}}
\newcommand{\eer}{\end{eqnarray}}

\def\beq{\begin{equation}}
\def\eeq{\end{equation}}

\def\ber{\begin{eqnarray}}
\def\eer{\end{eqnarray}}

\begin{document}

\title{Observed galaxy power spectrum in cubic Galileon model}

\author{Bikash R. Dinda}
\email{bikash@ctp-jamia.res.in} 
\affiliation{Centre for Theoretical Physics, Jamia Millia Islamia, New Delhi-110025, India}

\author{Md. Wali Hossain}
\email{wali.hossain@apctp.org} 
\affiliation{Asia Pacific Center for Theoretical Physics, Pohang 37673, Korea}

\author{Anjan A Sen}
\email{aasen@jmi.ac.in} 
\affiliation{Centre for Theoretical Physics, Jamia Millia Islamia, New Delhi-110025, India}

\begin{abstract}
In this paper, we study the effects of general relativistic corrections on the observed galaxy power spectrum in thawing class of cubic Galileon model with linear potential that preserves the shift symmetry. In this scenario, the observed galaxy power spectrum differs from the standard matter power spectrum mainly due to redshift space distortion (RSD) factor and relativistic effects. The RSD term enhances the matter power spectrum both at larger and smaller scales whereas the relativistic terms further enhance the matter power spectrum only at larger scales. In comparison with $\Lambda$CDM, the observed galaxy power spectrum is always suppressed at large scales in this scenario although this suppression is always small compared to the canonical quintessence scenario.
\end{abstract}

\maketitle
\date{\today}

\section{Introduction}

Observational cosmology is currently passing through a revolutionary phase. It all started since 1998 when we first discovered using Supernova Type-Ia observation that our Universe is going through an accelerated phase of expansion \citep{Riess:1998cb,Perlmutter:1998np}. Since then a wide variety of cosmological observations related to the cosmic microwave background radiation anisotropy \citep{Spergel:2003cb,Hinshaw:2003ex,Ade:2015xua,Ade:2015lrj}, baryon acoustic oscillations measurements in galaxy power spectrum \citep{2015A&A...574A..59D,Ata:2017dya} with unprecedented accuracies have confirmed this acceleration. All these observations also confirm that we live in a Universe with flat spatial section, having $25\%$ of the energy budget in the form of cold dark matter (cdm) and $5\%$ in baryons \citep{Ade:2015xua}. It also confirms another accelerated phase of expansion at very early stage of cosmological evolution termed as {\it inflation} \citep{Starobinsky:1980te,Guth:1980zm,Linde:1983gd,Linde:1981mu,Liddle:1999mq}.

The confirmation of the accelerated expansion in our Universe defies our understanding of the attractive nature of gravity that can only produce decelerated expansion in the Universe under the realm of general theory of relativity. To get this accelerated expansion, either we need to add unknown form of matter with repulsive gravitational force \citep{Copeland:2006wr,Sahni:1999gb,Padmanabhan:2006ag,Frieman:2008sn} that has no direct observational detection till date or we need to modify the Einstein gravity at large cosmological scales \citep{Clifton:2011jh,Hinterbichler:2011tt,DeFelice:2010aj,deRham:2014zqa,deRham:2012az,Horndeski:1974wa,Starobinsky:1980te,Dvali:2000hr,Hu:2007nk,Amendola:2006we,Nicolis:2008in,deRham:2010kj,deRham:2010tw,deRham:2011by,Heisenberg:2014kea,Deffayet:2010qz,DeFelice:2010nf} where this accelerated expansion has been observed. One of possibilities to get this repulsive gravity had already prescribed by Einstein himself although in a different context. This is the cosmological constant $\Lambda$, with an equation of state (EoS) $p = -\rho$. Together with the presence of cold dark matter, this is the concordance $\Lambda$CDM model that is the simplest way to explain late time acceleration of the Universe.  At late times, the energy content  in $\Lambda$ has to be around $70\%$ of the total energy density of the Universe to result accelerated Universe. To achieve this, the value of $\Lambda$ required is  embarrassingly low compared with what we expect from our current understanding of particle physics and results the issue of fine tuning \citep{Martin:2012bt}. The constant $\Lambda$ also demands that the expansion starts precisely at the present epoch, making this epoch a very special one in the entire cosmological evolution. Theoreticians describe this as cosmic coincidence problem \citep{Steinhardt:1999nw}.

The fine tuning is unavoidable as far our current understanding of particle physics and cosmology is concerned. But we can ameliorate the cosmic coincidence problem by replacing $\Lambda$ with an unknown component with negative pressure that is not constant but evolves with cosmic time. We call this dark energy \citep{Copeland:2006wr}. Motivation for considering dark energy also comes from the inconsistency in $\Lambda$CDM model with respect to couple of  recent observational results \citep{Riess:2016jrr,Hildebrandt:2016iqg,Heymans:2013fya,Bonvin:2016crt}.

Similar to inflaton, we can model dark energy as a scalar field slowly rolling over a sufficiently flat potential around present time \citep{Wetterich:1987fk,Wetterich:1987fm,Ratra:1987rm}. These scalar field dark energy models can further be classified to tracker \citep{Steinhardt:1999nw} and thawing models \citep{Scherrer:2007pu,Chiba:2009sj} depending on the form of its potentials and the subsequent evolution. Late time acceleration with scalar field dark energy model has been studied extensively by different authors \citep{Scherrer:2007pu,Chiba:2009sj,Steinhardt:1999nw,Caldwell:1997ii,Zlatev:1998tr,Amendola:1999er,Sahni:1999qe,Perrotta:1999am,Sahni:2001qp,Hossain:2014zma,Caldwell:1999ew,Elizalde:2004mq,Sen:2002nu,Sen:2002in,Gibbons:2002md,Garousi:2004uf,Copeland:2004hq,ArmendarizPicon:2000dh,ArmendarizPicon:2000ah,Rendall:2005fv,Bento:2002ps,Bento:2002yx}.

A large scale modification of gravity in the higher dimensional brane world scenario has been proposed by Dvali, Gabadadze and Porrati (DGP) \citep{Dvali:2000hr} to explain the late time cosmic acceleration. In the decoupling limit of DGP model, one can obtain a scalar degree of freedom containing a higher derivative term like $(\nabla \phi)^2 \Box \phi$ \citep{Luty:2003vm}.The Lagrangian for such scalar degree of freedom respects the Galilean shift symmetry and known as Galileon \citep{Nicolis:2008in}. Apart from the cubic Galilon and usual kinetic terms, the full Galileon Lagrangian has two more terms with higher derivatives \citep{Nicolis:2008in,Deffayet:2009wt}. Despite the presence of  higher derivative terms in the Lagrangian, the equation of motion of the Galileon field is second order \citep{Nicolis:2008in,Deffayet:2009wt} and the theory is free from Ostrogradsky ghosts \citep{Woodard:2006nt}. Vainshtein mechanism \citep{Vainshtein:1972sx}, first proposed to overcome the problem of van Dam-Veltman-Zakharov
(vDVZ) discontinuity  \citep{vanDam:1970vg,Zakharov:1970cc} in the linear theory of massive gravity \citep{Fierz:1939ix}, can also be implemented in Galileon theory \citep{Nicolis:2008in} to preserve the local physics. 

Till date, there have been plethora of investigations to constrain the background evolution of the scalar field dark energy models including the Galileon models \citep{Copeland:2006wr,Chow:2009fm,Silva:2009km,Kobayashi:2010wa,Kobayashi:2009wr,Gannouji:2010au,DeFelice:2010gb,DeFelice:2010pv,Ali:2010gr,Mota:2010bs,deRham:2010tw,deRham:2011by,Heisenberg:2014kea,Deffayet:2010qz,Ali:2012cv,Hossain:2012qm,Hossain:2017ica}. The inhomogeneity in the dark energy field has not been constrained till date as dark energy perturbations are only relevant on horizon scales and beyond and accurate measurements of observed galaxy power spectra on these very large scales has not been done yet. But with the scope of future optical and infrared/radio surveys like LSST, SKA, we shall have the opportunity to probe our Universe at horizon scales and beyond which in turn enable us to probe dark energy inhomogeneities. As cosmological constant does not contain any inhomogeneity whereas any other evolving dark energy contains inhomogeneities, any detection (or not detection) of dark energy inhomogeneity can decisively settle the issue of dark energy being cosmological constant or not.

To study the structure formation on horizon scales and beyond, one needs to consider the full general relativistic (GR)  treatment. There are also number of general relativistic corrections in the observed galaxy power spectra related to gravitational potential and peculiar velocity. The observed galaxy power spectra including necessary GR corrections has been studied earlier for $\Lambda$CDM model \citep{Yoo:2009au,Bonvin:2011bg,Challinor:2011bk} as well as for tracking \citep{Duniya:2015nva} and thawing scalar field models \citep{Dinda:2016ibo}. In this present work, we extend the similar study for the cubic Galileon models with linear potential that preserve the shift symmetry. Later we also consider other phenomenological potentials that breaks the shift symmetry.

The paper is organised as: in section II, we briefly describe the background evolution of the Universe in cubic Galileon model; in section III, we study the first order general relativistic perturbation in the Galileon field and study the deviations from $\Lambda$CDM model in some important perturbed quantities; in section IV, we study the observed galaxy power spectrum and its deviation for the cubic Galileon model from $ \Lambda $CDM; finally in section V, we present our conclusion.

\section{Background evolution}

We consider the lowest nontrivial order of the Galileon action {\it i.e.}, cubic Galileon action along with a potential \citep{Ali:2012cv,Hossain:2012qm,Hossain:2017ica}

\begin{equation}
S=\int d^4x\sqrt{-g}\Bigl [\frac{M^2_{\rm{pl}}}{2} R +  \frac{1}{2}(\nabla \phi)^2\Bigl(1+\frac{\alpha}{M^3}\Box \phi\Bigr) - V(\phi) \Bigr] + \mathcal{S}_m \, ,
\label{eq:action}
\end{equation}

\noindent
where $M_{\rm{pl}}=(8\pi G)^{-1/2}$ is the reduced Planck mass. $\alpha$ is a  dimensionless constant; for $\alpha=0$ this action~\eqref{eq:action} reduces to that of a standard quintessence action \citep{Wetterich:1987fk,Wetterich:1987fm,Ratra:1987rm}. $V(\phi)$ is the potential. $V(\phi) = c_{1}\phi$ preserve the shift symmetry and we mainly consider this potential for our subsequent study. ${\cal S}_{\rm m}$ is the action for the matter field. $M$ is a constant of mass dimension one; by a redefinition of the parameter $\alpha$, we can fix $M=M_{\rm{pl}}$. Action~\eqref{eq:action} can also be thought as a particular form of the Kinetic Gravity Braiding action \citep{Deffayet:2010qz}.
\\ 
\noindent
Variation of the action \eqref{eq:action} with respect to (w.r.t.) the metric tensor $g_{\mu\nu}$ and assuming a flat Friedman-Robertson-Walker (FRW) spacetime with scale factor $a(t)$, we get the Einstein's equations

\begin{equation}
3M_{\rm pl}^2H^2 = \rho_{\rm m}+\frac{\dot{\phi}^2}{2}\Bigl(1-6\frac{\alpha}{M_{\rm{pl}}^{3}} H\dot{\phi}\Bigr)+V{(\phi)},
\end{equation}

\begin{equation}
M_{\rm pl}^2(2\dot H + 3H^2) = -\frac{\dot{\phi}^2}{2}\Bigl(1+2\frac{\alpha}{M_{\rm{pl}}^{3}}\ddot{\phi}\Bigr)+V(\phi),
\end{equation}
 
\noindent
where overdot is the derivative w.r.t. the time and $H$ is the Hubble parameter. Varying the action w.r.t the field $\phi$, we get the equation of motion for the field

\begin{equation}
\ddot{\phi} + 3H\dot{\phi}-3\frac{\alpha}{M_{\rm pl}^{3}} \dot{\phi}\Bigl(3H^2\dot{\phi}+\dot{H}\dot{\phi}+2H\ddot{\phi}\Bigr)+ V_{\phi}=0,
\end{equation}

\noindent
where subscript $\phi$ is the derivative w.r.t the field $\phi$.

\section{Relativistic perturbations with the Galileon field}

In this paper we are mainly interested in the observed galaxy power spectrum on a scale where the perturbations are assumed to be linear i.e. we can use the linear perturbation theory with the full general relativistic treatment. In the linear perturbation theory the scalar, vector and tensor perturbations evolve independently. So, we can study the linear scalar perturbations independently with two scalar degrees of freedom. Here we work in the conformal Newtonian gauge where the perturbed space-time is given by

\begin{equation}
ds^{2} = (1+2\Psi) dt^{2} - a(t)^{2}(1-2\Phi) d\vec{r}.d\vec{r},
\label{eq:sptm}
\end{equation}

\noindent
where $\vec{r}$ is the comoving coordinate, $\Phi$ is the gravitational potential and for simplicity we choose anisotropic stress to be zero which corresponds to $ \Psi = \Phi $. So, we are left with one scalar degree of freedom which is $ \Phi $. In this perturbed space-time the linearized Einstein equations become \citep{2008PhRvD..78l3504U}:

\begin{eqnarray}
\vec{\nabla}^{2} \Phi - 3 a^{2} H (\dot{\Phi} + H \Phi) &=& 4 \pi G a^{2} \sum_{i} \delta \rho_{i} \, , 
\label{eq:ein_per_1}\\
\dot{\Phi} + H \Phi &=& 4 \pi G a \sum_{i} (\bar{\rho_{i}} + \bar{P_{i}}) v_{i} \, , 
\label{eq:ein_per_2}\\
\ddot{\Phi} + 4 H \dot{\Phi} + (2 \dot{H} + 3 H^{2}) \Phi &=& 4 \pi G \sum_{i} \delta P_{i} \, ,
\label{eq:ein_per_3}
\end{eqnarray}

\noindent
where the summation index $ i $ stands for either 'm' for matter or '$\phi$' for Galileon field, $ \mathcal{H} $ is the conformal Hubble parameter ($ \mathcal{H} = a H $), $bar$ represents the unperturbed quantity for the individual fluid $i$; $\delta \rho_{i} $, $ \delta P_{i} $ and $ v_{i} $ are the perturbations of the individual component's energy density, pressure and velocity field respectively. Combining Eqs.~\eqref{eq:ein_per_1} and \eqref{eq:ein_per_2} we get the relativistic Poisson equation which is given by

\begin{equation}
\vec{\nabla}^{2} \Phi = 4 \pi G a^{2} \sum_{i} \bar{\rho_{i}} \Delta_{i},
\end{equation}

\noindent
where we have introduced a quantity $ \Delta_{i} $ corresponding to the particular individual component which is given by $ \Delta_{i} = \delta_{i} + 3 \mathcal{H} (1 + w_{i}) v_{i} $ where $ \delta_{i} $ is defined as $ \delta \rho_{i} = \bar{\rho}_{i} \delta_{i} $. This gauge invariant quantity is the comoving energy density contrast for a particular component i.e. either for the matter or for the Galileon field.
\\
\noindent
Working in the space-time~\eqref{eq:sptm}, we can calculate components of the energy momentum tensor from the action \eqref{eq:action}. The first order perturbed energy density, pressure and velocity for the Galileon field $\phi$ are respectively given by \citep{Hossain:2017ica}

\begin{eqnarray}
\delta \rho_{\phi} &=& (1-9 \beta H \dot{\phi}) \dot{\phi} \dot{\delta \phi} + \beta \dot{\phi}^{2} \frac{\vec{\nabla}^{2} \delta \phi}{a^{2}} - (1-12 \beta H \dot{\phi}) \dot{\phi}^{2} \Phi + 3 \beta \dot{\phi}^{3} \dot{\Phi} + V_{\phi} \delta \phi,
\label{eq:del_rho_phi} \\
\delta P_{\phi} &=& \beta \dot{\phi}^{2} \ddot{\delta \phi} + (1+2 \beta \ddot{\phi}) \dot{\phi} \dot{\delta \phi} - (1+4 \beta \ddot{\phi}) \dot{\phi}^{2} \Phi - \beta \dot{\phi}^{3} \dot{\Phi} - V_{\phi} \delta \phi, 
\label{eq:del_p_phi}\\
a (\bar{\rho_{\phi}} + \bar{P_{\phi}}) v_{\phi} &=& \dot{\phi} \bigg[ \beta \dot{\phi} \dot{\delta \phi} + (1-3 \beta H \dot{\phi}) \delta \phi - \beta \dot{\phi}^{2} \Phi \bigg],
\label{eq:v_phi}
\end{eqnarray}

%\begin{equation}
%\delta P_{\phi} = \beta \dot{\phi}^{2} \ddot{\delta \phi} + (1+2 \beta \ddot{\phi}) \dot{\phi} \dot{\delta \phi} - (1+4 \beta \ddot{\phi}) \dot{\phi}^{2} \Phi - \beta \dot{\phi}^{3} \dot{\Phi} - V_{\phi} \delta \phi, 
%\label{eq:del_p_phi}
%\end{equation}

%\noindent
%and

%\begin{equation}
%a (\bar{\rho_{\phi}} + \bar{P_{\phi}}) v_{\phi} = \dot{\phi} \bigg[ \beta \dot{\phi} \dot{\delta \phi} + (1-3 \beta H \dot{\phi}) \delta \phi - \beta \dot{\phi}^{2} \Phi \bigg],
%\label{eq:v_phi}
%\end{equation}

\noindent
where $ \delta \phi $ is the first order perturbation to the background field $ \phi $ and $ \beta = \frac{\alpha}{M_{\rm pl}^{3}} $.
\\
\noindent
Now putting Eq.~\eqref{eq:del_p_phi} into Eq.~\eqref{eq:ein_per_3} we get evolution equation for the gravitational potential $ \Phi $. By varying the action \eqref{eq:action} we can calculate the Euler-Lagrangian equation order by order and in the first order perturbation we get evolution equation for the $ \delta \phi $. 
\\
\noindent
We now introduce following dimensionless quantities \citep{Ali:2012cv,Bikash:2016ica,Hossain:2012qm,Scherrer:2007pu}

\begin{eqnarray}
x &=& \frac{\Big{(} \dfrac{d \phi}{d N} \Big{)}}{\sqrt{6} M_{Pl}}, \hspace{1 cm}  y = \frac{\sqrt{V}}{\sqrt{3} H M_{Pl}}, \nonumber\\
\lambda &=& - M_{Pl} \frac{V_{\phi}}{V}, \hspace{1 cm}  \Gamma = V \frac{V_{\phi \phi}}{V_{\phi}^{2}}, \nonumber\\
\epsilon &=& -6 \beta H^{2} \Big{(} \dfrac{d \phi}{d N} \Big{)}, \hspace{1 cm} q=(\delta \phi)/\Big{(} \dfrac{d \phi}{d N} \Big{)}.
\label{eq:dimless_var}
\end{eqnarray}

\noindent
where $N=\ln(a)$ is the number of e-foldings. Using these dimensionless quantities, we form the following autonomous system of equations \citep{Ali:2012cv,Hossain:2012qm,Scherrer:2007pu}:

\begin{eqnarray}
\dfrac{d x}{d N} &=& \frac{3 x^3 \left(2+5 \epsilon +\epsilon^2\right)-3 x \left(2-\epsilon +y^2 (2+3 \epsilon )\right)+2 \sqrt{6} y^2 \lambda -\sqrt{6} x^2 y^2 \epsilon  \lambda }{4+4 \epsilon +x^2 \epsilon^2} \nonumber\\
\dfrac{d y}{d N} &=& -\frac{y \left(12 \left(-1+y^2\right) (1+\epsilon )-6 x^2 \left(2+4 \epsilon +\epsilon^2\right)+\sqrt{6} x^3 \epsilon^2 \lambda +2 \sqrt{6} x \left(2+\left(2+y^2\right) \epsilon \right) \lambda \right)}{8+8 \epsilon +2 x^2 \epsilon^2}, \nonumber\\
\dfrac{d \epsilon}{d N} &=& -\frac{\epsilon  \left(-3 x \left(-3+y^2\right) (2+\epsilon )+3 x^3 \left(2+3 \epsilon +\epsilon^2\right)-2 \sqrt{6} y^2 \lambda -\sqrt{6} x^2 y^2 \epsilon  \lambda \right)}{x \left(4+4 \epsilon +x^2 \epsilon^2\right)}, \nonumber\\
\dfrac{d \lambda}{d N} &=& \sqrt{6}x\lambda^2(1-\Gamma),\nonumber\\
\dfrac{d \mathcal{H}}{d N} &=& -\frac{1}{2} (1 + 3 w_{\phi} \Omega_{\phi}) \mathcal{H}, \nonumber\\
\dfrac{d \Phi}{d N} &=& \Phi_{1}, \nonumber\\
\dfrac{d q}{d N} &=& q_{1}, \nonumber\\
\dfrac{d \Phi_{1}}{d N} &=& A_{2}^{-1} [ x^2 (\epsilon  (4 \epsilon ^2 (-2 (J-3) x^2+L-3)+4 \epsilon  (-4 J+L+6 x^2-6)+L x^2 \epsilon ^3-48) \nonumber\\
&& -12 Q^2 (\epsilon  (\epsilon  (x^2 (2 \epsilon +3)+4)+8)+4)) ] \Phi \nonumber\\
&& - A_{1}^{-1} [ 2 (\epsilon +1) \left(A_4 x^2 \epsilon -2 A_3\right) ] q \nonumber\\
&& - A_{2}^{-1} [ 2 x^4 \epsilon ^2 \left(\epsilon  (J+2 \epsilon )+3 Q^2-3\right)+2 x^2 \left(\epsilon ^2 (8 J+10 \epsilon -11)+4 (J-6) \epsilon +12 Q^2 (\epsilon +1)^2-12\right)+40 (\epsilon +1)^2 ] \Phi_{1} \nonumber\\
&& + A_{2}^{-1} [ 2 x^2 \left(\epsilon  \left(2 J \left(\epsilon  \left(x^2 \epsilon -2\right)-4\right)+3 \epsilon  \left(x^2 \left(Q^2 (3 \epsilon +4)-2 (\epsilon +1)\right)+3 \epsilon +20\right)+84\right)+24\right) ] q_{1}, \nonumber\\
\dfrac{d q_{1}}{d N } &=& A_{2}^{-1} [ 8 J \left(\epsilon  \left(3 x^2 \epsilon +8\right)+4\right)-2 x^2 \epsilon ^3 \left(L+\left(6 Q^2-3\right) x^2+3\right)-8 \epsilon ^2 \left(L+3 \left(Q^2+2\right) x^2\right)-8 \epsilon  \left(L+3 x^2+9\right) ] \Phi \nonumber\\
&& + A_{1}^{-1} [ 2 A_3 \epsilon +4 A_4 (\epsilon +1) ] q \nonumber\\
&& + A_{2}^{-1} [ \epsilon  \left(16 J+\epsilon  \left(2 x^2 \left(-6 Q^2+7 \epsilon +16\right)+x^4 \epsilon ^2+28\right)+56\right)+64 ] \Phi_{1} \nonumber\\
&& + A_{2}^{-1} [ 2 J \left(\epsilon  \left(x^2 \left(\epsilon  \left(x^2 \epsilon -8\right)+4\right)-24\right)-16\right)-3 x^4 \epsilon ^2 \left(-2 Q^2 (3 \epsilon +1)+\epsilon  (\epsilon +6)+2\right) \nonumber\\
&& + 6 x^2 \left(Q^2 \left(6 \epsilon ^2+8 \epsilon +4\right)+\epsilon  \left(-\epsilon ^2+\epsilon -8\right)-4\right)-12 ((\epsilon -4) \epsilon -2) ] q_{1},
\label{eq:dynsys}
\end{eqnarray}

\noindent
where, $ L = \frac{k^{2}}{3 \mathcal{H}^{2}} $ and

\begin{eqnarray}
Q &=& \frac{y}{x} \nonumber\\
J &=& \sqrt{\frac{3}{2}} \lambda \frac{y^{2}}{x} \nonumber\\
\omega_{\phi} &=& \frac{p_{\phi}}{\rho_{\phi}} = \frac{\epsilon  (3 (\epsilon +8)-4 J)-12 Q^2 (\epsilon +1)+12}{3 \left(Q^2+\epsilon +1\right) \left(\epsilon  \left(x^2 \epsilon +4\right)+4\right)} \nonumber\\
\Omega_{\phi} &=& x^2 \left(Q^2+\epsilon +1\right) \nonumber\\
A_{1} &=& 4 + \epsilon (4 + x^{2} \epsilon) \nonumber\\
A_{2} &=& A_{1}^{2} \nonumber\\
A_{3} &=& - Q^{-2} A_{1}^{-3} x^{2} [Q^2 ( 4 J^2 \epsilon  (\epsilon  (x^6 \epsilon ^3+4 x^4 \epsilon  (\epsilon +1)-4 x^2 (7 \epsilon +6)+8 )+16 ) \nonumber\\
&& + 6 J (\epsilon  (-x^6 \epsilon ^3 (5 \epsilon +4)+x^4 \epsilon  (\epsilon  ((\epsilon -24) \epsilon -40)-16)+16 x^2 (\epsilon +1) (2 \epsilon  (\epsilon +6)+5)-8 (\epsilon  (\epsilon +16)+26) )-64 ) \nonumber\\
&& + 9 (x^6 \epsilon ^3 ( 3 \epsilon  (\epsilon +2)^2+4 )+x^4 \epsilon  (\epsilon  (\epsilon  (\epsilon  (23 \epsilon +112)+156)+80)+16)-x^2 (\epsilon  (\epsilon  (\epsilon  (\epsilon  (9 \epsilon +94)+380)+480)+208)+32) \nonumber\\
&& - 2 \epsilon ^3 (3 \epsilon +26)+96 \epsilon +32 ) )+2 \Gamma  J^2 \epsilon  (x^2 \epsilon -2 ) (\epsilon  (x^2 \epsilon +4 )+4 )^2 \nonumber\\
&& +3 Q^4 x^2 (\epsilon  (8 J (\epsilon  (x^2 (\epsilon +1) (\epsilon  (x^2 \epsilon +8 )+4 )-2 (7 \epsilon +12) )-8 ) \nonumber\\
&& - 3 x^4 \epsilon ^2 (3 \epsilon  (\epsilon +2) (\epsilon +3)+8)-6 x^2 (\epsilon  (\epsilon  (\epsilon  (15 \epsilon +88)+132)+72)+16)+12 (\epsilon  (\epsilon  (26-3 \epsilon )+60)+36) )+96 ) \nonumber\\
&& + 9 Q^6 x^4 \epsilon  (\epsilon  (\epsilon  (x^2 (3 \epsilon  (\epsilon +2)+4)+42 \epsilon +92 )+64 )+16 )], \nonumber\\
\end{eqnarray}

\begin{eqnarray}
A_{4} &=& Q^{-2} (1+\epsilon)^{-1} A_{1}^{-3} [-2 J^2 x^2 \epsilon  (2 Q^2 (\epsilon  (x^2 (\epsilon  (\epsilon +2) (x^2 \epsilon +8 )+8 )-44 \epsilon -80 )-32 ) \nonumber\\
&& + \Gamma  (3 \epsilon +2) (\epsilon  (x^2 \epsilon +4)+4)^2)-4 J Q^2 (x^4 \epsilon  (\epsilon  (\epsilon  (\epsilon  ((L+21) \epsilon -45)-192)-168)+6 Q^2 (\epsilon  (\epsilon  (13 \epsilon +34)+28)+8)-48 ) \nonumber\\
&& + 2 x^2 (\epsilon  (\epsilon  (\epsilon  (4 L (\epsilon +1)+75 \epsilon +390)+612)+360)-24 Q^2 (2 \epsilon +1) (\epsilon +1)^2+72 )+16 (\epsilon +1)^2 ((L+6) \epsilon +3) \nonumber\\
&& + 3 x^6 \epsilon ^3 (Q^2 (\epsilon +1) (\epsilon +4)+\epsilon  ((\epsilon -1) \epsilon -7)-4 ) )+Q^2 (9 (x^2 (16 Q^2 (3 \epsilon ^2+\epsilon +2 ) (\epsilon +1)^2 \nonumber\\
&& + \epsilon  (\epsilon  (\epsilon  (3 \epsilon  (\epsilon +16)+284)+456)+240)+32 )+x^6 \epsilon ^2 (Q^4 (3 \epsilon ^3-12 \epsilon -8 )+Q^2 (3 \epsilon +2) (\epsilon  (\epsilon  (\epsilon +3)+8)+8) \nonumber\\
&& - 2 (\epsilon +1) (\epsilon  (\epsilon  (2 \epsilon +7)+10)+4) )-x^4 (16 Q^4 (\epsilon +1)^2 (\epsilon  (3 \epsilon +4)+2) \nonumber\\
&& -2 Q^2 (\epsilon  (\epsilon  (\epsilon  (\epsilon  (27 \epsilon +184)+384)+352)+160)+32) \nonumber\\
&& + \epsilon  (\epsilon  (\epsilon  (\epsilon  (\epsilon +7) (3 \epsilon +50)+624)+496)+192)+32 )+48 \epsilon  (\epsilon +1)^2 ) \nonumber\\
&& - L (\epsilon  (x^2 \epsilon +4 )+4 )^2 (\epsilon  (\epsilon  (x^2 (-3 Q^2+2 \epsilon +6 )+5 )+8 )+12 ) )].
\end{eqnarray}

\noindent
Note that for simplicity of the notations, in the above set of equations, we have kept the same notations for $\Phi$ and $q$ in the Fourier space corresponding to the same quantities in the real space.

\noindent
By putting Eq.~\eqref{eq:del_rho_phi} into Eq.~\eqref{eq:ein_per_1} and going to the Fourier space we get the matter density contrast given by

\begin{eqnarray}
\delta_{m} &=& -\frac{1}{\Omega_{m}} \bigg[(2-x^2\epsilon) \frac{d \Phi}{d N} + 2 \Big{(} 1+L-x^2(1+2\epsilon) \Big{)} \Phi \nonumber\\
&+& x^2(2+3\epsilon) \frac{d q}{d N} + x^2 \Big{(} (2+3 \epsilon)A - 2 J + L \epsilon \Big{)} q \bigg] \, .
\label{eq:delm}
\end{eqnarray}

\noindent
Similarly, by putting Eq.~\eqref{eq:v_phi} into Eq.~\eqref{eq:ein_per_2} and going to the Fourier space we get the pecular velocity for the matter given by

\begin{equation}
y_{m} = 3 \mathcal{H} v_{m} = \frac{1}{\Omega_{m}} \bigg[2 \frac{d \Phi}{d N} + (2-x^2\epsilon) \Phi + x^2\epsilon \frac{d q}{d N} -x^2 \Big{(} 6+\epsilon (3-A) \Big{)} q \bigg],
\label{eq:ym}
\end{equation}

\noindent
where

\begin{equation}
A = \frac{\big{(} \frac{d^2 \phi}{d N^2} \big{)}}{\big{(} \frac{d \phi}{d N} \big{)}} = \frac{-3 B \epsilon -2 B+2 J+6 \epsilon }{2 (\epsilon +1)},
\end{equation}

\noindent
with $ B = 1.5(1-\omega_{\phi} \Omega_{\phi}) $. Now we can calculate comoving matter energy density contrast from Eqs.~\eqref{eq:delm} and \eqref{eq:ym} by using the definition $ \Delta_{m} = \delta_{m} + y_{m} $.

%%%%%%%%%%%%%%%%%%%%%%%%%%%%%%%%%%
\begin{center}
\begin{figure*}
\begin{tabular}{c@{\quad}c}
\epsfig{file=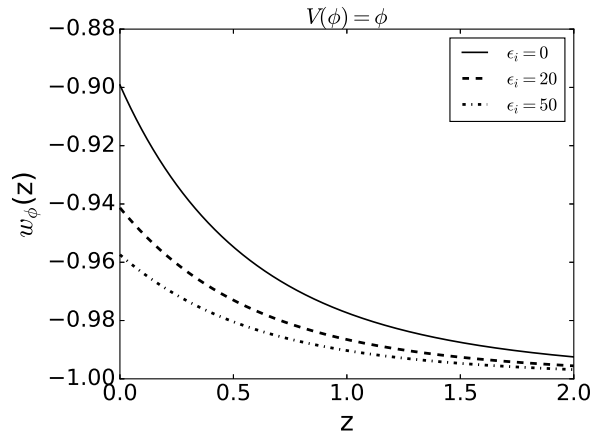,width=8.5 cm}
\end{tabular}
\caption{Behaviour of the Equation of state for the Galileon field $w_{\phi}$ as a function of redshift for different $ \epsilon_{i} $ with linear potential with $\Omega_{m0} = 0.28$.
}
\label{fig:w_phi}
\end{figure*}
\end{center}
%%%%%%%%%%%%%%%%%%%%%%%%%%%%%%%%%%%

\subsection{Initial conditions}  
  
To solve the autonomous system~\eqref{eq:dynsys}, we need initial conditions for the background quantities ($x, y, \epsilon, \lambda, \mathcal{H}$) as well as for the perturbed quantities ($\Phi, \dfrac{d\Phi}{dN}, q, \dfrac{d q}{dN} $). We fix the initial condition at $z=1000$ in early matter dominated era where the dark energy contribution is negligible. In this work, we focus on the thawing class of the Galileon models where the Galileon field $ \phi $ is initially frozen at $ w_{\phi} \sim -1 $ in early matter dominated era due to large Hubble friction. The condition $w_{\phi} \sim -1$ automatically transformed to the condition $x_{i} \sim 0$ through Eq.~\eqref{eq:dimless_var}. So, we fix $ x_{i} = 10^{-8}$. The solutions of the system of evolution equations Eq.~\eqref{eq:dynsys} are not 
sensitive to the initial value of $ x $ as long $x_{i} \ll 1$. Since the dark energy density parameter $ \Omega_{\phi} $ is related to $ x $ and $ y $ (see  Eq. (15)), we can relate the initial condition in $ y $ to the boundary condition in $ \Omega_{\phi} $. So, we fix $ y_{i} $ in such a way that the value of $ \Omega_{\phi} $ at present becomes $ 0.72 $. The initial slope of the potential is determined by the initial value of $ \lambda $. $ \lambda_{i} $ determines the evolution of the equation of state (EoS) of the Galileon field. For $ \lambda_{i} \ll 1 $, the EoS of the Galileon field does not deviate much from its initial frozen value $ -1 $ and always stays very close to the cosmological constant behaviour. For large value of $ \lambda_{i} $, the Galileon field thaws away sufficiently from its initial frozen state and can have sufficient deviations from the cosmological constant behviour. For all the models we fix $ \lambda_{i} = 0.7 $. Next, the initial condition $ \mathcal{H}_{i} $ is taken such a way that at present $ {\mathcal{H}_{0}} = H_{0} = 100 h~ \rm km/s/Mpc$  with $ h = 0.7 $. The initial condition for $ \epsilon_{i} $ remains as a parameter (note that this parameter is related to the parameter $\alpha$ in the action (1) and represents the contribution from Galileon term).
\\
\noindent
Initially at redshift $ z = 1000 $ there is hardly any contribution from the Galileon field in the evolution equations. So, we set $ q_{i} =0 $ and $ q_{1}|_{i} = \dfrac{d q}{dN} \Big{|}_{i} = 0 $. Next, we can find the initial condition for $ \dfrac{d \Phi}{dN} $ using the fact that during the matter dominated era $ \Phi $ is constant i.e. $ \Phi_{1}|_{i} = \dfrac{d \Phi}{dN} \Big{|}_{i} = 0 $. Also during matter dominated era, one can find that $\Delta_{m} \sim a$ and using the Poisson equation, we get the initial condition in $ \Phi $ which is given by

\begin{equation}
\Phi_{i} = - \frac{3}{2} \frac{\mathcal{H}^{2}_{i}}{k^2} a_{i}.
\end{equation}

%%%%%%%%%%%%%%%%%%%%%%%%%%%%%%%%%%
\begin{center}
\begin{figure*}
\begin{tabular}{c@{\quad}c}
\epsfig{file=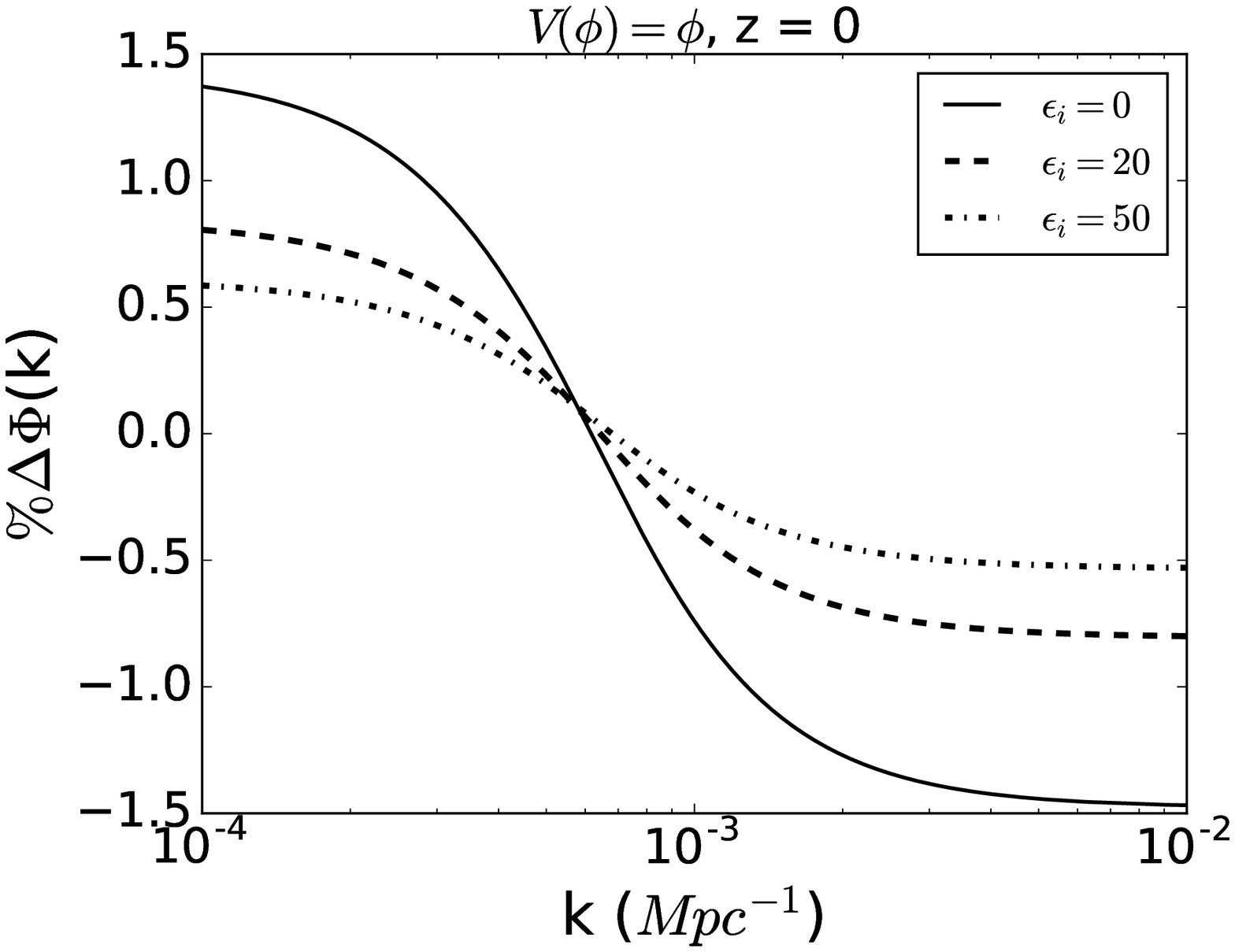,width=7.5 cm}
\epsfig{file=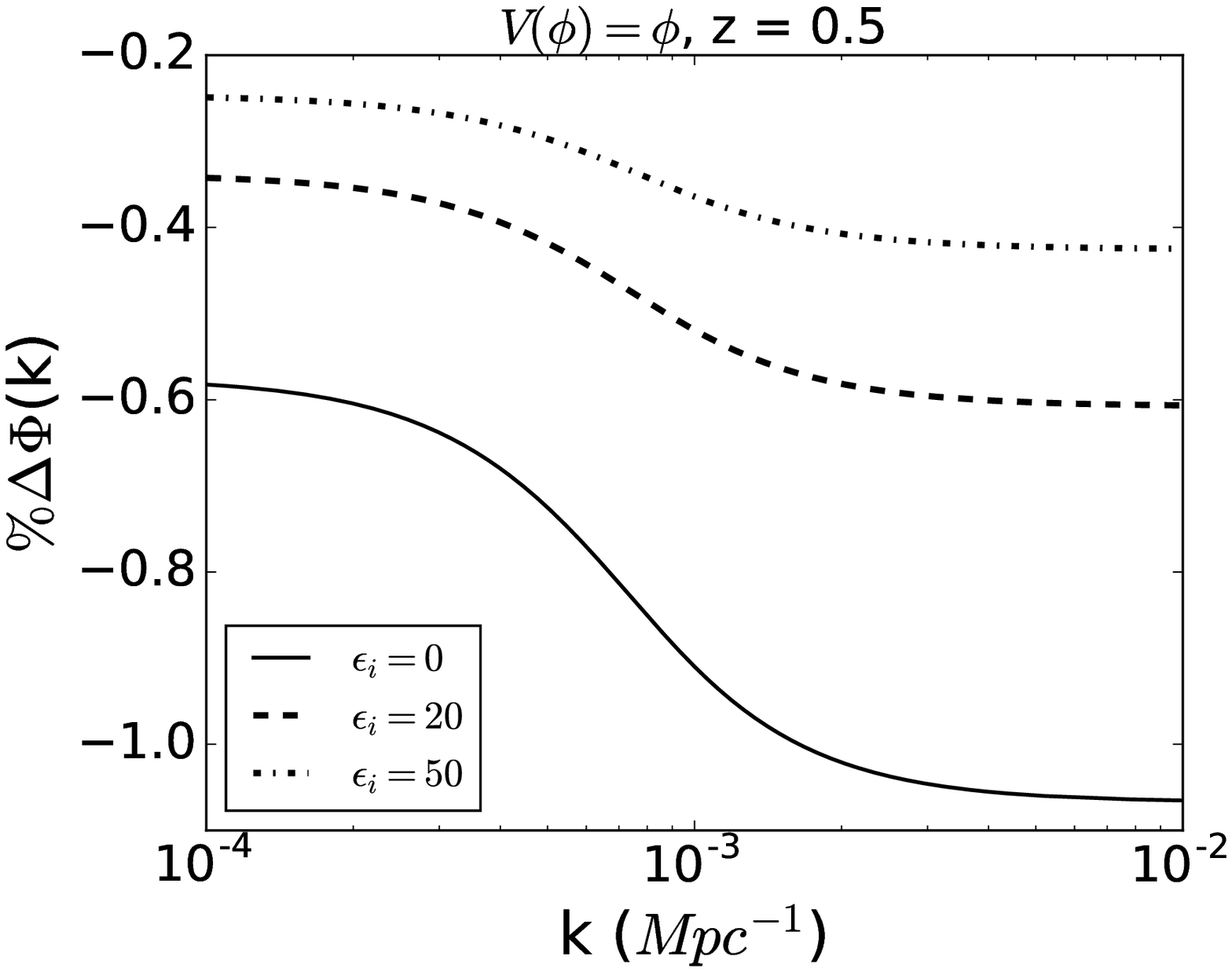,width=7.5 cm}\\
\epsfig{file=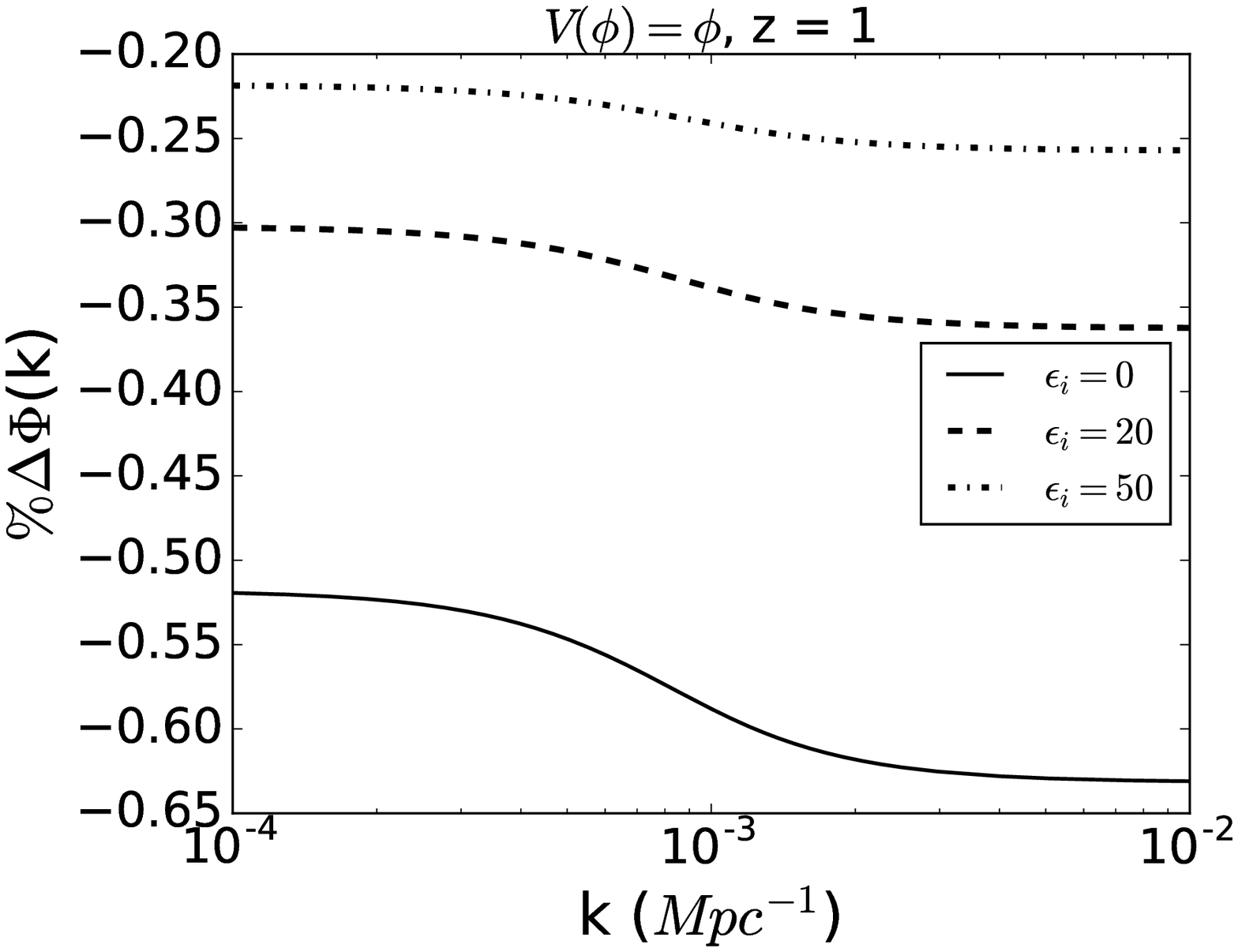,width=7.5 cm}
\epsfig{file=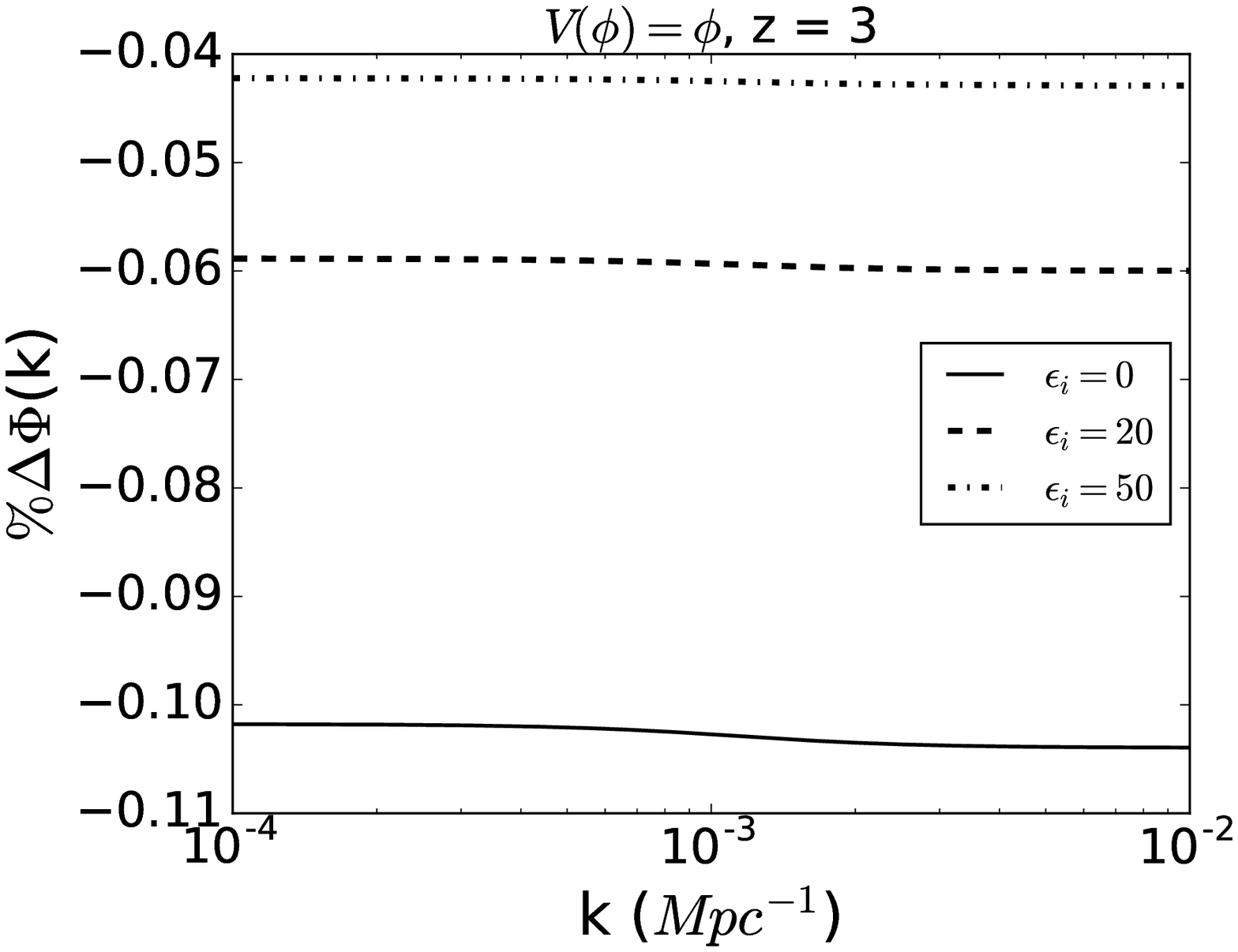,width=7.5 cm}
\end{tabular}
\caption{Percentage deviation in $\Phi$ from $ \Lambda $CDM model for different $ \epsilon_{i} $ with linear potential.
}
\label{fig:delphi}
\end{figure*}
\end{center}
%%%%%%%%%%%%%%%%%%%%%%%%%%%%%%%%%%%

%%%%%%%%%%%%%%%%%%%%%%%%%%%%%%%%%%
\begin{center}
\begin{figure*}
\begin{tabular}{c@{\quad}c}
\epsfig{file=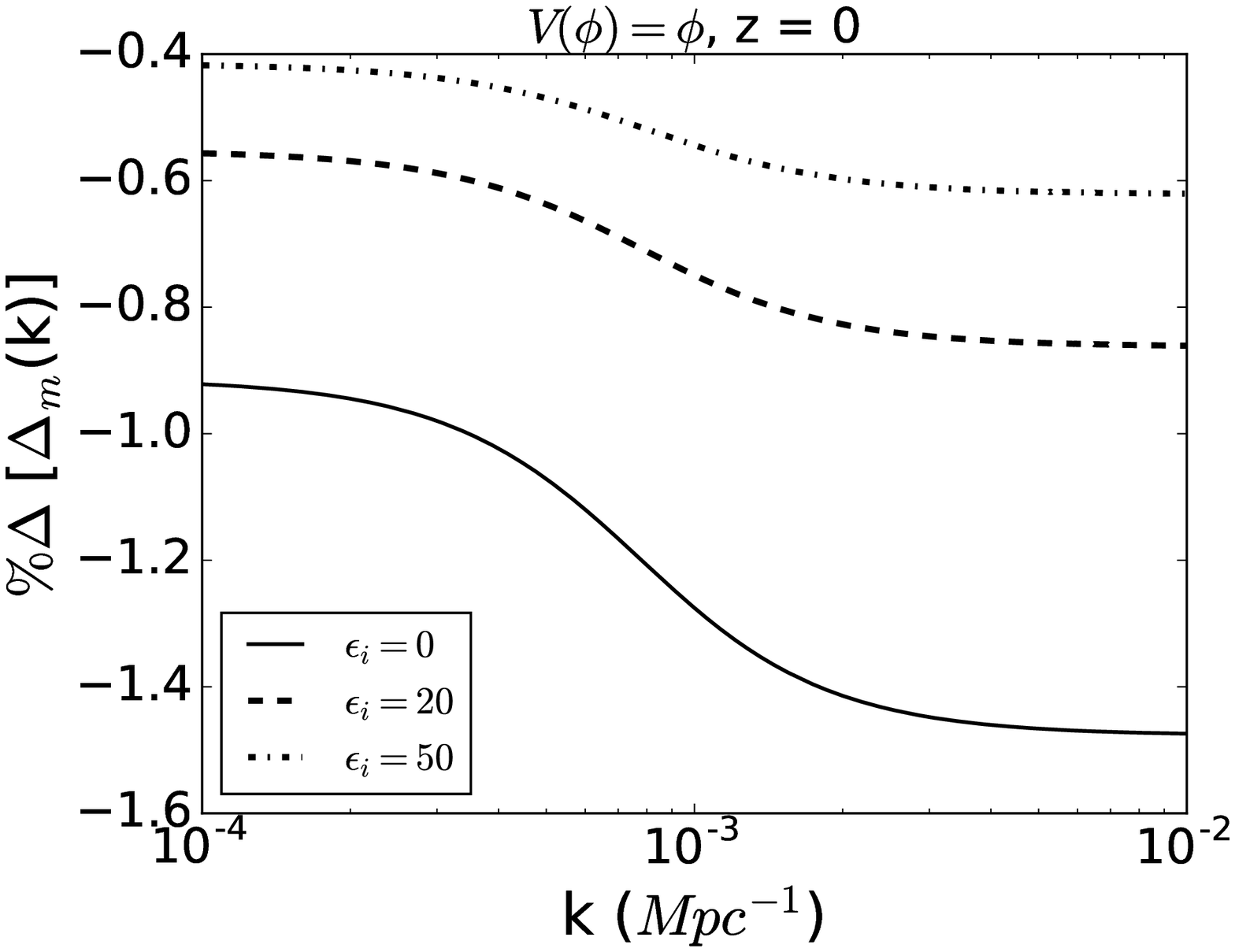,width=7.5 cm}
\epsfig{file=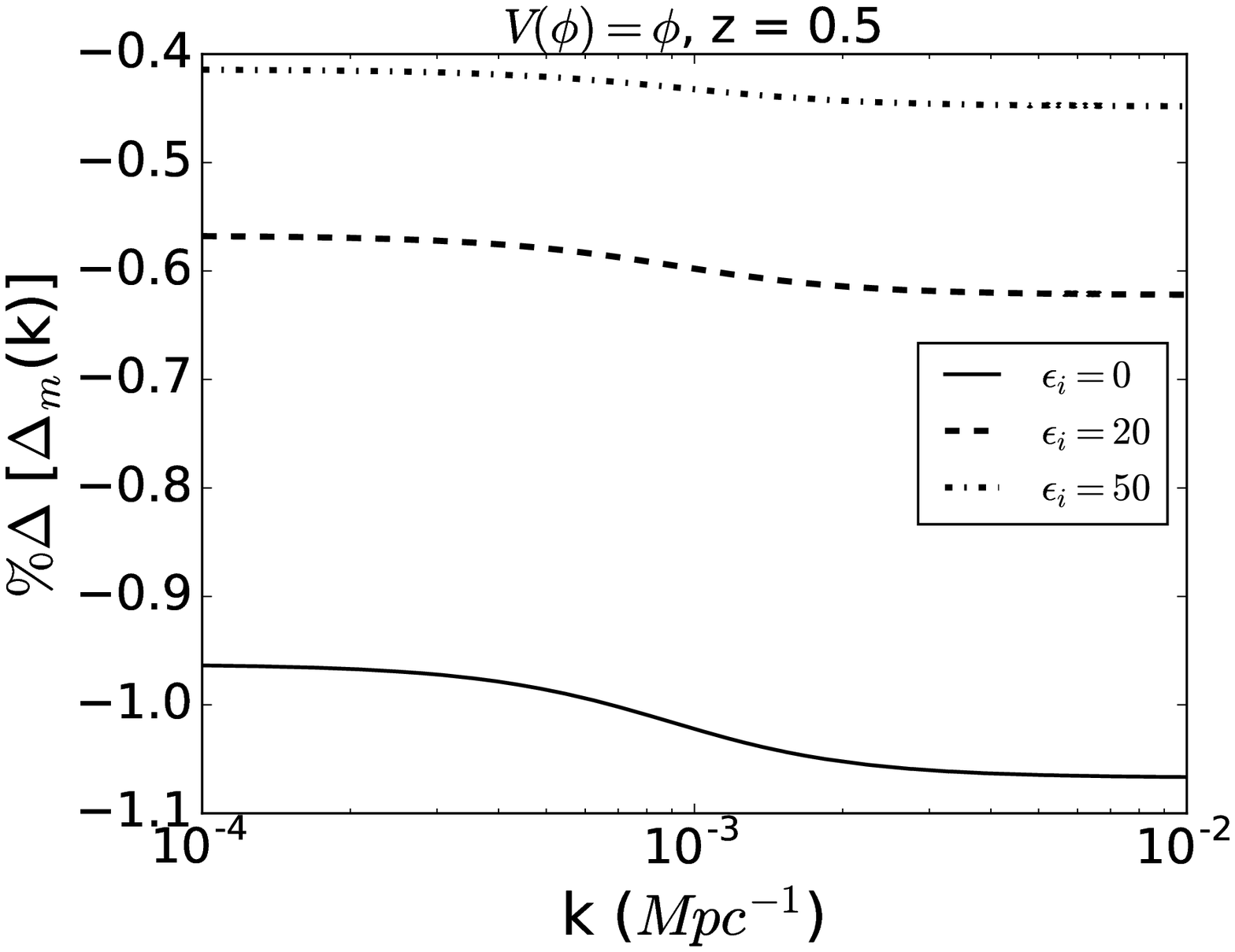,width=7.5 cm}\\
\epsfig{file=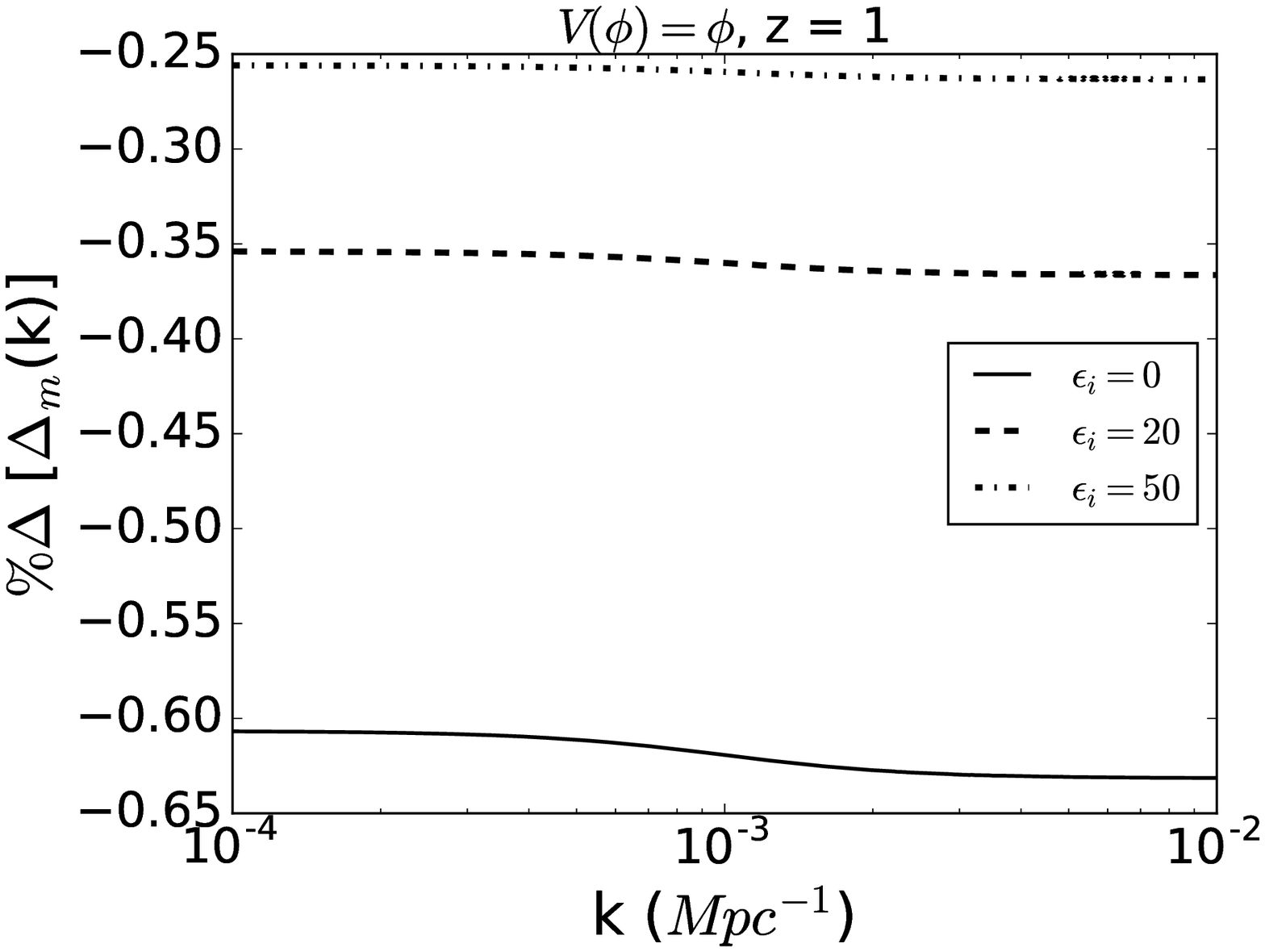,width=7.5 cm}
\epsfig{file=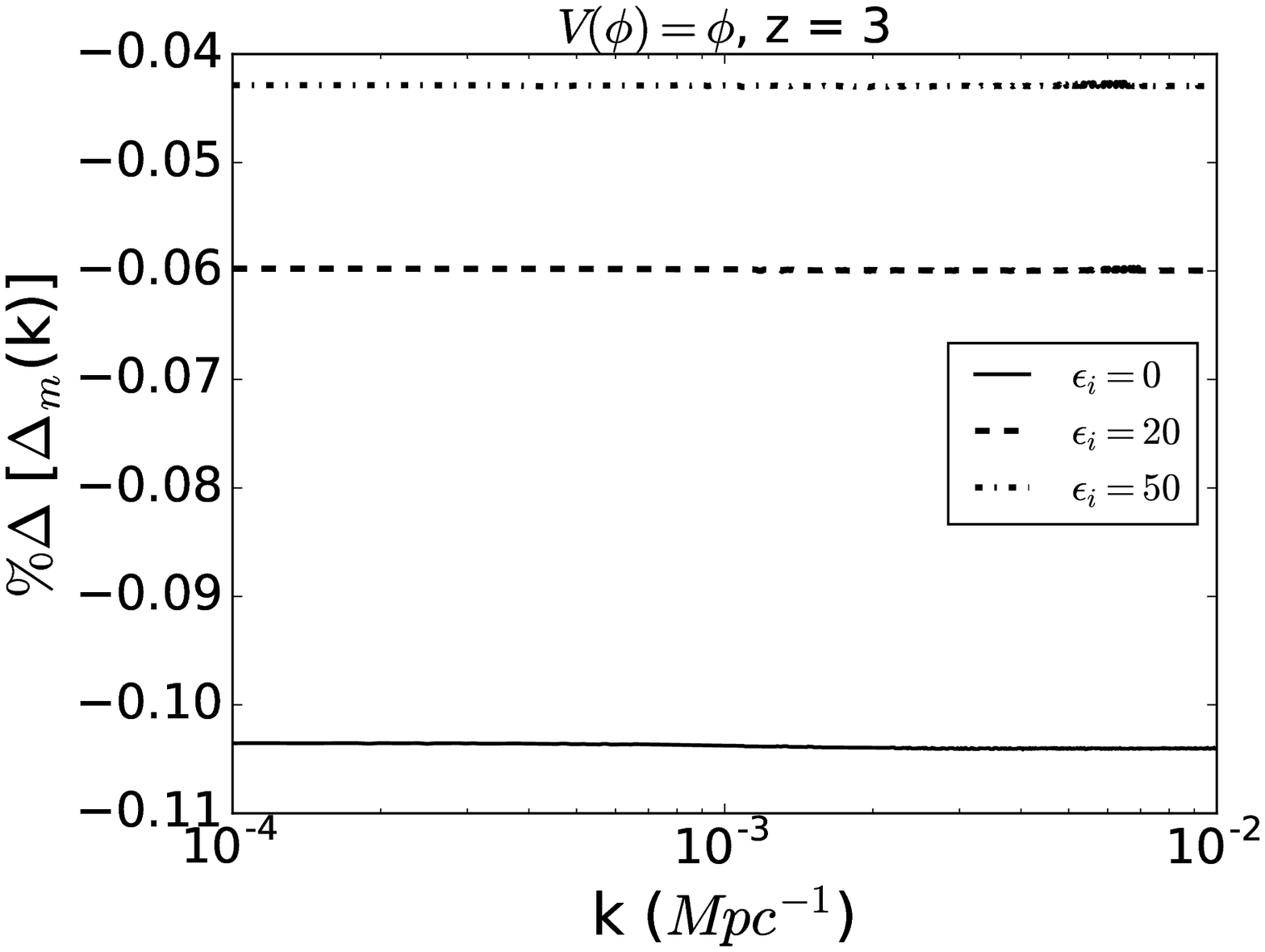,width=7.5 cm}
\end{tabular}
\caption{Percentage deviation in comoving density contrast $ \Delta_{m} $ from $ \Lambda $CDM model for different $ \epsilon_{i} $ with linear potential.
}
\label{fig:deldelm}
\end{figure*}
\end{center}
%%%%%%%%%%%%%%%%%%%%%%%%%%%%%%%%%%%

%%%%%%%%%%%%%%%%%%%%%%%%%%%%%%%%%%
\begin{center}
\begin{figure*}
\begin{tabular}{c@{\quad}c}
\epsfig{file=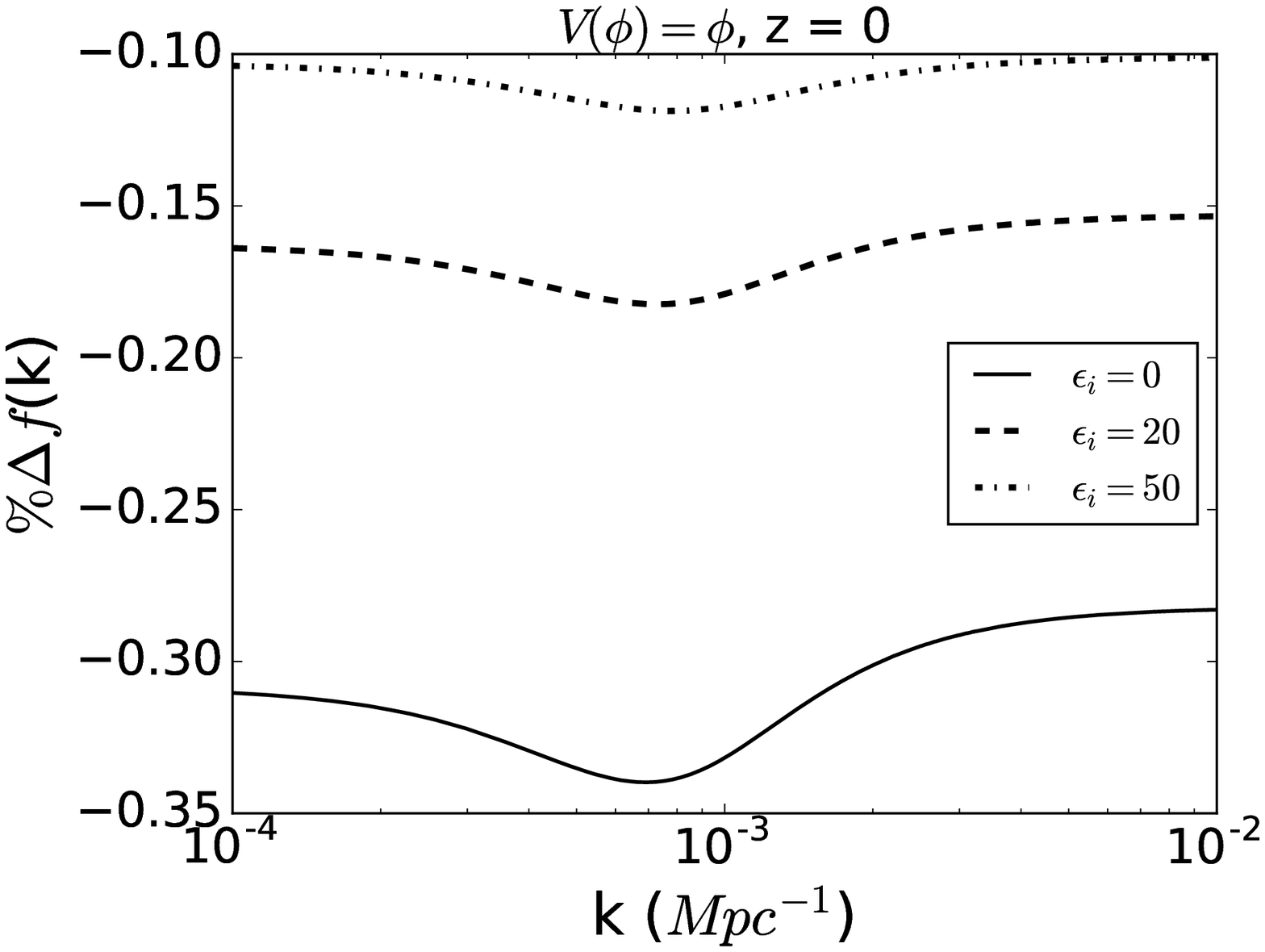,width=7.5 cm}
\epsfig{file=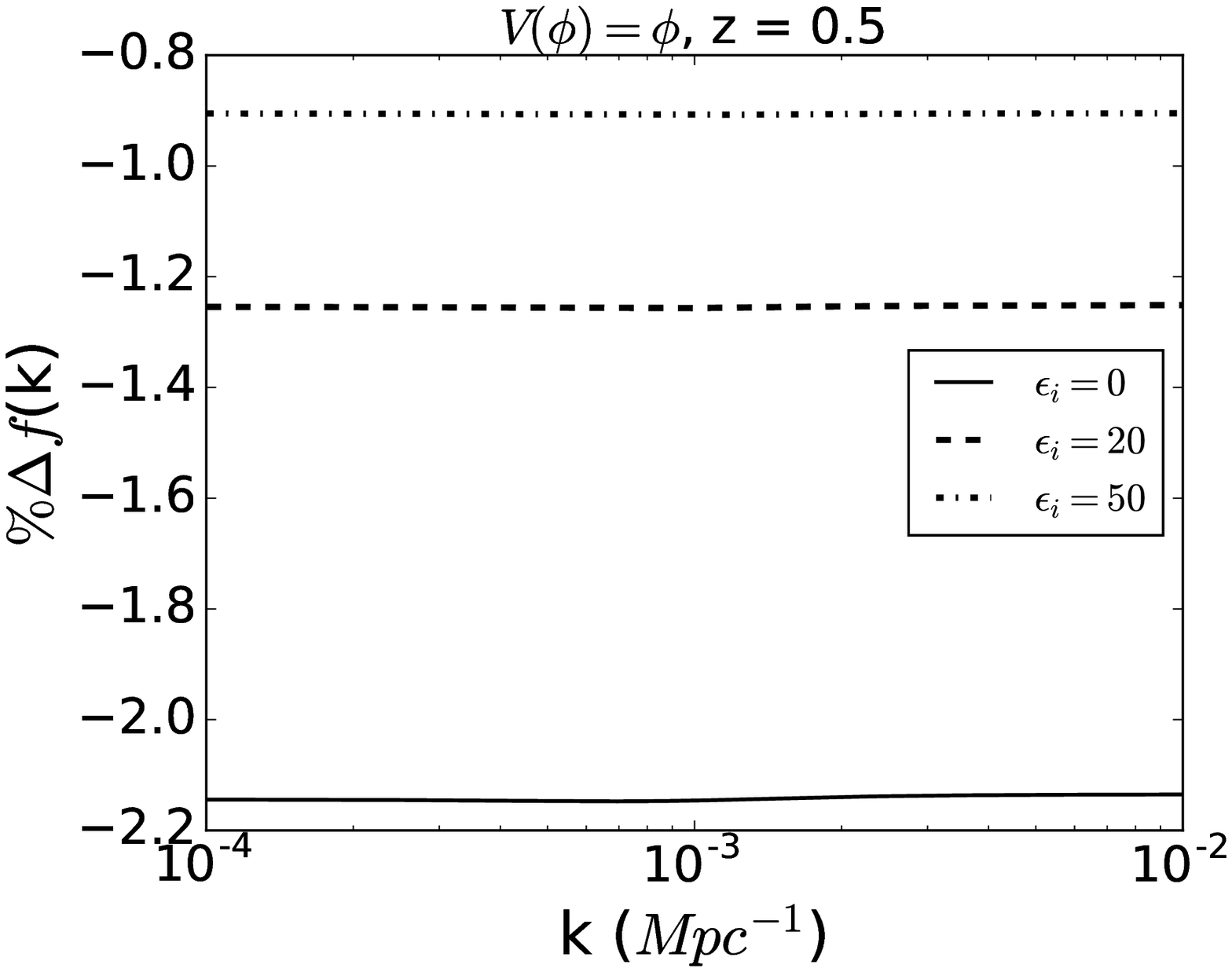,width=7.5 cm}\\
\epsfig{file=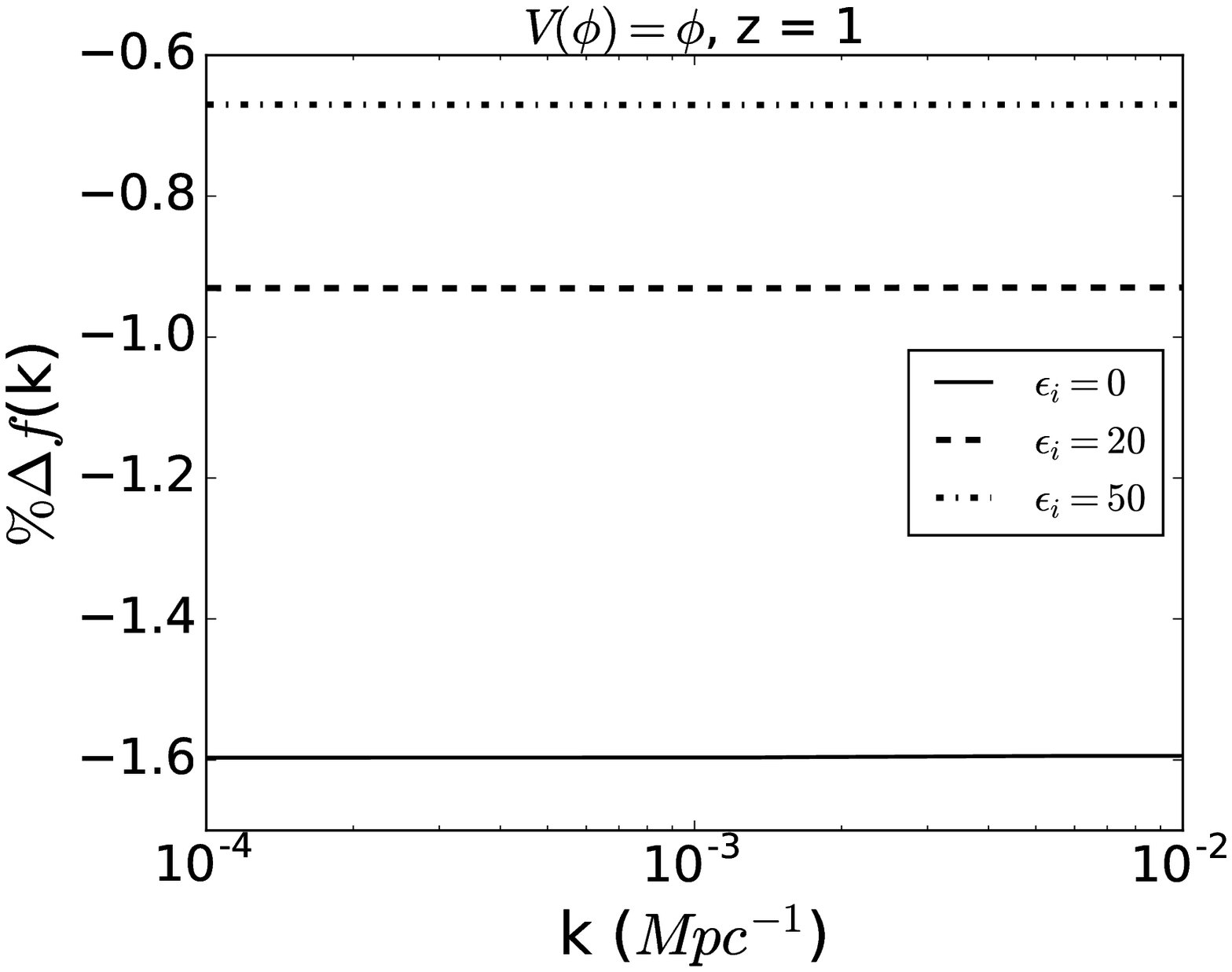,width=7.5 cm}
\epsfig{file=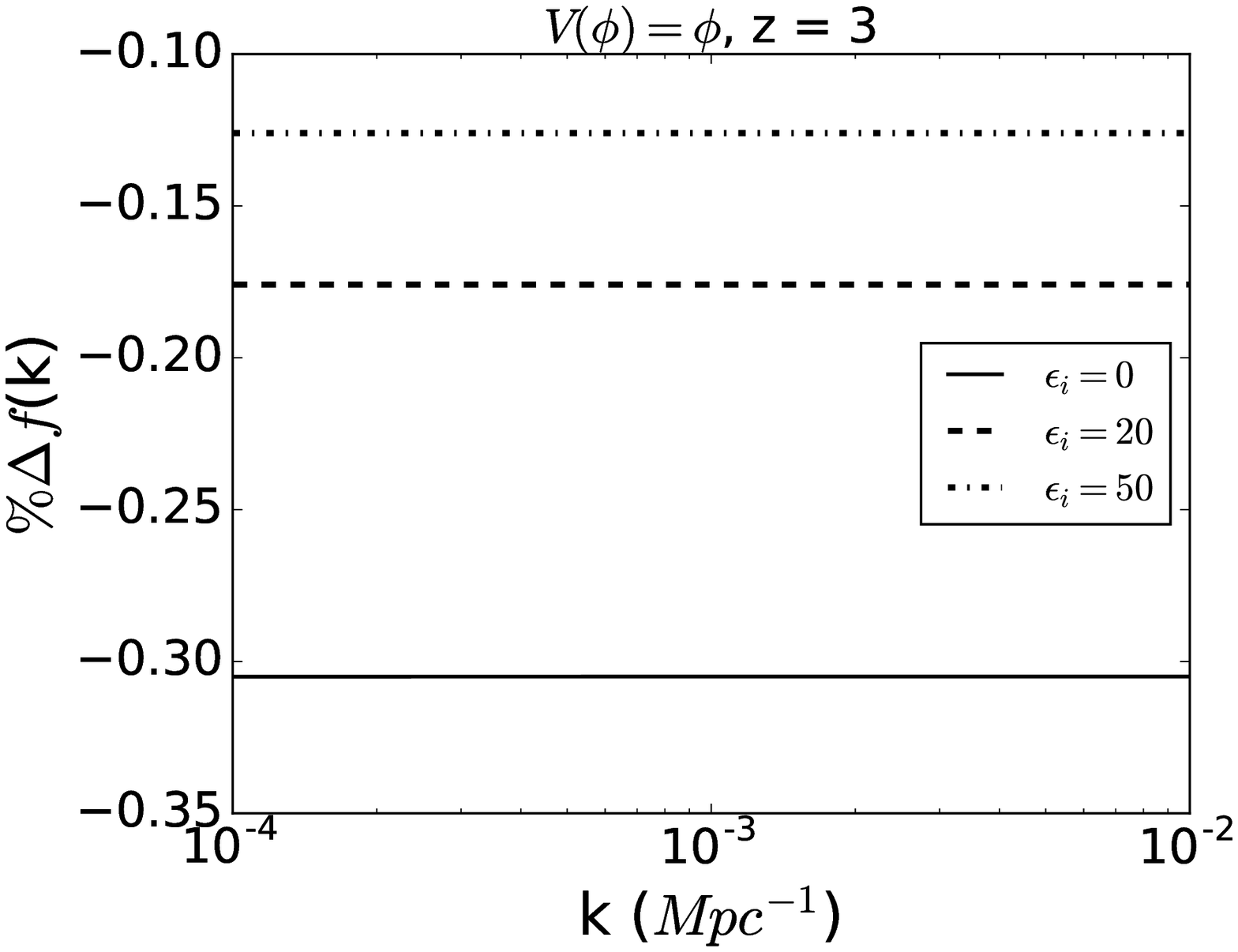,width=7.5 cm}
\end{tabular}
\caption{Percentage deviation in $f$ from $ \Lambda $CDM model for different $ \epsilon_{i} $ with linear potential.
}
\label{fig:delf}
\end{figure*}
\end{center}
%%%%%%%%%%%%%%%%%%%%%%%%%%%%%%%%%%%

\subsection{Behaviour of different cosmological parameters}

By using the above mentioned initial conditions we solve the system of autonomous equations given in Eq.~\eqref{eq:dynsys} for three different initial conditions ($ \epsilon_{i} = 0 $, $ 20 $ and $ 50 $) with linear potential and study various cosmological parameters. In all subsequent sections, we study these three cases except for the last figure where two other polynomial potentials (squared and inverse-squared) have been introduced to see the differences in different potentials.

In Fig.~\ref{fig:w_phi}, we show the evolution of the EoS for these three cases. As we consider thawing class of Galileon models, the EoS of all three cases initially starts from nearly $ -1 $ and slowly increases towards higher values at late times. At present ($z=0$), the EoS of the models $ \epsilon_{i} = 0 $, $ 20 $ and $ 50 $ reaches to the values nearly $ -0.9 $, $ -0.94 $ and $ -0.96 $ respectively. It shows that the models with higher $ \epsilon_{i} $ values deviate lesser from $ \Lambda $CDM behaviour. So with similar initial conditions, Galileon models are closer to $\Lambda$CDM  than the standard quintessence models.

In Fig.~\ref{fig:delphi}, we study the deviations in the gravitational potential $ \Phi $ for all three different initial conditions and compare them with the $ \Lambda $CDM model. In this plot and in all  subsequent plots, we define $ \% \Delta X = (\frac{X_{\rm de}}{X_{\Lambda \rm CDM}}-1) \times 100 $ for any quantity $ X $. At lower redshifts the deviations are enhanced on larger scale whereas the deviations are suppressed on smaller scales. At higher redshift the deviations are always suppressed and the suppression decreases with increasing redshift. The differences in the deviations between larger and smaller scales decrease with increasing redshift which means the scale dependency of the deviations decreases with increasing redshift. This behaviour is not surprising because of the fact that the dark energy perturbation is only relevant on large scales and at lower redshifts. So, whatever deviation is present on smaller scales, is due the differences in the background expansion only. Similarly on higher redshifts, the effect of dark energy is negligible. As the matter perturbation is scale independent, on higher redshifts, the deviation from $\Lambda$CDM is also scale independent.

In Fig.~\ref{fig:deldelm}, we study the deviations in $ \Delta_{\rm m} $ for all three cases compared to the $ \Lambda $CDM. There is always suppression in the deviations for all the models compared to the $ \Lambda $CDM on all scales and for all redshifts. This suppression decreases with increasing redshifts. Another point is that the suppressions are always smaller on larger scales compared to smaller scales. And these differences between two scales decrease with increasing redshifts because of the same reason that the scale dependence comes only through the dark energy perturbation which plays an extra role only on large scales. 

Next, we introduce a quantity $f=-\frac{k^2 v_{\rm m}}{{\mathcal{H}}\Delta_{\rm m}}$ which is related to the velocity perturbation and gives rise to the redshift space distortion \citep{Duniya:2016ibg}. The reason to introduce this quantity is that it plays an important role to the observed galaxy power spectrum which is discussed in the next section. So, before going to the discussion of the observed galaxy power spectrum it is important to study the behaviour of $ f $. In Fig.~\ref{fig:delf}, we study the deviations in $ f $ for all the models compared to the $ \Lambda $CDM. The deviations are always suppressed and the suppressions are almost scale independent except at very low redshifts because of the same reason due to the dark energy perturbation discussed above. One interesting point to notice that the suppressions at first increase with increasing redshifts and is  maximum at redshift $ z \sim 0.5 $ and then decrease with increasing redshifts for $ z \gtrsim 0.5 $. We should stress that this maximum occurring  at $z\sim 0.5$ is due to our certain choice of parameters. For other choices, this redshift value where the maximum occurs will change, but there will always be a maximum deviation at some particular redshift.

\section{The observed galaxy power spectrum}

To describe the inhomogeneous Universe and its evolution, the main quantity of interest is the matter density perturbation whose evolution we study through cosmological perturbation theory (here linear perturbation theory using Eq.~\eqref{eq:delm}). However we can not directly measure the matter perturbation. We actually observe the tracers of this matter inhomogeneity such as galaxies. By studying the distribution of the galaxies in the Universe we can probe the underlying structure formation history of the Universe. Since the fluctuation in the galaxy number density is related to the matter density perturbation, we can study the underlying dark matter density fluctuation on different scales by observing various features in the galaxy distribution. As because dark energy also plays an important role to the structure formation, we can also use the galaxy distribution to distinguish different dark energy models or modified gravity models. 

Theoretically, the galaxy density contrast $ \delta_{\rm g} $ and the matter density contrast $ \delta_{\rm m} $ can be related by a simple relation $ \delta_{\rm g} = b \delta_{\rm m} $ by introducing a bias parameter $b$; however this relation is gauge dependent on super-Hubble or near super-Hubble scales. So, to have a physical bias we have to use comoving density perturbation in the bias relation. On large scale, the rest frame of the dark matter and galaxies coincide and in this frame we use the gauge independent relation as $ \Delta_{\rm g} (k,z) = b(z) \Delta_{\rm m} (k,z) $ by assuming a linear bias with Gaussian initial conditions and this relation is valid on all linear scales. However this $ \Delta_{\rm g} $ is  not an observable quantity on large scale because of some extra relativistic effects such as light cone and redshift effects \citep{Duniya:2016ibg,Duniya:2013eta,Duniya:2015nva,Challinor:2011bk}.

In late eighties, Kaiser \citep{Kaiser:1987qv} showed that we do not observe galaxy distribution in real space but in redshift space. In addition to the matter density perturbation, the peculiar velocities of the galaxies also affect the galaxy distribution in redshift space. This effect is known as the Kaiser redshift space distortion which is a measure of the large scale velocity fields. The Kaiser redshift space distortion term contains valuable information to the large scale structure formation.

In addition,  the gravitational potential in the metric can affect the photon geodesics by integration along the path. This effect is known as the magnification bias \citep{Moessner:1997qs} i.e. the observed galaxy distribution is also affected by the gravitational lensing. This gravitational lensing can allow us to detect the faint galaxies too through the magnification due to lensing effect.

In recent years, people have shown that on large scales in the observed galaxy distribution there are some other effects which are purely general relativistic. These effects are influenced by the gravitational potential, velocity fields and the matter density perturbations on the observed number density of galaxies on large scales \citep{Yoo:2009au,Bonvin:2011bg,Bonvin:2014owa,Challinor:2011bk,Jeong:2011as,Yoo:2012se,Bertacca:2012tp,Duniya:2016ibg,Duniya:2015dpa}. This effects are negligible in the sub-Hubble limit compared to the other effects like Kaiser redshift space distortion. Since we can not neglect these general relativistic effects on large scales, these effects can be important to probe dark energy perturbation as well as to distinguish different dark energy models.

By incorporating all the above mentioned effects, the galaxy number overdensity $\Delta^{\rm obs}$ (across the sky and at different redshifts and angles) can be written as \citep{Duniya:2016ibg,Duniya:2013eta,Duniya:2015nva,Challinor:2011bk}

\begin{equation}
\Delta^{\rm obs} = \left[{b + f \mu^2} + \mathcal{A} \left(\frac{\mathcal{H}}{k}\right)^2 +  i\mu\mathcal{B} \left(\frac{\mathcal{H}}{k}\right)\right]\Delta_{\rm m}\, ,
\label{eq:del_obs}
\end{equation}

\noindent
where $b$ is the bias parameter which relates the galaxy density contrast to the underlying dark matter density contrast, $f$ is the redshift space distortion parameter which is mentioned in the previous section, $\mu  = -\frac{\vec{n}.\vec{k}}{k}$ with $\vec{n}$ denotes the direction of the observation, $ \vec{k}$ denotes the wave vector whose magnitude is $k$. The parameters $\mathcal{A}$ and $\mathcal{B}$ are given by
 
\begin{equation}
\mathcal{A} = 3f + \left(\frac{k}{\mathcal{H}}\right)^2 \Big{[} 3 + \frac{\mathcal{H}'}{\mathcal{H}^{2}} + \frac{\Phi'}{\mathcal{H} \Phi} \Big{]} \frac{\Phi}{\Delta_{\rm m}},
\label{eq:curlA}
\end{equation}

\begin{equation}
\mathcal{B} = - \Big{[} 2 + \frac{\mathcal{H}'}{\mathcal{H}^{2}} \Big{]} f.
\end{equation}

\begin{figure*}
    \centering
    {
        \includegraphics[scale=0.55]{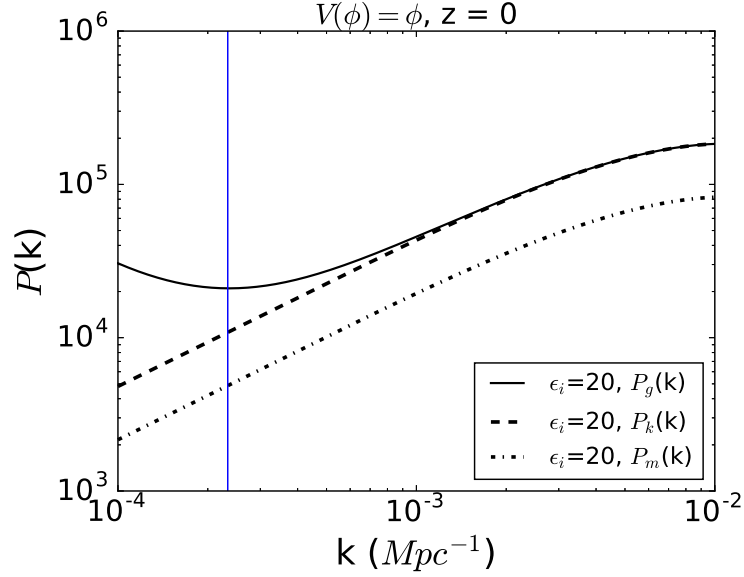}
    }
\caption{Dashed-dotted, dashed and continuous lines are for the usual matter power spectrum $P_{\rm m}$ (Eq.~\eqref{eq:ps1}), the galaxy power spectrum taking only Kaiser term $P_{\rm k}$ (Eq.~\eqref{eq:ps2}) and the full observed galaxy power spectrum $P_{\rm g}$ (Eq.~\eqref{eq:ps}) respectively for $ \epsilon_{i} = 20 $. The vertical blue line is the horizon scales at $z=0$.}
\label{fig:ps}
\end{figure*}

\noindent
Here we have assumed scale independent bias which is a valid assumption on large scales where we use linear perturbation theory. We have also assumed a constant comoving galaxy number density where galaxy evolution bias is absent and we have taken the unit magnification bias \citep{Duniya:2015nva}. We have neglected other terms like the time-delay, ISW and weak lensing integrated terms. For simplicity we put $ b = 1 $ throughout all the subsequent calculations. The right hand side (r.h.s) of the Eq.~\eqref{eq:del_obs} contains four terms. The first term is related to the galaxy bias, the second term is the Kaiser redshift space distortion term and other two terms are completely due to the general relativistic effects. The quantity ${\mathcal{A}}$ in the third term is related to the peculiar velocity fields and the gravitational potential. The quantity ${\mathcal{B}}$ in the fourth term is related to the Doppler effect.
\\
\noindent
Using definition of the power spectrum and Eq.~\eqref{eq:del_obs}, we can relate the matter power spectrum to the observed galaxy overdensity power spectrum $ P_{\rm g} $ (the real part) given by \citep{Duniya:2016ibg,Duniya:2013eta,Duniya:2015nva,Jeong:2011as}

\begin{equation}
P_{\rm g}(k,z)= \left[\left(b + f \mu^{2}\right)^{2} + 2\left(b + f \mu^{2}\right) \left(\frac{\mathcal{A}}{{Y}^{2}} \right) + \frac{{\mathcal{A}}^{2}}{{Y}^{4}} + \mu^{2} \left(\frac{\mathcal{B}^{2}}{{Y}^{2}} \right)\right]P_{\rm m}(k,z)\, ,
\label{eq:ps}
\end{equation}

\noindent
where $Y=\frac{k}{\mathcal{H}} $ and $P_{\rm m}$ is the matter power spectrum given by

\begin{equation}
P_{\rm m}(k,z) = A k^{n_{\rm s}-4} T(k)^{2}\left(\frac{|\Delta_{m}(k,z)|}{|\Phi(k,0)|}\right)^{2} \, ,
\label{eq:ps1}
\end{equation}

\noindent
which is valid on all scales. One can check that Eq.~\eqref{eq:ps1} reduces to the standard definition of the matter power spectrum on sub-Hubble scales given by $ P_{\rm m}(k,z) \propto k^{n_{\rm s}} T(k)^{2}\left(\frac{|\delta_{\rm m}(k,z)|}{|\delta_{\rm m}(k,0)|}\right)^{2} $. The constant $ A $ is determined by the $ \sigma_{8} $ normalisation. Here we use the Eisenstein-Hu transfer function for $T(k)$. In the $ \sigma_{8} $ normalisation we put scalar spectral index of primordial power spectrum $n_{\rm s} = 0.96$, $\sigma_{8} = 0.8$, $h=0.7$, $\Omega_{\rm m0} = 0.28$ and $\Omega_{\rm b0} = 0.05$.
\\
\noindent
In Fig.~\ref{fig:ps}, we have plotted the line of sight ($\mu =1$) observed galaxy power spectrum for the linear potential at $ z = 0 $ using Eq.~\eqref{eq:ps}. In all the subsequent plots we put $ \mu = 1 $ and by $ P_{\rm k} $ we mean the observed galaxy power spectrum keeping only the bias and the Kaiser redshift space distortion terms i.e. without $\mathcal{A}$ and $\mathcal{B}$ terms. So, in all the subsequent plots 
\begin{equation}
 P_{\rm k}(k,z) = \left(b + f \mu^{2}\right)^{2} P_{\rm m}(k,z) \, .
 \label{eq:ps2}
\end{equation}
From Fig.~\ref{fig:ps} we see that the observed galaxy power spectrum without GR corrections ({\it i.e.}, $ P_{\rm k} $) enhances from the standard matter power spectrum $ P_{\rm m} $ and the enhancement is present on all scales i.e. $ P_{\rm k} $ shifts with an almost constant factor to  higher values compared to $ P_{\rm m} $ on all scales. When the GR corrections ($\mathcal{A}$ and $\mathcal{B}$ terms) are included, the full observed galaxy power spectrum $ P_{\rm g} $ enhances further from $ P_{\rm k} $ on larger scales only, which is quite obvious because the relativistic effects are negligible on sub-Hubble scales. In Fig.~\ref{fig:ps}, the vertical line is the exact horizon scale ($ k = a H $) at $ z=0 $.

\begin{center}
\begin{figure*}
\begin{tabular}{c@{\quad}c}
\epsfig{file=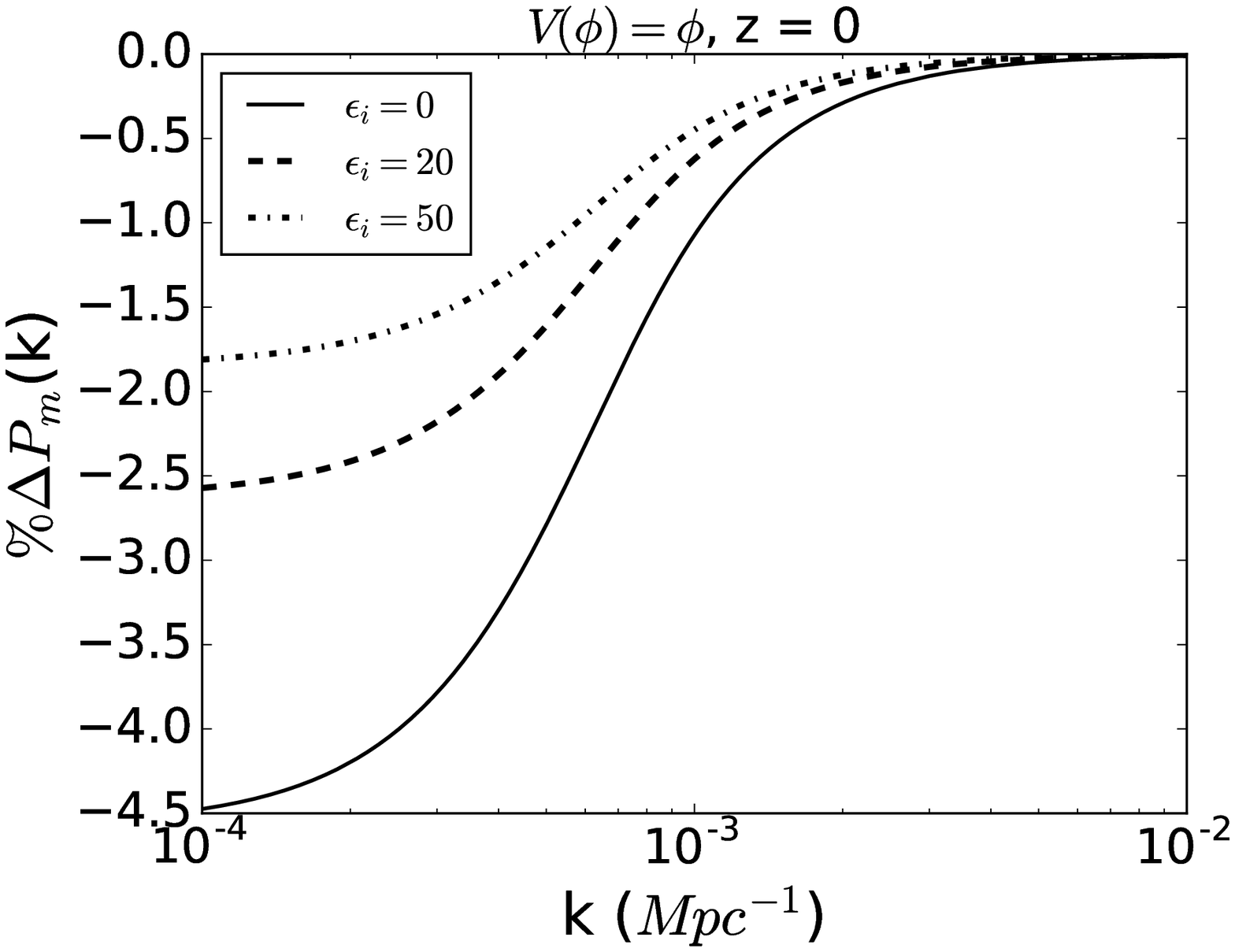,width=5.8 cm}
\epsfig{file=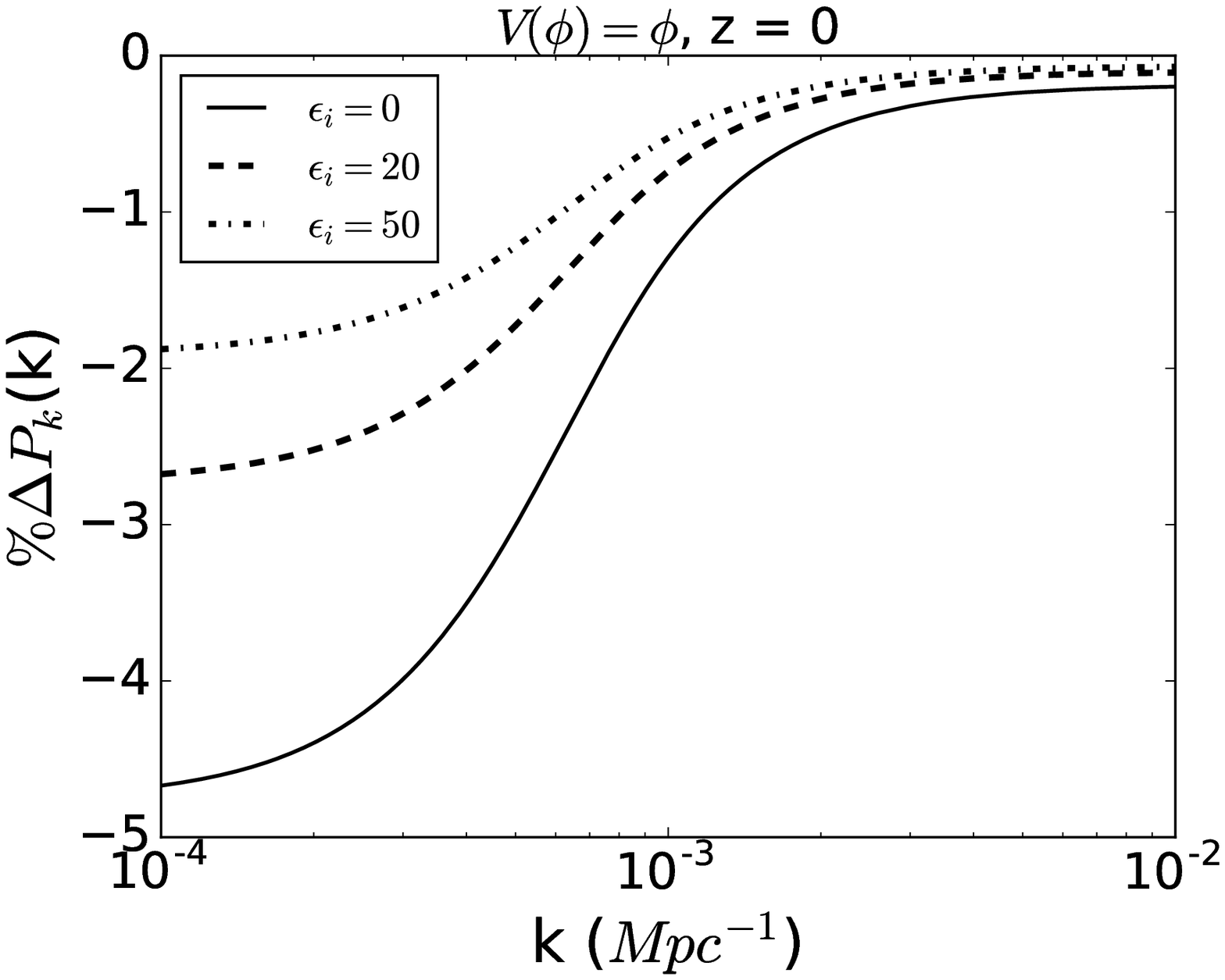,width=5.8 cm}
\epsfig{file=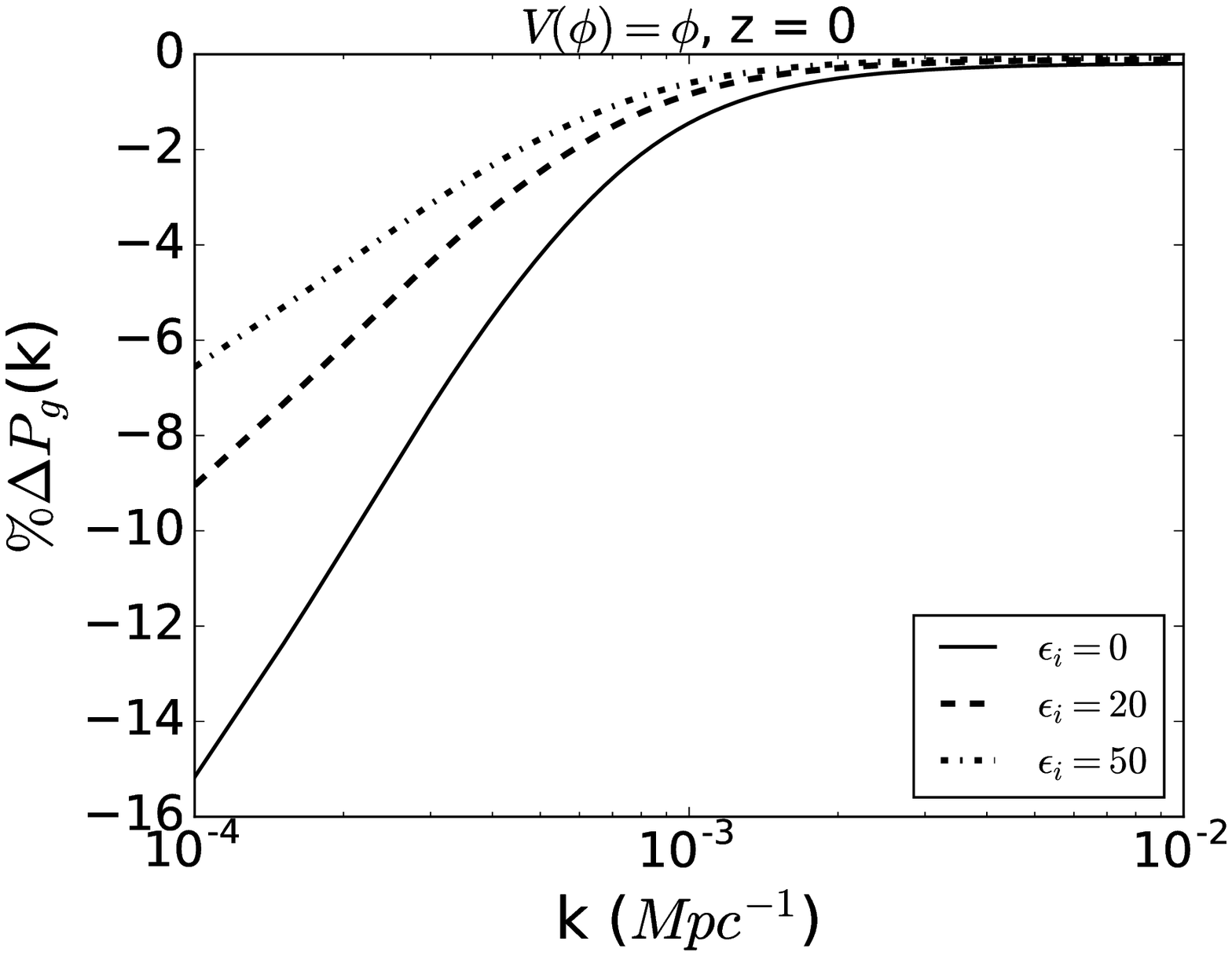,width=5.8 cm}\\
\epsfig{file=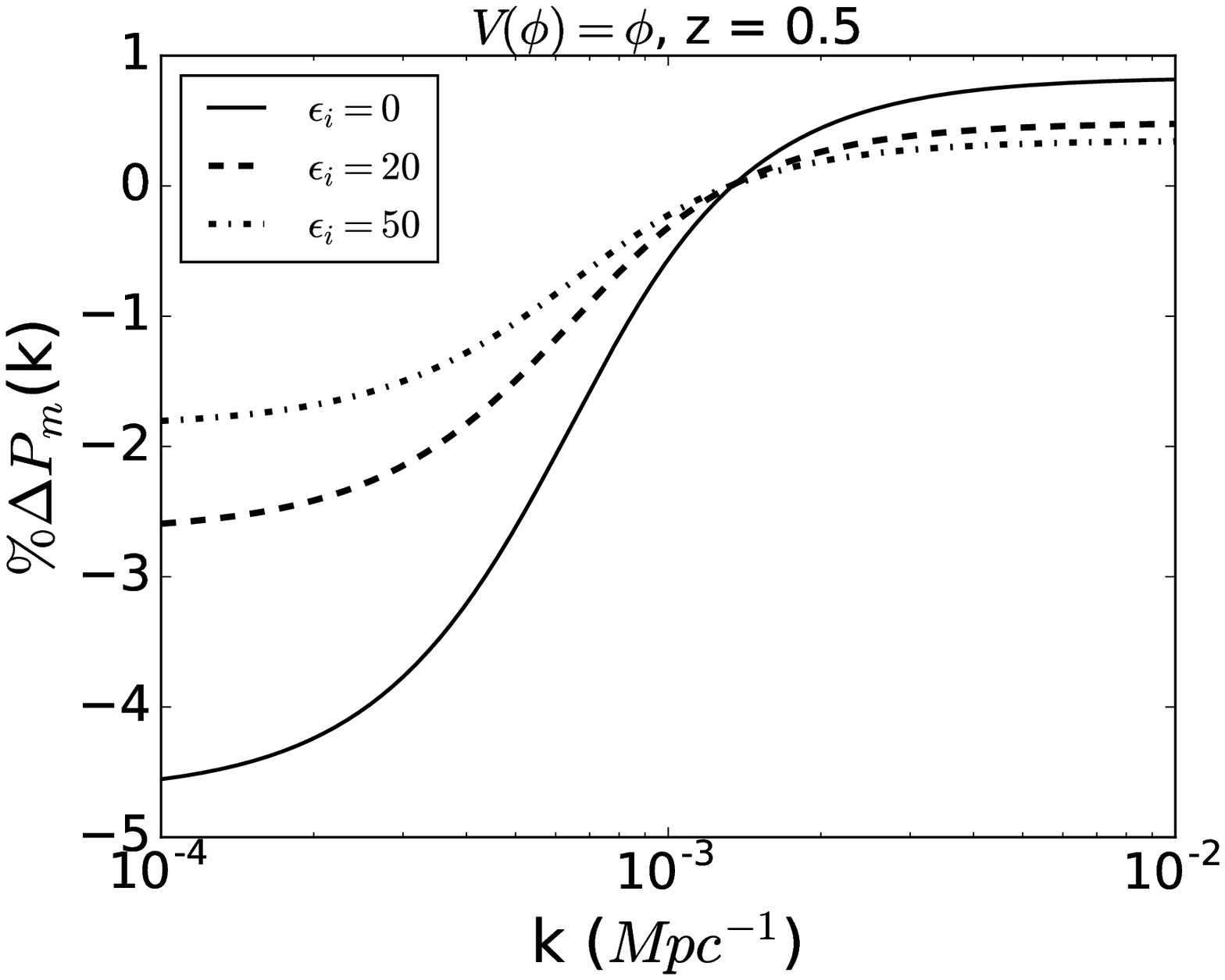,width=5.8 cm}
\epsfig{file=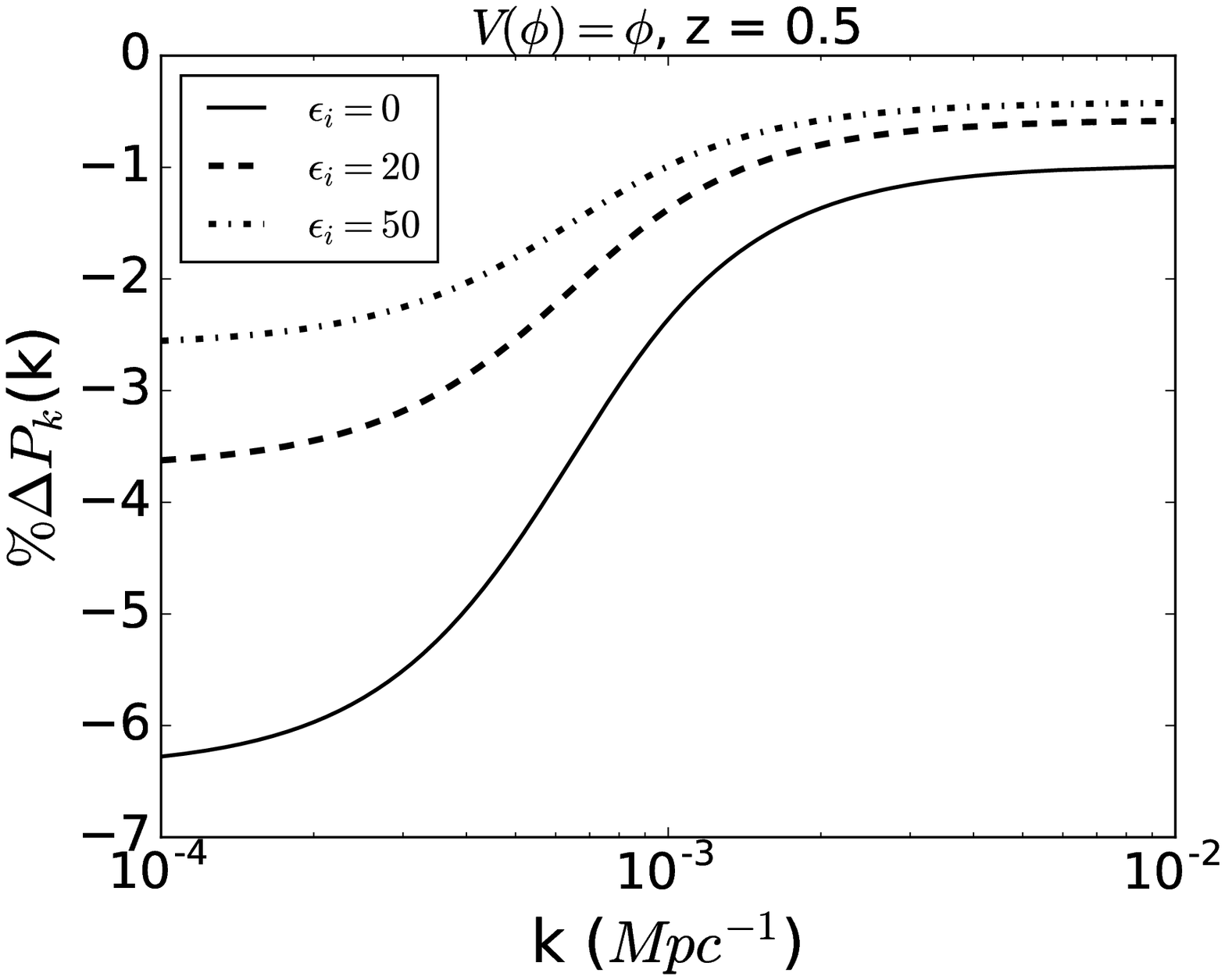,width=5.8 cm}
\epsfig{file=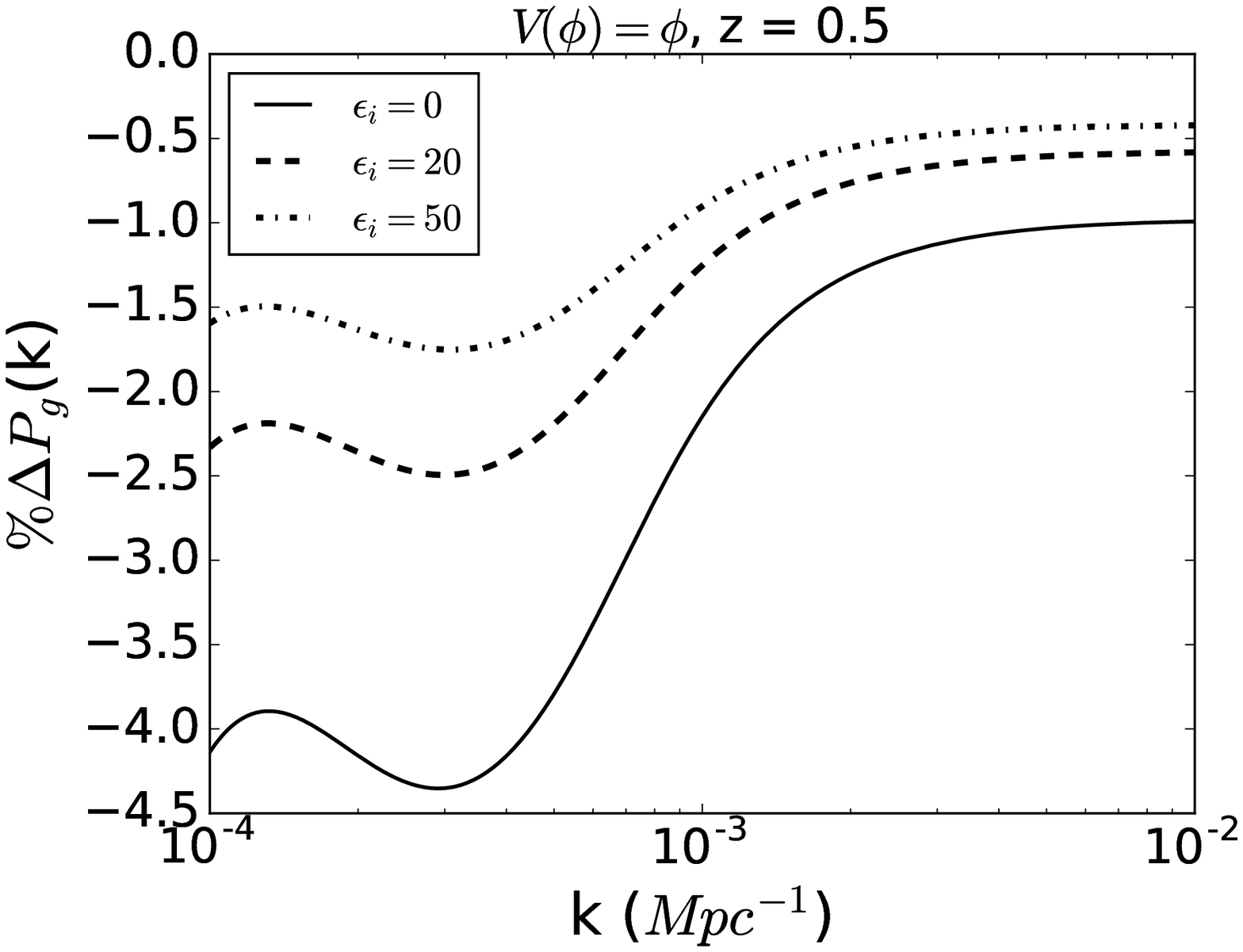,width=5.8 cm}\\
\epsfig{file=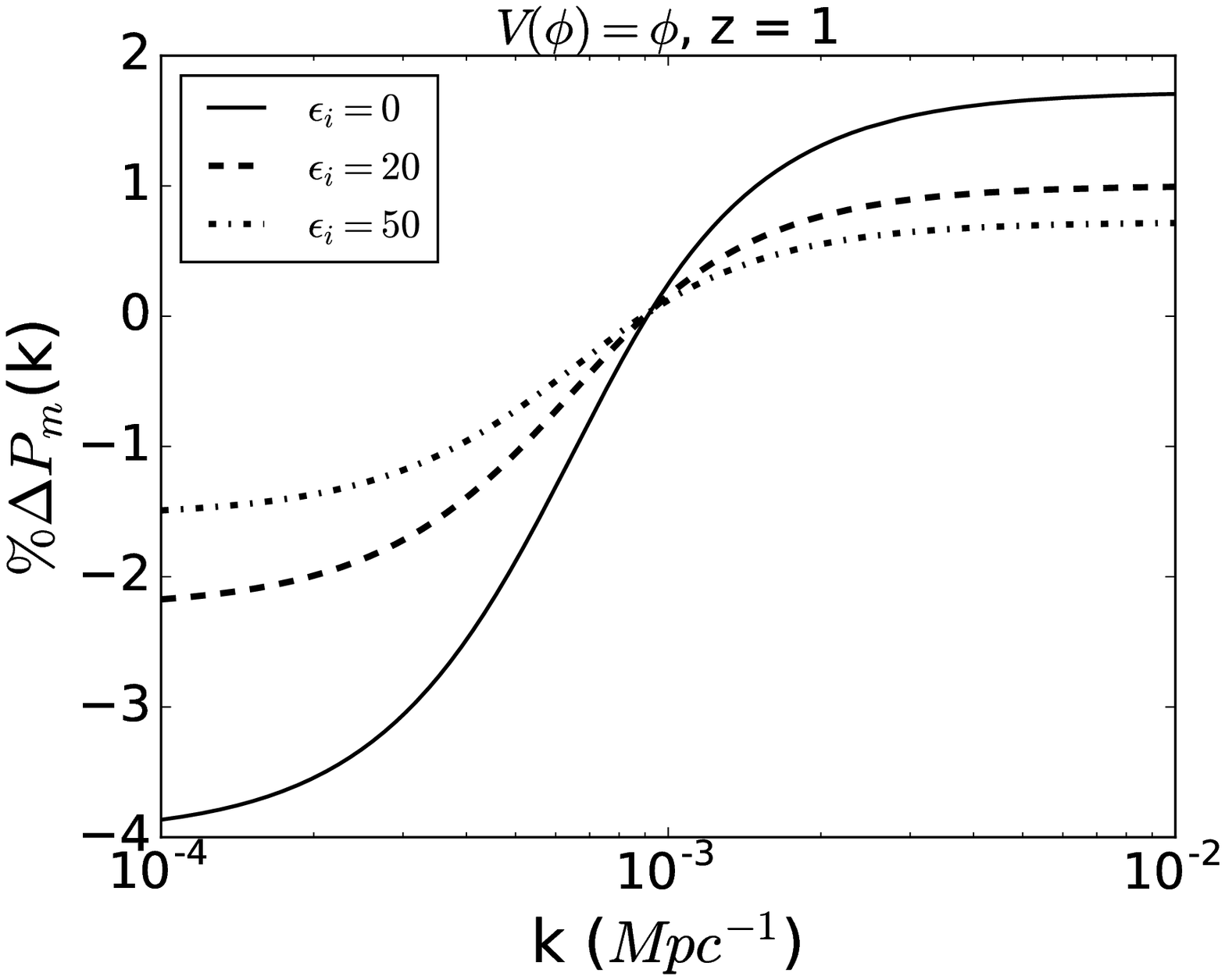,width=5.8 cm}
\epsfig{file=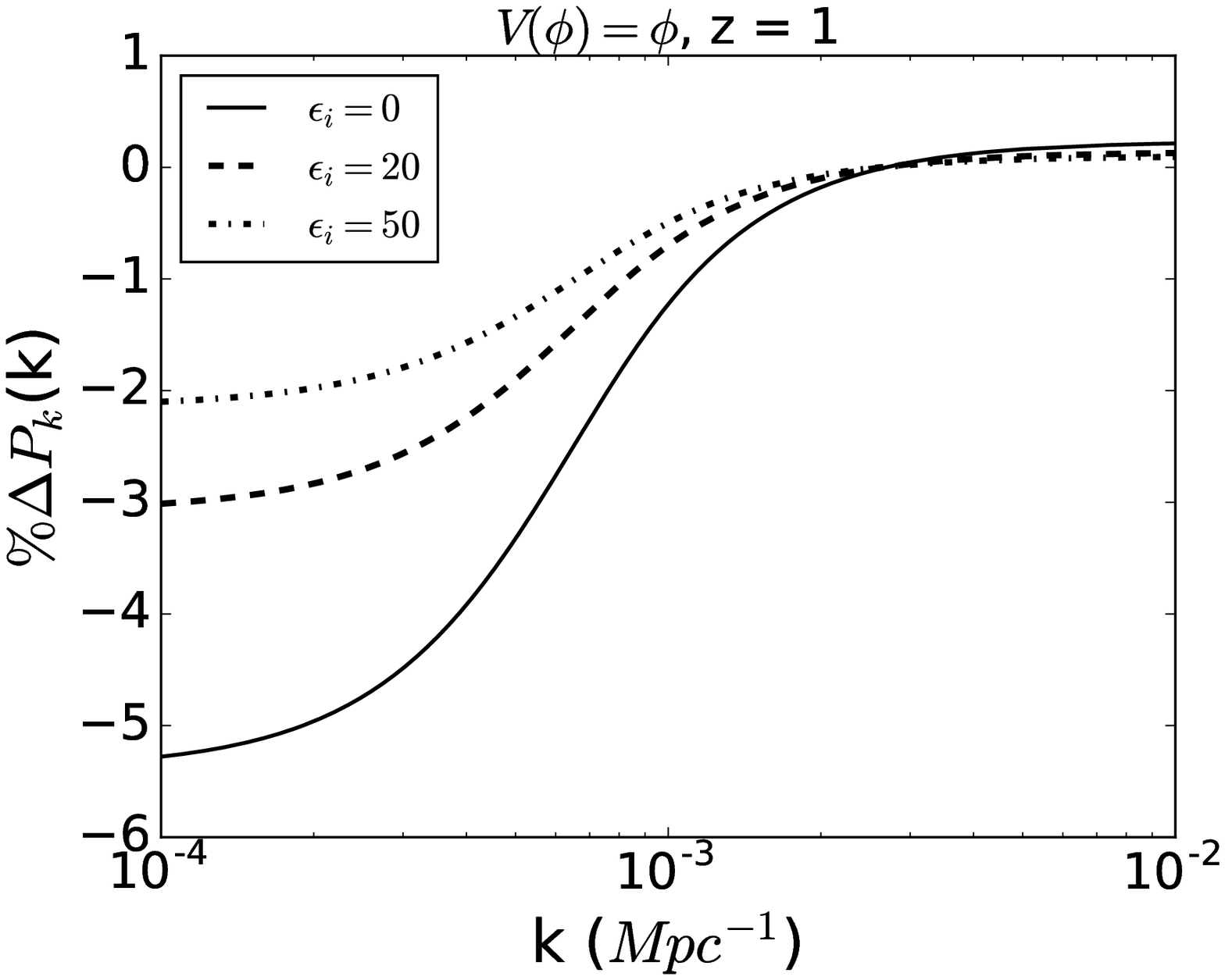,width=5.8 cm}
\epsfig{file=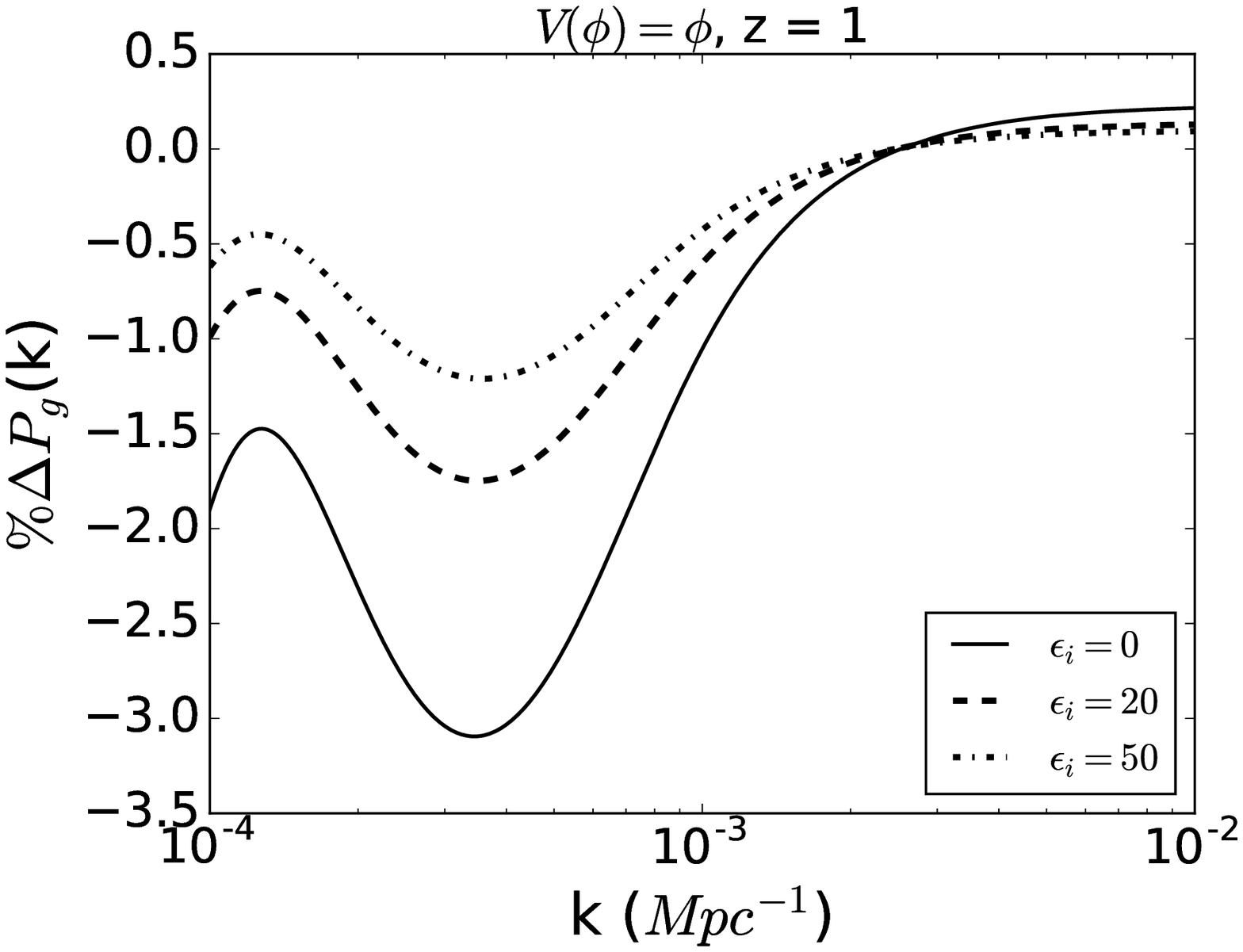,width=5.8 cm}\\
\epsfig{file=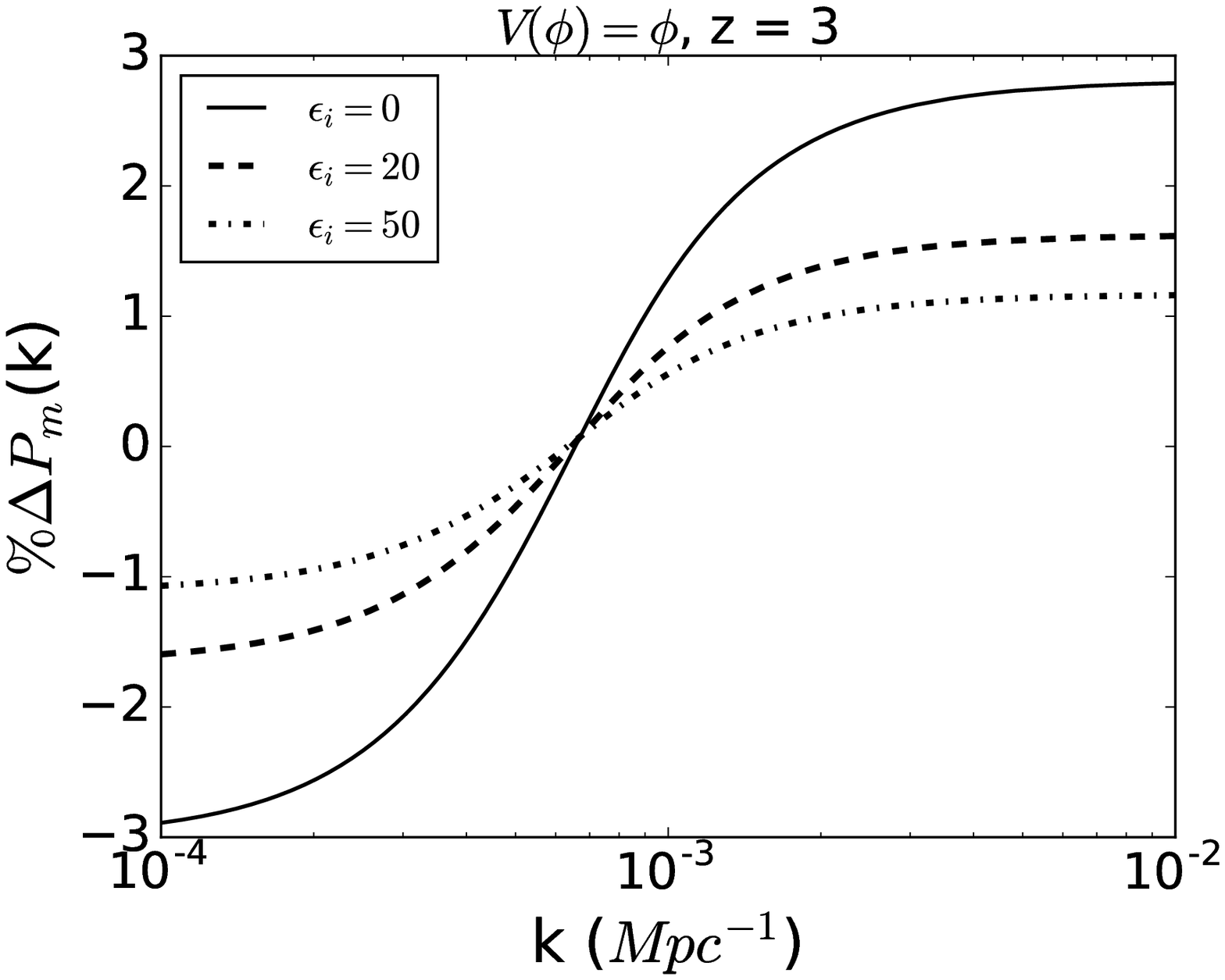,width=5.8 cm}
\epsfig{file=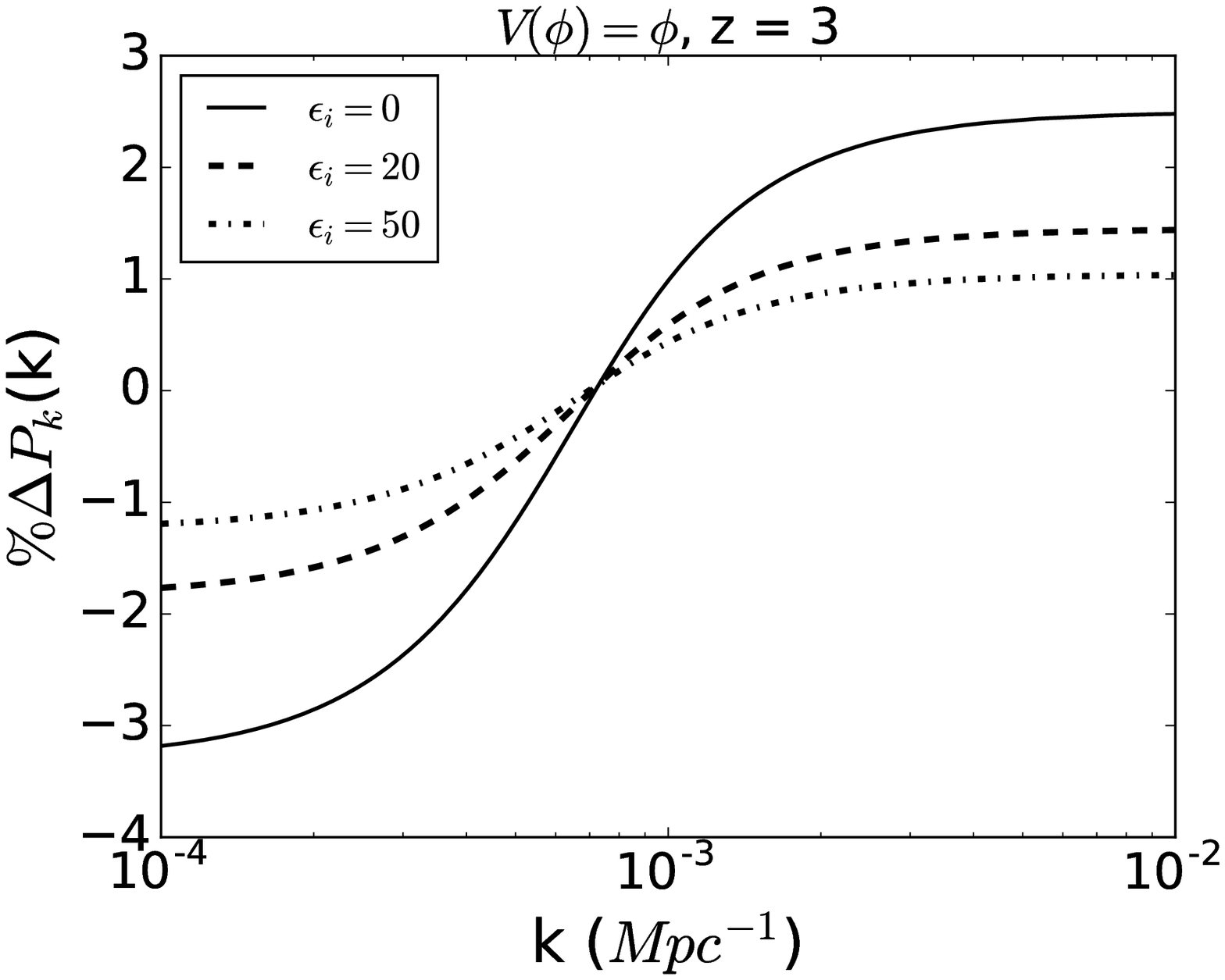,width=5.8 cm}
\epsfig{file=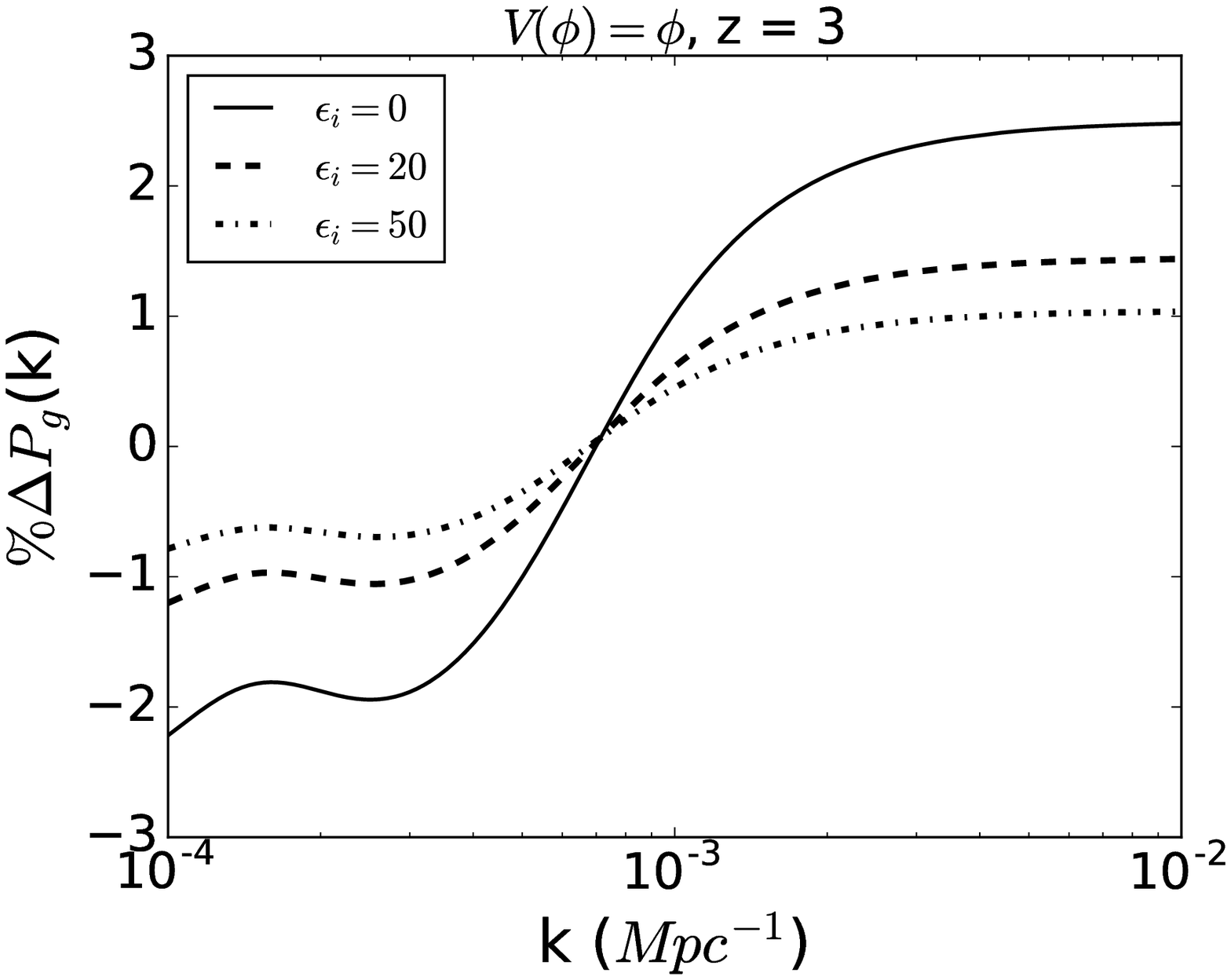,width=5.8 cm}
\end{tabular}
\caption{Percentage deviation in $ P (\rm k) $ from $ \Lambda$CDM model for different $ \epsilon_{i} $ with linear potential as a function $k$: negative values in y-axis means they are all suppressed from $ \Lambda $CDM. Left most plots are for standard matter power spectra $P_{\rm m}$ given by Eq.~\eqref{eq:ps1}, middle plots are for power spectra with Kaiser redshift space distortion term included and the right ones for full observed galaxy power spectra $P_{\rm g}$ given by Eq.~\eqref{eq:ps} with GR corrections.
}
\label{fig:del_ps}
\end{figure*}
\end{center}

Next we study the deviations in $ P_{\rm m} $, $ P_{\rm k} $ and $ P_{\rm g} $ for Galileon models from $ \Lambda $CDM in Fig.~\ref{fig:del_ps}. Firstly, the deviations in $ P_{\rm m} $ for different models from $ \Lambda $CDM comes through $ \Delta_{\rm m}(k,z) $ and $ \Phi(k,0) $ through Eq.~\eqref{eq:ps1}. So, the deviation in $ P_{\rm m} $  from $ \Lambda $CDM is due to these two competing terms. In Fig.~\ref{fig:deldelm}, we have already shown that the deviation in $\Delta_{m}$ is not substantial. So the main contribution comes from the difference in gravitational potential $\Phi(k,0)$. On large scales, this has an extra contribution from dark energy perturbation and hence this result the suppression in $P_{\rm m}$ on large scales from $\Lambda$CDM model. This is shown in the left most panel in Fig.~\ref{fig:del_ps}.

%%%%%%%%%%%%%%%%%%%%%%%%%%%%%%%%%%
%%%%%%%%%%%%%%%%%%%%%%%%%%%%%%%%%%%

Compared to the deviations in $ P_{\rm m} $, the deviations in $ P_{\rm k} $ comes due to the extra contribution from $ f $ . In Fig.~\ref{fig:delf} we have already seen that the deviations in $ f $ are marginal and also has a maximum at $ z \approx 0.5 $ and except for very low redshifts, the deviations are almost scale independent.  Hence the deviation in  $ P_{\rm k} $ is mostly similar to that in 
$ P_{\rm m} $. Only around $ z\sim 0.5$, it is bit higher than $ P_{\rm m} $ due to the maximum contribution from $f$. This is shown in middle panel of  Fig.~\ref{fig:del_ps}.

In Fig.~\ref{fig:ps} we have seen that, due to the extra GR correction, $ P_{\rm g} $ deviates from $ P_{\rm k} $ only on larger scales otherwise they are almost same on smaller scales. So, the deviations in $ P_{\rm g} $ follow the exact deviations in $P_{\rm k}$ on smaller scales which is clear from the middle and right panels of Fig.~\ref{fig:del_ps}. On larger scales, however, there is a extra effect due to GR correction terms as described through terms $\mathcal{A}$ and $\mathcal{B}$. Due to this, there is a large suppression from $\Lambda$CDM on large scales and smaller redshifts. This is shown in the right panel of  Fig.~\ref{fig:del_ps}.

One can also notice that in all the figures, the deviations are always higher for $\epsilon_{i} =0$ compared to non zero values for $\epsilon_{i}$. Given the fact that non zero $\epsilon_{i}$ represents Galileon models  and $\epsilon_{i} = 0$ represents standard quintessence, one can conclude that Galileon models are harder to distinguigh from $\Lambda$CDM model compared to standard quintessence.

\begin{center}
\begin{figure*}
\begin{tabular}{c@{\quad}c}
\epsfig{file=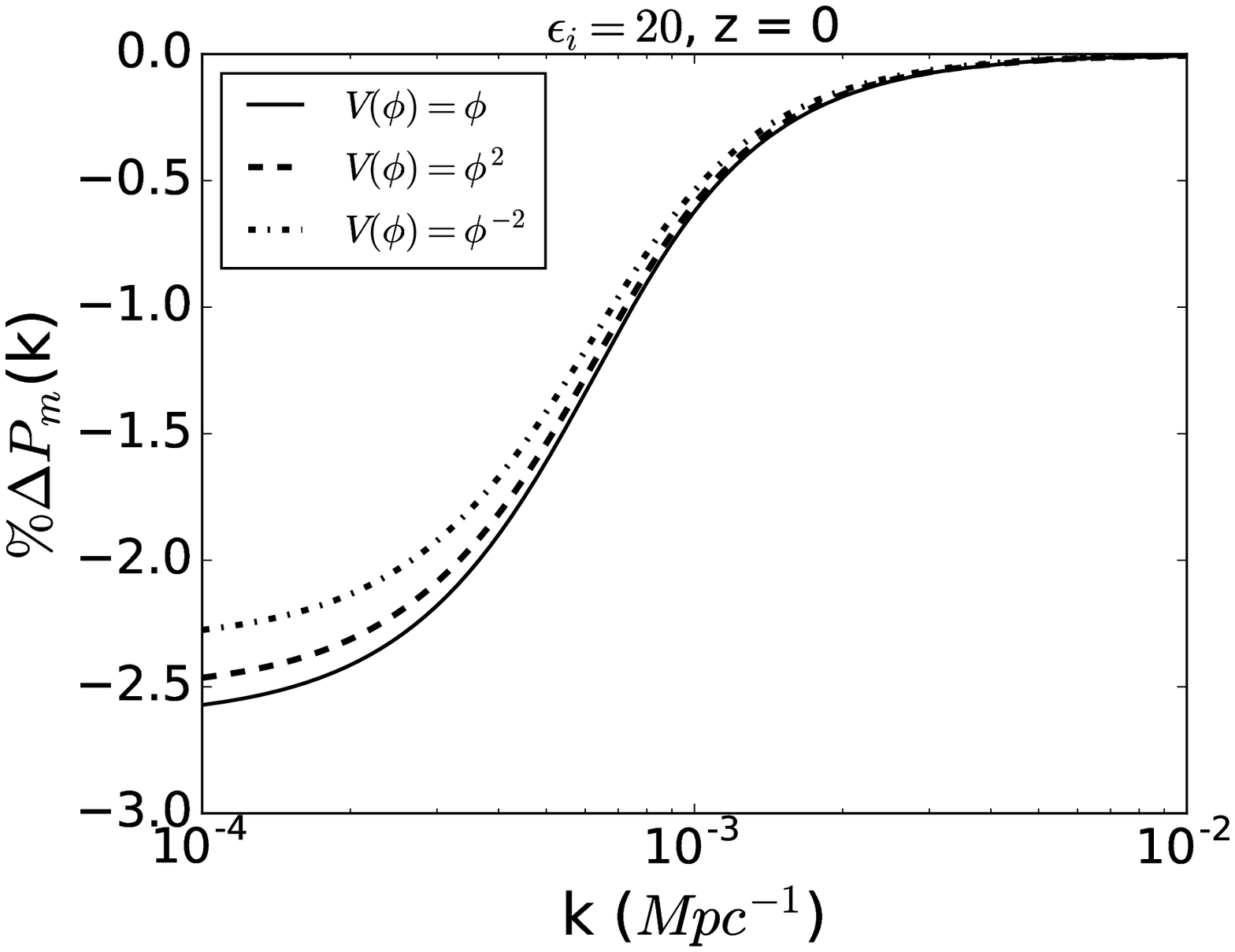,width=5.8 cm}
\epsfig{file=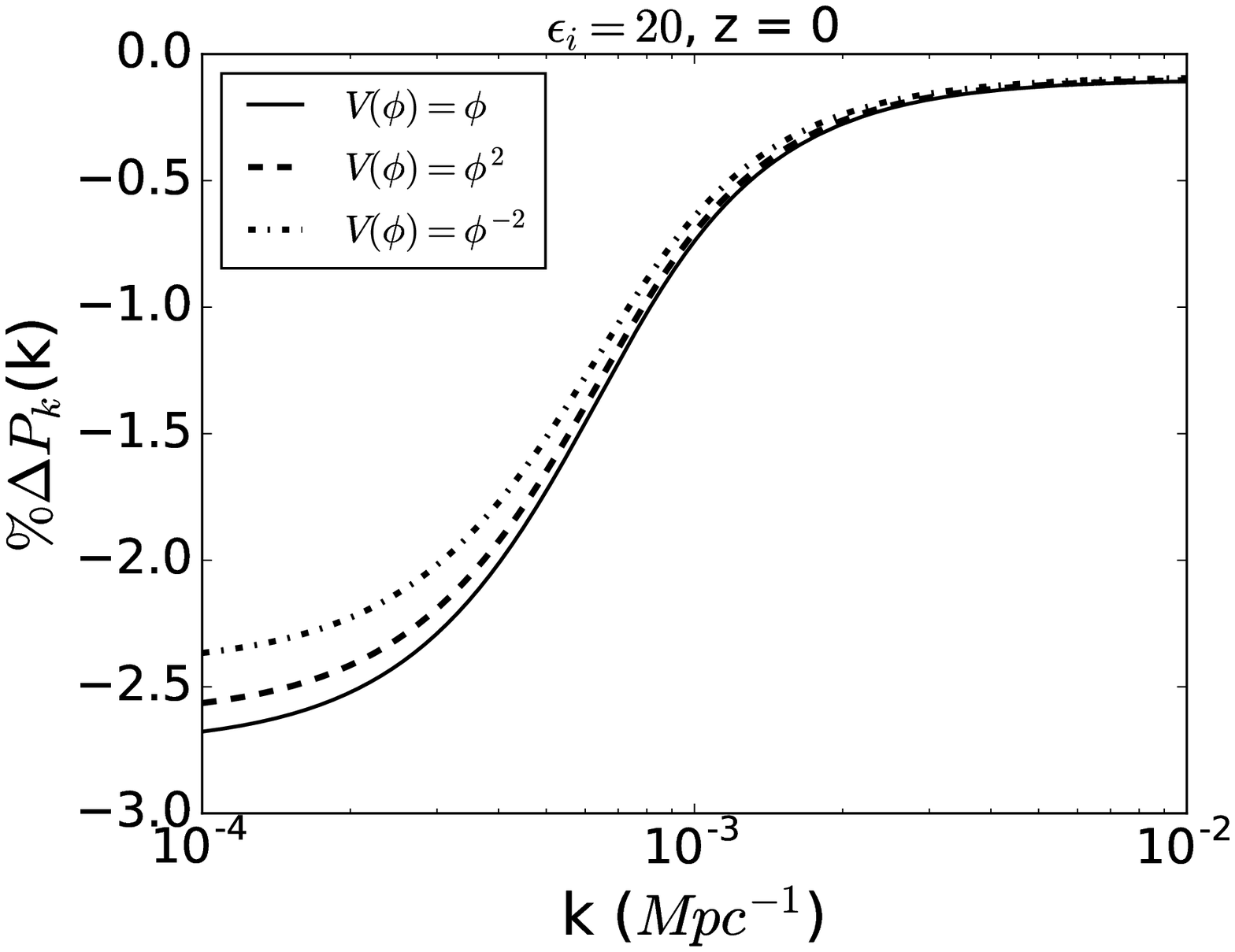,width=5.8 cm}
\epsfig{file=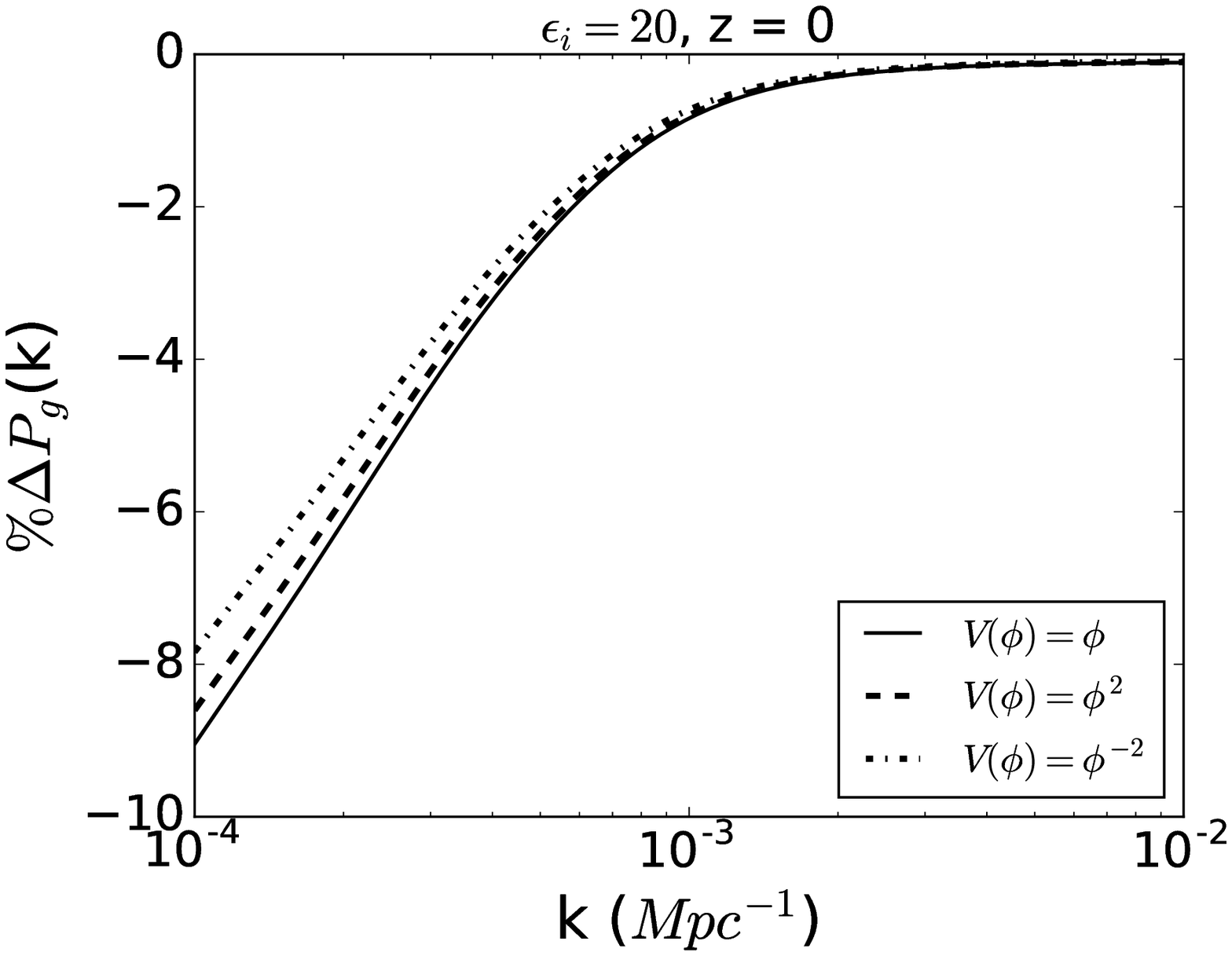,width=5.8 cm}
\end{tabular}
\caption{Percentage deviation in $ P(k) $ from $\Lambda$CDM model for different potentials as a function of $k$ at $z=0$ and for $\epsilon_{i} = 20$: negative values in y-axis means they are all suppressed from $\Lambda$CDM.}
\label{fig:del_ps_z}
\end{figure*}
\end{center}

Finally in Fig.~\ref{fig:del_ps_z}, we consider other phenomenological potentials like squared and inverse-squared potentials and compare them with linear potential. It is shown that the linear potential has marginally higher deviation from $\Lambda$CDM compared to the other potentials.

%%%%%%%%%%%%%%%%%%%%%%%%%%%%%%%%%%
%%%%%%%%%%%%%%%%%%%%%%%%%%%%%%%%%%

%%%%%%%%%%%%%%%%%%%%
\section{Summary and Conclusion}
%%%%%%%%%%%%%%%%%%%%

In this paper, we study  the observed galaxy power power spectra in cubic Galileon model with a linear potential which preserves the shift symmetry.  In this scenario potential is responsible for late time acceleration. Although there is a higher derivative term in the action, the equation motion is still second order and the theory is free from ghost. We have considered thawing dynamics of the Galileon field. We form a single autonomous system involving both the background evolution and the linear perturbation equation for matter and dark energy. 

We show that the deviation from $\Lambda$CDM in comoving matter density contrast $\Delta_{m}$ and growth rate $f$ is not substantial for cubic Galileon models. The gravitational potential gets slightly enhanced on large scales compared to $\Lambda$CDM due to the added contribution from the perturbed Galileon field.   

The observed galaxy power spectrum contains several correction terms related to redshift space distortion as well as other relativistic corrections that are present on large scales only. Due to the presence of the these terms, on large scales there are substantial deviation from the $\Lambda$CDM model in observed galaxy power spectrum $P_{g}$. But compared to standard quintessence, these deviation are small in Galileon model. This makes Galileon models hard to distinguish from $\Lambda$CDM even on larger scales.

We also consider some phenomenological potentials like squared and inverse-squared potentials which break shift symmetry and show that the deviations from $\Lambda$CDM in observed galaxy power spectrum for these potentials are always less than the linear potential which preserves the shift symmetry. 

In future, we aim to extend this study to massive gravity \citep{deRham:2011by} and generalized Proca theories for gravity \citep{DeFelice:2016yws}.

\section{Acknowledgements}
B.R.D. thanks CSIR, Govt. of India for financial support through SRF scheme (No:09/466(0157)/2012-EMR-I). We also acknowledge the usage of HOPE- a Python Just-In-Time compiler for astrophysical computations \citep{2015A&C....10....1A}.

\bibliography{ref}

\begin{thebibliography}{100}
\expandafter\ifx\csname natexlab\endcsname\relax\def\natexlab#1{#1}\fi
\expandafter\ifx\csname bibnamefont\endcsname\relax
  \def\bibnamefont#1{#1}\fi
\expandafter\ifx\csname bibfnamefont\endcsname\relax
  \def\bibfnamefont#1{#1}\fi
\expandafter\ifx\csname citenamefont\endcsname\relax
  \def\citenamefont#1{#1}\fi
\expandafter\ifx\csname url\endcsname\relax
  \def\url#1{\texttt{#1}}\fi
\expandafter\ifx\csname urlprefix\endcsname\relax\def\urlprefix{URL }\fi
\providecommand{\bibinfo}[2]{#2}
\providecommand{\eprint}[2][]{\url{#2}}

\bibitem[{\citenamefont{Riess et~al.}(1998)}]{Riess:1998cb}
\bibinfo{author}{\bibfnamefont{A.~G.} \bibnamefont{Riess}} \bibnamefont{et~al.}
  (\bibinfo{collaboration}{Supernova Search Team}), \bibinfo{journal}{Astron.
  J.} \textbf{\bibinfo{volume}{116}}, \bibinfo{pages}{1009}
  (\bibinfo{year}{1998}), \eprint{astro-ph/9805201}.

\bibitem[{\citenamefont{Perlmutter et~al.}(1999)}]{Perlmutter:1998np}
\bibinfo{author}{\bibfnamefont{S.}~\bibnamefont{Perlmutter}}
  \bibnamefont{et~al.} (\bibinfo{collaboration}{Supernova Cosmology Project}),
  \bibinfo{journal}{Astrophys. J.} \textbf{\bibinfo{volume}{517}},
  \bibinfo{pages}{565} (\bibinfo{year}{1999}), \eprint{astro-ph/9812133}.

\bibitem[{\citenamefont{Spergel et~al.}(2003)}]{Spergel:2003cb}
\bibinfo{author}{\bibfnamefont{D.~N.} \bibnamefont{Spergel}}
  \bibnamefont{et~al.} (\bibinfo{collaboration}{WMAP}),
  \bibinfo{journal}{Astrophys. J. Suppl.} \textbf{\bibinfo{volume}{148}},
  \bibinfo{pages}{175} (\bibinfo{year}{2003}), \eprint{astro-ph/0302209}.

\bibitem[{\citenamefont{Hinshaw et~al.}(2003)}]{Hinshaw:2003ex}
\bibinfo{author}{\bibfnamefont{G.}~\bibnamefont{Hinshaw}} \bibnamefont{et~al.}
  (\bibinfo{collaboration}{WMAP}), \bibinfo{journal}{Astrophys. J. Suppl.}
  \textbf{\bibinfo{volume}{148}}, \bibinfo{pages}{135} (\bibinfo{year}{2003}),
  \eprint{astro-ph/0302217}.

\bibitem[{\citenamefont{Ade et~al.}(2015)}]{Ade:2015xua}
\bibinfo{author}{\bibfnamefont{P.~A.~R.} \bibnamefont{Ade}}
  \bibnamefont{et~al.} (\bibinfo{collaboration}{Planck}),
  \bibinfo{journal}{ArXiv e-prints}  (\bibinfo{year}{2015}),
  \eprint{1502.01589}.

\bibitem[{\citenamefont{Ade et~al.}(2016)}]{Ade:2015lrj}
\bibinfo{author}{\bibfnamefont{P.~A.~R.} \bibnamefont{Ade}}
  \bibnamefont{et~al.} (\bibinfo{collaboration}{Planck}),
  \bibinfo{journal}{Astron. Astrophys.} \textbf{\bibinfo{volume}{594}},
  \bibinfo{pages}{A20} (\bibinfo{year}{2016}), \eprint{1502.02114}.

\bibitem[{\citenamefont{{Delubac} et~al.}(2015)\citenamefont{{Delubac},
  {Bautista}, {Busca}, {Rich}, {Kirkby}, {Bailey}, {Font-Ribera}, {Slosar},
  {Lee}, {Pieri} et~al.}}]{2015A&A...574A..59D}
\bibinfo{author}{\bibfnamefont{T.}~\bibnamefont{{Delubac}}},
  \bibinfo{author}{\bibfnamefont{J.~E.} \bibnamefont{{Bautista}}},
  \bibinfo{author}{\bibfnamefont{N.~G.} \bibnamefont{{Busca}}},
  \bibinfo{author}{\bibfnamefont{J.}~\bibnamefont{{Rich}}},
  \bibinfo{author}{\bibfnamefont{D.}~\bibnamefont{{Kirkby}}},
  \bibinfo{author}{\bibfnamefont{S.}~\bibnamefont{{Bailey}}},
  \bibinfo{author}{\bibfnamefont{A.}~\bibnamefont{{Font-Ribera}}},
  \bibinfo{author}{\bibfnamefont{A.}~\bibnamefont{{Slosar}}},
  \bibinfo{author}{\bibfnamefont{K.-G.} \bibnamefont{{Lee}}},
  \bibinfo{author}{\bibfnamefont{M.~M.} \bibnamefont{{Pieri}}},
  \bibnamefont{et~al.}, \bibinfo{journal}{Astron. Astrophys.}
  \textbf{\bibinfo{volume}{574}}, \bibinfo{eid}{A59} (\bibinfo{year}{2015}),
  \eprint{1404.1801}.

\bibitem[{\citenamefont{Ata et~al.}(2017)}]{Ata:2017dya}
\bibinfo{author}{\bibfnamefont{M.}~\bibnamefont{Ata}} \bibnamefont{et~al.}
  (\bibinfo{year}{2017}), \eprint{1705.06373}.

\bibitem[{\citenamefont{Starobinsky}(1980)}]{Starobinsky:1980te}
\bibinfo{author}{\bibfnamefont{A.~A.} \bibnamefont{Starobinsky}},
  \bibinfo{journal}{Phys. Lett.} \textbf{\bibinfo{volume}{B91}},
  \bibinfo{pages}{99} (\bibinfo{year}{1980}).

\bibitem[{\citenamefont{Guth}(1981)}]{Guth:1980zm}
\bibinfo{author}{\bibfnamefont{A.~H.} \bibnamefont{Guth}},
  \bibinfo{journal}{Phys. Rev.} \textbf{\bibinfo{volume}{D23}},
  \bibinfo{pages}{347} (\bibinfo{year}{1981}).

\bibitem[{\citenamefont{Linde}(1983)}]{Linde:1983gd}
\bibinfo{author}{\bibfnamefont{A.~D.} \bibnamefont{Linde}},
  \bibinfo{journal}{Phys. Lett.} \textbf{\bibinfo{volume}{B129}},
  \bibinfo{pages}{177} (\bibinfo{year}{1983}).

\bibitem[{\citenamefont{Linde}(1982)}]{Linde:1981mu}
\bibinfo{author}{\bibfnamefont{A.~D.} \bibnamefont{Linde}},
  \bibinfo{journal}{Phys. Lett.} \textbf{\bibinfo{volume}{B108}},
  \bibinfo{pages}{389} (\bibinfo{year}{1982}).

\bibitem[{\citenamefont{Liddle}(1999)}]{Liddle:1999mq}
\bibinfo{author}{\bibfnamefont{A.~R.} \bibnamefont{Liddle}}, in
  \emph{\bibinfo{booktitle}{{Proceedings, Summer School in High-energy physics
  and cosmology: Trieste, Italy, June 29-July 17, 1998}}}
  (\bibinfo{year}{1999}), pp. \bibinfo{pages}{260--295},
  \eprint{astro-ph/9901124},
  \urlprefix\url{http://alice.cern.ch/format/showfull?sysnb=0301651}.

\bibitem[{\citenamefont{Copeland et~al.}(2006)\citenamefont{Copeland, Sami, and
  Tsujikawa}}]{Copeland:2006wr}
\bibinfo{author}{\bibfnamefont{E.~J.} \bibnamefont{Copeland}},
  \bibinfo{author}{\bibfnamefont{M.}~\bibnamefont{Sami}}, \bibnamefont{and}
  \bibinfo{author}{\bibfnamefont{S.}~\bibnamefont{Tsujikawa}},
  \bibinfo{journal}{Int. J. Mod. Phys.} \textbf{\bibinfo{volume}{D15}},
  \bibinfo{pages}{1753} (\bibinfo{year}{2006}), \eprint{hep-th/0603057}.

\bibitem[{\citenamefont{Sahni and Starobinsky}(2000)}]{Sahni:1999gb}
\bibinfo{author}{\bibfnamefont{V.}~\bibnamefont{Sahni}} \bibnamefont{and}
  \bibinfo{author}{\bibfnamefont{A.~A.} \bibnamefont{Starobinsky}},
  \bibinfo{journal}{Int. J. Mod. Phys.} \textbf{\bibinfo{volume}{D9}},
  \bibinfo{pages}{373} (\bibinfo{year}{2000}), \eprint{astro-ph/9904398}.

\bibitem[{\citenamefont{Padmanabhan}(2006)}]{Padmanabhan:2006ag}
\bibinfo{author}{\bibfnamefont{T.}~\bibnamefont{Padmanabhan}},
  \bibinfo{journal}{AIP Conf. Proc.} \textbf{\bibinfo{volume}{861}},
  \bibinfo{pages}{179} (\bibinfo{year}{2006}), \bibinfo{note}{[,179(2006)]},
  \eprint{astro-ph/0603114}.

\bibitem[{\citenamefont{Frieman et~al.}(2008)\citenamefont{Frieman, Turner, and
  Huterer}}]{Frieman:2008sn}
\bibinfo{author}{\bibfnamefont{J.}~\bibnamefont{Frieman}},
  \bibinfo{author}{\bibfnamefont{M.}~\bibnamefont{Turner}}, \bibnamefont{and}
  \bibinfo{author}{\bibfnamefont{D.}~\bibnamefont{Huterer}},
  \bibinfo{journal}{Ann. Rev. Astron. Astrophys.}
  \textbf{\bibinfo{volume}{46}}, \bibinfo{pages}{385} (\bibinfo{year}{2008}),
  \eprint{0803.0982}.

\bibitem[{\citenamefont{Clifton et~al.}(2012)\citenamefont{Clifton, Ferreira,
  Padilla, and Skordis}}]{Clifton:2011jh}
\bibinfo{author}{\bibfnamefont{T.}~\bibnamefont{Clifton}},
  \bibinfo{author}{\bibfnamefont{P.~G.} \bibnamefont{Ferreira}},
  \bibinfo{author}{\bibfnamefont{A.}~\bibnamefont{Padilla}}, \bibnamefont{and}
  \bibinfo{author}{\bibfnamefont{C.}~\bibnamefont{Skordis}},
  \bibinfo{journal}{Phys. Rept.} \textbf{\bibinfo{volume}{513}},
  \bibinfo{pages}{1} (\bibinfo{year}{2012}), \eprint{1106.2476}.

\bibitem[{\citenamefont{Hinterbichler}(2012)}]{Hinterbichler:2011tt}
\bibinfo{author}{\bibfnamefont{K.}~\bibnamefont{Hinterbichler}},
  \bibinfo{journal}{Rev. Mod. Phys.} \textbf{\bibinfo{volume}{84}},
  \bibinfo{pages}{671} (\bibinfo{year}{2012}), \eprint{1105.3735}.

\bibitem[{\citenamefont{De~Felice and
  Tsujikawa}(2010{\natexlab{a}})}]{DeFelice:2010aj}
\bibinfo{author}{\bibfnamefont{A.}~\bibnamefont{De~Felice}} \bibnamefont{and}
  \bibinfo{author}{\bibfnamefont{S.}~\bibnamefont{Tsujikawa}},
  \bibinfo{journal}{Living Rev. Rel.} \textbf{\bibinfo{volume}{13}},
  \bibinfo{pages}{3} (\bibinfo{year}{2010}{\natexlab{a}}), \eprint{1002.4928}.

\bibitem[{\citenamefont{de~Rham}(2014)}]{deRham:2014zqa}
\bibinfo{author}{\bibfnamefont{C.}~\bibnamefont{de~Rham}},
  \bibinfo{journal}{Living Rev. Rel.} \textbf{\bibinfo{volume}{17}},
  \bibinfo{pages}{7} (\bibinfo{year}{2014}), \eprint{1401.4173}.

\bibitem[{\citenamefont{de~Rham}(2012)}]{deRham:2012az}
\bibinfo{author}{\bibfnamefont{C.}~\bibnamefont{de~Rham}},
  \bibinfo{journal}{Comptes Rendus Physique} \textbf{\bibinfo{volume}{13}},
  \bibinfo{pages}{666} (\bibinfo{year}{2012}), \eprint{1204.5492}.

\bibitem[{\citenamefont{Horndeski}(1974)}]{Horndeski:1974wa}
\bibinfo{author}{\bibfnamefont{G.~W.} \bibnamefont{Horndeski}},
  \bibinfo{journal}{Int. J. Theor. Phys.} \textbf{\bibinfo{volume}{10}},
  \bibinfo{pages}{363} (\bibinfo{year}{1974}).

\bibitem[{\citenamefont{Dvali et~al.}(2000)\citenamefont{Dvali, Gabadadze, and
  Porrati}}]{Dvali:2000hr}
\bibinfo{author}{\bibfnamefont{G.~R.} \bibnamefont{Dvali}},
  \bibinfo{author}{\bibfnamefont{G.}~\bibnamefont{Gabadadze}},
  \bibnamefont{and} \bibinfo{author}{\bibfnamefont{M.}~\bibnamefont{Porrati}},
  \bibinfo{journal}{Phys. Lett.} \textbf{\bibinfo{volume}{B485}},
  \bibinfo{pages}{208} (\bibinfo{year}{2000}), \eprint{hep-th/0005016}.

\bibitem[{\citenamefont{Hu and Sawicki}(2007)}]{Hu:2007nk}
\bibinfo{author}{\bibfnamefont{W.}~\bibnamefont{Hu}} \bibnamefont{and}
  \bibinfo{author}{\bibfnamefont{I.}~\bibnamefont{Sawicki}},
  \bibinfo{journal}{Phys. Rev.} \textbf{\bibinfo{volume}{D76}},
  \bibinfo{pages}{064004} (\bibinfo{year}{2007}), \eprint{0705.1158}.

\bibitem[{\citenamefont{Amendola et~al.}(2007)\citenamefont{Amendola, Gannouji,
  Polarski, and Tsujikawa}}]{Amendola:2006we}
\bibinfo{author}{\bibfnamefont{L.}~\bibnamefont{Amendola}},
  \bibinfo{author}{\bibfnamefont{R.}~\bibnamefont{Gannouji}},
  \bibinfo{author}{\bibfnamefont{D.}~\bibnamefont{Polarski}}, \bibnamefont{and}
  \bibinfo{author}{\bibfnamefont{S.}~\bibnamefont{Tsujikawa}},
  \bibinfo{journal}{Phys. Rev.} \textbf{\bibinfo{volume}{D75}},
  \bibinfo{pages}{083504} (\bibinfo{year}{2007}), \eprint{gr-qc/0612180}.

\bibitem[{\citenamefont{Nicolis et~al.}(2009)\citenamefont{Nicolis, Rattazzi,
  and Trincherini}}]{Nicolis:2008in}
\bibinfo{author}{\bibfnamefont{A.}~\bibnamefont{Nicolis}},
  \bibinfo{author}{\bibfnamefont{R.}~\bibnamefont{Rattazzi}}, \bibnamefont{and}
  \bibinfo{author}{\bibfnamefont{E.}~\bibnamefont{Trincherini}},
  \bibinfo{journal}{Phys. Rev.} \textbf{\bibinfo{volume}{D79}},
  \bibinfo{pages}{064036} (\bibinfo{year}{2009}), \eprint{0811.2197}.

\bibitem[{\citenamefont{de~Rham
  et~al.}(2011{\natexlab{a}})\citenamefont{de~Rham, Gabadadze, and
  Tolley}}]{deRham:2010kj}
\bibinfo{author}{\bibfnamefont{C.}~\bibnamefont{de~Rham}},
  \bibinfo{author}{\bibfnamefont{G.}~\bibnamefont{Gabadadze}},
  \bibnamefont{and} \bibinfo{author}{\bibfnamefont{A.~J.}
  \bibnamefont{Tolley}}, \bibinfo{journal}{Phys. Rev. Lett.}
  \textbf{\bibinfo{volume}{106}}, \bibinfo{pages}{231101}
  (\bibinfo{year}{2011}{\natexlab{a}}), \eprint{1011.1232}.

\bibitem[{\citenamefont{de~Rham
  et~al.}(2011{\natexlab{b}})\citenamefont{de~Rham, Gabadadze, Heisenberg, and
  Pirtskhalava}}]{deRham:2010tw}
\bibinfo{author}{\bibfnamefont{C.}~\bibnamefont{de~Rham}},
  \bibinfo{author}{\bibfnamefont{G.}~\bibnamefont{Gabadadze}},
  \bibinfo{author}{\bibfnamefont{L.}~\bibnamefont{Heisenberg}},
  \bibnamefont{and}
  \bibinfo{author}{\bibfnamefont{D.}~\bibnamefont{Pirtskhalava}},
  \bibinfo{journal}{Phys. Rev.} \textbf{\bibinfo{volume}{D83}},
  \bibinfo{pages}{103516} (\bibinfo{year}{2011}{\natexlab{b}}),
  \eprint{1010.1780}.

\bibitem[{\citenamefont{de~Rham and Heisenberg}(2011)}]{deRham:2011by}
\bibinfo{author}{\bibfnamefont{C.}~\bibnamefont{de~Rham}} \bibnamefont{and}
  \bibinfo{author}{\bibfnamefont{L.}~\bibnamefont{Heisenberg}},
  \bibinfo{journal}{Phys. Rev.} \textbf{\bibinfo{volume}{D84}},
  \bibinfo{pages}{043503} (\bibinfo{year}{2011}), \eprint{1106.3312}.

\bibitem[{\citenamefont{Heisenberg et~al.}(2014)\citenamefont{Heisenberg,
  Kimura, and Yamamoto}}]{Heisenberg:2014kea}
\bibinfo{author}{\bibfnamefont{L.}~\bibnamefont{Heisenberg}},
  \bibinfo{author}{\bibfnamefont{R.}~\bibnamefont{Kimura}}, \bibnamefont{and}
  \bibinfo{author}{\bibfnamefont{K.}~\bibnamefont{Yamamoto}},
  \bibinfo{journal}{Phys. Rev.} \textbf{\bibinfo{volume}{D89}},
  \bibinfo{pages}{103008} (\bibinfo{year}{2014}), \eprint{1403.2049}.

\bibitem[{\citenamefont{Deffayet et~al.}(2010)\citenamefont{Deffayet, Pujolas,
  Sawicki, and Vikman}}]{Deffayet:2010qz}
\bibinfo{author}{\bibfnamefont{C.}~\bibnamefont{Deffayet}},
  \bibinfo{author}{\bibfnamefont{O.}~\bibnamefont{Pujolas}},
  \bibinfo{author}{\bibfnamefont{I.}~\bibnamefont{Sawicki}}, \bibnamefont{and}
  \bibinfo{author}{\bibfnamefont{A.}~\bibnamefont{Vikman}},
  \bibinfo{journal}{JCAP} \textbf{\bibinfo{volume}{1010}}, \bibinfo{pages}{026}
  (\bibinfo{year}{2010}), \eprint{1008.0048}.

\bibitem[{\citenamefont{De~Felice and Tsujikawa}(2011)}]{DeFelice:2010nf}
\bibinfo{author}{\bibfnamefont{A.}~\bibnamefont{De~Felice}} \bibnamefont{and}
  \bibinfo{author}{\bibfnamefont{S.}~\bibnamefont{Tsujikawa}},
  \bibinfo{journal}{Phys. Rev.} \textbf{\bibinfo{volume}{D84}},
  \bibinfo{pages}{124029} (\bibinfo{year}{2011}), \eprint{1008.4236}.

\bibitem[{\citenamefont{Martin}(2012)}]{Martin:2012bt}
\bibinfo{author}{\bibfnamefont{J.}~\bibnamefont{Martin}},
  \bibinfo{journal}{Comptes Rendus Physique} \textbf{\bibinfo{volume}{13}},
  \bibinfo{pages}{566} (\bibinfo{year}{2012}), \eprint{1205.3365}.

\bibitem[{\citenamefont{Steinhardt et~al.}(1999)\citenamefont{Steinhardt, Wang,
  and Zlatev}}]{Steinhardt:1999nw}
\bibinfo{author}{\bibfnamefont{P.~J.} \bibnamefont{Steinhardt}},
  \bibinfo{author}{\bibfnamefont{L.-M.} \bibnamefont{Wang}}, \bibnamefont{and}
  \bibinfo{author}{\bibfnamefont{I.}~\bibnamefont{Zlatev}},
  \bibinfo{journal}{Phys. Rev.} \textbf{\bibinfo{volume}{D59}},
  \bibinfo{pages}{123504} (\bibinfo{year}{1999}), \eprint{astro-ph/9812313}.

\bibitem[{\citenamefont{Riess et~al.}(2016)}]{Riess:2016jrr}
\bibinfo{author}{\bibfnamefont{A.~G.} \bibnamefont{Riess}}
  \bibnamefont{et~al.}, \bibinfo{journal}{ArXiv e-prints}
  (\bibinfo{year}{2016}), \eprint{1604.01424}.

\bibitem[{\citenamefont{Hildebrandt et~al.}(2017)}]{Hildebrandt:2016iqg}
\bibinfo{author}{\bibfnamefont{H.}~\bibnamefont{Hildebrandt}}
  \bibnamefont{et~al.}, \bibinfo{journal}{Mon. Not. Roy. Astron. Soc.}
  \textbf{\bibinfo{volume}{465}}, \bibinfo{pages}{1454} (\bibinfo{year}{2017}),
  \eprint{1606.05338}.

\bibitem[{\citenamefont{Heymans et~al.}(2013)}]{Heymans:2013fya}
\bibinfo{author}{\bibfnamefont{C.}~\bibnamefont{Heymans}} \bibnamefont{et~al.},
  \bibinfo{journal}{Mon. Not. Roy. Astron. Soc.}
  \textbf{\bibinfo{volume}{432}}, \bibinfo{pages}{2433} (\bibinfo{year}{2013}),
  \eprint{1303.1808}.

\bibitem[{\citenamefont{Bonvin et~al.}(2016)}]{Bonvin:2016crt}
\bibinfo{author}{\bibfnamefont{V.}~\bibnamefont{Bonvin}} \bibnamefont{et~al.},
  \bibinfo{journal}{ArXiv e-prints}  (\bibinfo{year}{2016}),
  \eprint{1607.01790}.

\bibitem[{\citenamefont{Wetterich}(1988{\natexlab{a}})}]{Wetterich:1987fk}
\bibinfo{author}{\bibfnamefont{C.}~\bibnamefont{Wetterich}},
  \bibinfo{journal}{Nucl. Phys.} \textbf{\bibinfo{volume}{B302}},
  \bibinfo{pages}{645} (\bibinfo{year}{1988}{\natexlab{a}}).

\bibitem[{\citenamefont{Wetterich}(1988{\natexlab{b}})}]{Wetterich:1987fm}
\bibinfo{author}{\bibfnamefont{C.}~\bibnamefont{Wetterich}},
  \bibinfo{journal}{Nucl. Phys.} \textbf{\bibinfo{volume}{B302}},
  \bibinfo{pages}{668} (\bibinfo{year}{1988}{\natexlab{b}}).

\bibitem[{\citenamefont{Ratra and Peebles}(1988)}]{Ratra:1987rm}
\bibinfo{author}{\bibfnamefont{B.}~\bibnamefont{Ratra}} \bibnamefont{and}
  \bibinfo{author}{\bibfnamefont{P.~J.~E.} \bibnamefont{Peebles}},
  \bibinfo{journal}{Phys. Rev.} \textbf{\bibinfo{volume}{D37}},
  \bibinfo{pages}{3406} (\bibinfo{year}{1988}).

\bibitem[{\citenamefont{Scherrer and Sen}(2008)}]{Scherrer:2007pu}
\bibinfo{author}{\bibfnamefont{R.~J.} \bibnamefont{Scherrer}} \bibnamefont{and}
  \bibinfo{author}{\bibfnamefont{A.~A.} \bibnamefont{Sen}},
  \bibinfo{journal}{Phys. Rev.} \textbf{\bibinfo{volume}{D77}},
  \bibinfo{pages}{083515} (\bibinfo{year}{2008}), \eprint{0712.3450}.

\bibitem[{\citenamefont{Chiba}(2009)}]{Chiba:2009sj}
\bibinfo{author}{\bibfnamefont{T.}~\bibnamefont{Chiba}},
  \bibinfo{journal}{Phys. Rev.} \textbf{\bibinfo{volume}{D79}},
  \bibinfo{pages}{083517} (\bibinfo{year}{2009}), \bibinfo{note}{[Erratum:
  Phys. Rev.D80,109902(2009)]}, \eprint{0902.4037}.

\bibitem[{\citenamefont{Caldwell et~al.}(1998)\citenamefont{Caldwell, Dave, and
  Steinhardt}}]{Caldwell:1997ii}
\bibinfo{author}{\bibfnamefont{R.~R.} \bibnamefont{Caldwell}},
  \bibinfo{author}{\bibfnamefont{R.}~\bibnamefont{Dave}}, \bibnamefont{and}
  \bibinfo{author}{\bibfnamefont{P.~J.} \bibnamefont{Steinhardt}},
  \bibinfo{journal}{Phys. Rev. Lett.} \textbf{\bibinfo{volume}{80}},
  \bibinfo{pages}{1582} (\bibinfo{year}{1998}), \eprint{astro-ph/9708069}.

\bibitem[{\citenamefont{Zlatev et~al.}(1999)\citenamefont{Zlatev, Wang, and
  Steinhardt}}]{Zlatev:1998tr}
\bibinfo{author}{\bibfnamefont{I.}~\bibnamefont{Zlatev}},
  \bibinfo{author}{\bibfnamefont{L.-M.} \bibnamefont{Wang}}, \bibnamefont{and}
  \bibinfo{author}{\bibfnamefont{P.~J.} \bibnamefont{Steinhardt}},
  \bibinfo{journal}{Phys. Rev. Lett.} \textbf{\bibinfo{volume}{82}},
  \bibinfo{pages}{896} (\bibinfo{year}{1999}), \eprint{astro-ph/9807002}.

\bibitem[{\citenamefont{Amendola}(2000)}]{Amendola:1999er}
\bibinfo{author}{\bibfnamefont{L.}~\bibnamefont{Amendola}},
  \bibinfo{journal}{Phys. Rev.} \textbf{\bibinfo{volume}{D62}},
  \bibinfo{pages}{043511} (\bibinfo{year}{2000}), \eprint{astro-ph/9908023}.

\bibitem[{\citenamefont{Sahni and Wang}(2000)}]{Sahni:1999qe}
\bibinfo{author}{\bibfnamefont{V.}~\bibnamefont{Sahni}} \bibnamefont{and}
  \bibinfo{author}{\bibfnamefont{L.-M.} \bibnamefont{Wang}},
  \bibinfo{journal}{Phys. Rev.} \textbf{\bibinfo{volume}{D62}},
  \bibinfo{pages}{103517} (\bibinfo{year}{2000}), \eprint{astro-ph/9910097}.

\bibitem[{\citenamefont{Perrotta et~al.}(1999)\citenamefont{Perrotta,
  Baccigalupi, and Matarrese}}]{Perrotta:1999am}
\bibinfo{author}{\bibfnamefont{F.}~\bibnamefont{Perrotta}},
  \bibinfo{author}{\bibfnamefont{C.}~\bibnamefont{Baccigalupi}},
  \bibnamefont{and}
  \bibinfo{author}{\bibfnamefont{S.}~\bibnamefont{Matarrese}},
  \bibinfo{journal}{Phys. Rev.} \textbf{\bibinfo{volume}{D61}},
  \bibinfo{pages}{023507} (\bibinfo{year}{1999}), \eprint{astro-ph/9906066}.

\bibitem[{\citenamefont{Sahni et~al.}(2002)\citenamefont{Sahni, Sami, and
  Souradeep}}]{Sahni:2001qp}
\bibinfo{author}{\bibfnamefont{V.}~\bibnamefont{Sahni}},
  \bibinfo{author}{\bibfnamefont{M.}~\bibnamefont{Sami}}, \bibnamefont{and}
  \bibinfo{author}{\bibfnamefont{T.}~\bibnamefont{Souradeep}},
  \bibinfo{journal}{Phys. Rev.} \textbf{\bibinfo{volume}{D65}},
  \bibinfo{pages}{023518} (\bibinfo{year}{2002}), \eprint{gr-qc/0105121}.

\bibitem[{\citenamefont{Wali~Hossain et~al.}(2015)\citenamefont{Wali~Hossain,
  Myrzakulov, Sami, and Saridakis}}]{Hossain:2014zma}
\bibinfo{author}{\bibfnamefont{M.}~\bibnamefont{Wali~Hossain}},
  \bibinfo{author}{\bibfnamefont{R.}~\bibnamefont{Myrzakulov}},
  \bibinfo{author}{\bibfnamefont{M.}~\bibnamefont{Sami}}, \bibnamefont{and}
  \bibinfo{author}{\bibfnamefont{E.~N.} \bibnamefont{Saridakis}},
  \bibinfo{journal}{Int. J. Mod. Phys.} \textbf{\bibinfo{volume}{D24}},
  \bibinfo{pages}{1530014} (\bibinfo{year}{2015}), \eprint{1410.6100}.

\bibitem[{\citenamefont{Caldwell}(2002)}]{Caldwell:1999ew}
\bibinfo{author}{\bibfnamefont{R.~R.} \bibnamefont{Caldwell}},
  \bibinfo{journal}{Phys. Lett.} \textbf{\bibinfo{volume}{B545}},
  \bibinfo{pages}{23} (\bibinfo{year}{2002}), \eprint{astro-ph/9908168}.

\bibitem[{\citenamefont{Elizalde et~al.}(2004)\citenamefont{Elizalde, Nojiri,
  and Odintsov}}]{Elizalde:2004mq}
\bibinfo{author}{\bibfnamefont{E.}~\bibnamefont{Elizalde}},
  \bibinfo{author}{\bibfnamefont{S.}~\bibnamefont{Nojiri}}, \bibnamefont{and}
  \bibinfo{author}{\bibfnamefont{S.~D.} \bibnamefont{Odintsov}},
  \bibinfo{journal}{Phys. Rev.} \textbf{\bibinfo{volume}{D70}},
  \bibinfo{pages}{043539} (\bibinfo{year}{2004}), \eprint{hep-th/0405034}.

\bibitem[{\citenamefont{Sen}(2002{\natexlab{a}})}]{Sen:2002nu}
\bibinfo{author}{\bibfnamefont{A.}~\bibnamefont{Sen}}, \bibinfo{journal}{JHEP}
  \textbf{\bibinfo{volume}{04}}, \bibinfo{pages}{048}
  (\bibinfo{year}{2002}{\natexlab{a}}), \eprint{hep-th/0203211}.

\bibitem[{\citenamefont{Sen}(2002{\natexlab{b}})}]{Sen:2002in}
\bibinfo{author}{\bibfnamefont{A.}~\bibnamefont{Sen}}, \bibinfo{journal}{JHEP}
  \textbf{\bibinfo{volume}{07}}, \bibinfo{pages}{065}
  (\bibinfo{year}{2002}{\natexlab{b}}), \eprint{hep-th/0203265}.

\bibitem[{\citenamefont{Gibbons}(2002)}]{Gibbons:2002md}
\bibinfo{author}{\bibfnamefont{G.~W.} \bibnamefont{Gibbons}},
  \bibinfo{journal}{Phys. Lett.} \textbf{\bibinfo{volume}{B537}},
  \bibinfo{pages}{1} (\bibinfo{year}{2002}), \eprint{hep-th/0204008}.

\bibitem[{\citenamefont{Garousi et~al.}(2004)\citenamefont{Garousi, Sami, and
  Tsujikawa}}]{Garousi:2004uf}
\bibinfo{author}{\bibfnamefont{M.~R.} \bibnamefont{Garousi}},
  \bibinfo{author}{\bibfnamefont{M.}~\bibnamefont{Sami}}, \bibnamefont{and}
  \bibinfo{author}{\bibfnamefont{S.}~\bibnamefont{Tsujikawa}},
  \bibinfo{journal}{Phys. Rev.} \textbf{\bibinfo{volume}{D70}},
  \bibinfo{pages}{043536} (\bibinfo{year}{2004}), \eprint{hep-th/0402075}.

\bibitem[{\citenamefont{Copeland et~al.}(2005)\citenamefont{Copeland, Garousi,
  Sami, and Tsujikawa}}]{Copeland:2004hq}
\bibinfo{author}{\bibfnamefont{E.~J.} \bibnamefont{Copeland}},
  \bibinfo{author}{\bibfnamefont{M.~R.} \bibnamefont{Garousi}},
  \bibinfo{author}{\bibfnamefont{M.}~\bibnamefont{Sami}}, \bibnamefont{and}
  \bibinfo{author}{\bibfnamefont{S.}~\bibnamefont{Tsujikawa}},
  \bibinfo{journal}{Phys. Rev.} \textbf{\bibinfo{volume}{D71}},
  \bibinfo{pages}{043003} (\bibinfo{year}{2005}), \eprint{hep-th/0411192}.

\bibitem[{\citenamefont{Armendariz-Picon
  et~al.}(2000)\citenamefont{Armendariz-Picon, Mukhanov, and
  Steinhardt}}]{ArmendarizPicon:2000dh}
\bibinfo{author}{\bibfnamefont{C.}~\bibnamefont{Armendariz-Picon}},
  \bibinfo{author}{\bibfnamefont{V.~F.} \bibnamefont{Mukhanov}},
  \bibnamefont{and} \bibinfo{author}{\bibfnamefont{P.~J.}
  \bibnamefont{Steinhardt}}, \bibinfo{journal}{Phys. Rev. Lett.}
  \textbf{\bibinfo{volume}{85}}, \bibinfo{pages}{4438} (\bibinfo{year}{2000}),
  \eprint{astro-ph/0004134}.

\bibitem[{\citenamefont{Armendariz-Picon
  et~al.}(2001)\citenamefont{Armendariz-Picon, Mukhanov, and
  Steinhardt}}]{ArmendarizPicon:2000ah}
\bibinfo{author}{\bibfnamefont{C.}~\bibnamefont{Armendariz-Picon}},
  \bibinfo{author}{\bibfnamefont{V.~F.} \bibnamefont{Mukhanov}},
  \bibnamefont{and} \bibinfo{author}{\bibfnamefont{P.~J.}
  \bibnamefont{Steinhardt}}, \bibinfo{journal}{Phys. Rev.}
  \textbf{\bibinfo{volume}{D63}}, \bibinfo{pages}{103510}
  (\bibinfo{year}{2001}), \eprint{astro-ph/0006373}.

\bibitem[{\citenamefont{Rendall}(2006)}]{Rendall:2005fv}
\bibinfo{author}{\bibfnamefont{A.~D.} \bibnamefont{Rendall}},
  \bibinfo{journal}{Class. Quant. Grav.} \textbf{\bibinfo{volume}{23}},
  \bibinfo{pages}{1557} (\bibinfo{year}{2006}), \eprint{gr-qc/0511158}.

\bibitem[{\citenamefont{Bento et~al.}(2002)\citenamefont{Bento, Bertolami, and
  Sen}}]{Bento:2002ps}
\bibinfo{author}{\bibfnamefont{M.~C.} \bibnamefont{Bento}},
  \bibinfo{author}{\bibfnamefont{O.}~\bibnamefont{Bertolami}},
  \bibnamefont{and} \bibinfo{author}{\bibfnamefont{A.~A.} \bibnamefont{Sen}},
  \bibinfo{journal}{Phys. Rev.} \textbf{\bibinfo{volume}{D66}},
  \bibinfo{pages}{043507} (\bibinfo{year}{2002}), \eprint{gr-qc/0202064}.

\bibitem[{\citenamefont{Bento et~al.}(2003)\citenamefont{Bento, Bertolami, and
  Sen}}]{Bento:2002yx}
\bibinfo{author}{\bibfnamefont{M.~d.~C.} \bibnamefont{Bento}},
  \bibinfo{author}{\bibfnamefont{O.}~\bibnamefont{Bertolami}},
  \bibnamefont{and} \bibinfo{author}{\bibfnamefont{A.~A.} \bibnamefont{Sen}},
  \bibinfo{journal}{Phys. Rev.} \textbf{\bibinfo{volume}{D67}},
  \bibinfo{pages}{063003} (\bibinfo{year}{2003}), \eprint{astro-ph/0210468}.

\bibitem[{\citenamefont{Luty et~al.}(2003)\citenamefont{Luty, Porrati, and
  Rattazzi}}]{Luty:2003vm}
\bibinfo{author}{\bibfnamefont{M.~A.} \bibnamefont{Luty}},
  \bibinfo{author}{\bibfnamefont{M.}~\bibnamefont{Porrati}}, \bibnamefont{and}
  \bibinfo{author}{\bibfnamefont{R.}~\bibnamefont{Rattazzi}},
  \bibinfo{journal}{JHEP} \textbf{\bibinfo{volume}{09}}, \bibinfo{pages}{029}
  (\bibinfo{year}{2003}), \eprint{hep-th/0303116}.

\bibitem[{\citenamefont{Deffayet et~al.}(2009)\citenamefont{Deffayet,
  Esposito-Farese, and Vikman}}]{Deffayet:2009wt}
\bibinfo{author}{\bibfnamefont{C.}~\bibnamefont{Deffayet}},
  \bibinfo{author}{\bibfnamefont{G.}~\bibnamefont{Esposito-Farese}},
  \bibnamefont{and} \bibinfo{author}{\bibfnamefont{A.}~\bibnamefont{Vikman}},
  \bibinfo{journal}{Phys. Rev.} \textbf{\bibinfo{volume}{D79}},
  \bibinfo{pages}{084003} (\bibinfo{year}{2009}), \eprint{0901.1314}.

\bibitem[{\citenamefont{Woodard}(2007)}]{Woodard:2006nt}
\bibinfo{author}{\bibfnamefont{R.~P.} \bibnamefont{Woodard}},
  \bibinfo{journal}{Lect. Notes Phys.} \textbf{\bibinfo{volume}{720}},
  \bibinfo{pages}{403} (\bibinfo{year}{2007}), \eprint{astro-ph/0601672}.

\bibitem[{\citenamefont{Vainshtein}(1972)}]{Vainshtein:1972sx}
\bibinfo{author}{\bibfnamefont{A.~I.} \bibnamefont{Vainshtein}},
  \bibinfo{journal}{Phys. Lett.} \textbf{\bibinfo{volume}{B39}},
  \bibinfo{pages}{393} (\bibinfo{year}{1972}).

\bibitem[{\citenamefont{van Dam and Veltman}(1970)}]{vanDam:1970vg}
\bibinfo{author}{\bibfnamefont{H.}~\bibnamefont{van Dam}} \bibnamefont{and}
  \bibinfo{author}{\bibfnamefont{M.~J.~G.} \bibnamefont{Veltman}},
  \bibinfo{journal}{Nucl. Phys.} \textbf{\bibinfo{volume}{B22}},
  \bibinfo{pages}{397} (\bibinfo{year}{1970}).

\bibitem[{\citenamefont{Zakharov}(1970)}]{Zakharov:1970cc}
\bibinfo{author}{\bibfnamefont{V.~I.} \bibnamefont{Zakharov}},
  \bibinfo{journal}{JETP Lett.} \textbf{\bibinfo{volume}{12}},
  \bibinfo{pages}{312} (\bibinfo{year}{1970}), \bibinfo{note}{[Pisma Zh. Eksp.
  Teor. Fiz.12,447(1970)]}.

\bibitem[{\citenamefont{Fierz and Pauli}(1939)}]{Fierz:1939ix}
\bibinfo{author}{\bibfnamefont{M.}~\bibnamefont{Fierz}} \bibnamefont{and}
  \bibinfo{author}{\bibfnamefont{W.}~\bibnamefont{Pauli}},
  \bibinfo{journal}{Proc. Roy. Soc. Lond.} \textbf{\bibinfo{volume}{A173}},
  \bibinfo{pages}{211} (\bibinfo{year}{1939}).

\bibitem[{\citenamefont{Chow and Khoury}(2009)}]{Chow:2009fm}
\bibinfo{author}{\bibfnamefont{N.}~\bibnamefont{Chow}} \bibnamefont{and}
  \bibinfo{author}{\bibfnamefont{J.}~\bibnamefont{Khoury}},
  \bibinfo{journal}{Phys. Rev.} \textbf{\bibinfo{volume}{D80}},
  \bibinfo{pages}{024037} (\bibinfo{year}{2009}), \eprint{0905.1325}.

\bibitem[{\citenamefont{Silva and Koyama}(2009)}]{Silva:2009km}
\bibinfo{author}{\bibfnamefont{F.~P.} \bibnamefont{Silva}} \bibnamefont{and}
  \bibinfo{author}{\bibfnamefont{K.}~\bibnamefont{Koyama}},
  \bibinfo{journal}{Phys. Rev.} \textbf{\bibinfo{volume}{D80}},
  \bibinfo{pages}{121301} (\bibinfo{year}{2009}), \eprint{0909.4538}.

\bibitem[{\citenamefont{Kobayashi}(2010)}]{Kobayashi:2010wa}
\bibinfo{author}{\bibfnamefont{T.}~\bibnamefont{Kobayashi}},
  \bibinfo{journal}{Phys. Rev.} \textbf{\bibinfo{volume}{D81}},
  \bibinfo{pages}{103533} (\bibinfo{year}{2010}), \eprint{1003.3281}.

\bibitem[{\citenamefont{Kobayashi et~al.}(2010)\citenamefont{Kobayashi,
  Tashiro, and Suzuki}}]{Kobayashi:2009wr}
\bibinfo{author}{\bibfnamefont{T.}~\bibnamefont{Kobayashi}},
  \bibinfo{author}{\bibfnamefont{H.}~\bibnamefont{Tashiro}}, \bibnamefont{and}
  \bibinfo{author}{\bibfnamefont{D.}~\bibnamefont{Suzuki}},
  \bibinfo{journal}{Phys. Rev.} \textbf{\bibinfo{volume}{D81}},
  \bibinfo{pages}{063513} (\bibinfo{year}{2010}), \eprint{0912.4641}.

\bibitem[{\citenamefont{Gannouji and Sami}(2010)}]{Gannouji:2010au}
\bibinfo{author}{\bibfnamefont{R.}~\bibnamefont{Gannouji}} \bibnamefont{and}
  \bibinfo{author}{\bibfnamefont{M.}~\bibnamefont{Sami}},
  \bibinfo{journal}{Phys. Rev.} \textbf{\bibinfo{volume}{D82}},
  \bibinfo{pages}{024011} (\bibinfo{year}{2010}), \eprint{1004.2808}.

\bibitem[{\citenamefont{De~Felice et~al.}(2010)\citenamefont{De~Felice,
  Mukohyama, and Tsujikawa}}]{DeFelice:2010gb}
\bibinfo{author}{\bibfnamefont{A.}~\bibnamefont{De~Felice}},
  \bibinfo{author}{\bibfnamefont{S.}~\bibnamefont{Mukohyama}},
  \bibnamefont{and}
  \bibinfo{author}{\bibfnamefont{S.}~\bibnamefont{Tsujikawa}},
  \bibinfo{journal}{Phys. Rev.} \textbf{\bibinfo{volume}{D82}},
  \bibinfo{pages}{023524} (\bibinfo{year}{2010}), \eprint{1006.0281}.

\bibitem[{\citenamefont{De~Felice and
  Tsujikawa}(2010{\natexlab{b}})}]{DeFelice:2010pv}
\bibinfo{author}{\bibfnamefont{A.}~\bibnamefont{De~Felice}} \bibnamefont{and}
  \bibinfo{author}{\bibfnamefont{S.}~\bibnamefont{Tsujikawa}},
  \bibinfo{journal}{Phys. Rev. Lett.} \textbf{\bibinfo{volume}{105}},
  \bibinfo{pages}{111301} (\bibinfo{year}{2010}{\natexlab{b}}),
  \eprint{1007.2700}.

\bibitem[{\citenamefont{Ali et~al.}(2010)\citenamefont{Ali, Gannouji, and
  Sami}}]{Ali:2010gr}
\bibinfo{author}{\bibfnamefont{A.}~\bibnamefont{Ali}},
  \bibinfo{author}{\bibfnamefont{R.}~\bibnamefont{Gannouji}}, \bibnamefont{and}
  \bibinfo{author}{\bibfnamefont{M.}~\bibnamefont{Sami}},
  \bibinfo{journal}{Phys. Rev.} \textbf{\bibinfo{volume}{D82}},
  \bibinfo{pages}{103015} (\bibinfo{year}{2010}), \eprint{1008.1588}.

\bibitem[{\citenamefont{Mota et~al.}(2010)\citenamefont{Mota, Sandstad, and
  Zlosnik}}]{Mota:2010bs}
\bibinfo{author}{\bibfnamefont{D.~F.} \bibnamefont{Mota}},
  \bibinfo{author}{\bibfnamefont{M.}~\bibnamefont{Sandstad}}, \bibnamefont{and}
  \bibinfo{author}{\bibfnamefont{T.}~\bibnamefont{Zlosnik}},
  \bibinfo{journal}{JHEP} \textbf{\bibinfo{volume}{12}}, \bibinfo{pages}{051}
  (\bibinfo{year}{2010}), \eprint{1009.6151}.

\bibitem[{\citenamefont{Ali et~al.}(2012)\citenamefont{Ali, Gannouji, Hossain,
  and Sami}}]{Ali:2012cv}
\bibinfo{author}{\bibfnamefont{A.}~\bibnamefont{Ali}},
  \bibinfo{author}{\bibfnamefont{R.}~\bibnamefont{Gannouji}},
  \bibinfo{author}{\bibfnamefont{M.~W.} \bibnamefont{Hossain}},
  \bibnamefont{and} \bibinfo{author}{\bibfnamefont{M.}~\bibnamefont{Sami}},
  \bibinfo{journal}{Phys. Lett.} \textbf{\bibinfo{volume}{B718}},
  \bibinfo{pages}{5} (\bibinfo{year}{2012}), \eprint{1207.3959}.

\bibitem[{\citenamefont{Hossain and Sen}(2012)}]{Hossain:2012qm}
\bibinfo{author}{\bibfnamefont{M.~W.} \bibnamefont{Hossain}} \bibnamefont{and}
  \bibinfo{author}{\bibfnamefont{A.~A.} \bibnamefont{Sen}},
  \bibinfo{journal}{Phys. Lett.} \textbf{\bibinfo{volume}{B713}},
  \bibinfo{pages}{140} (\bibinfo{year}{2012}), \eprint{1201.6192}.

\bibitem[{\citenamefont{Hossain}(2017)}]{Hossain:2017ica}
\bibinfo{author}{\bibfnamefont{M.~W.} \bibnamefont{Hossain}},
  \bibinfo{journal}{ArXiv e-prints}  (\bibinfo{year}{2017}),
  \eprint{1704.07956}.

\bibitem[{\citenamefont{Yoo et~al.}(2009)\citenamefont{Yoo, Fitzpatrick, and
  Zaldarriaga}}]{Yoo:2009au}
\bibinfo{author}{\bibfnamefont{J.}~\bibnamefont{Yoo}},
  \bibinfo{author}{\bibfnamefont{A.~L.} \bibnamefont{Fitzpatrick}},
  \bibnamefont{and}
  \bibinfo{author}{\bibfnamefont{M.}~\bibnamefont{Zaldarriaga}},
  \bibinfo{journal}{Phys. Rev.} \textbf{\bibinfo{volume}{D80}},
  \bibinfo{pages}{083514} (\bibinfo{year}{2009}), \eprint{0907.0707}.

\bibitem[{\citenamefont{Bonvin and Durrer}(2011)}]{Bonvin:2011bg}
\bibinfo{author}{\bibfnamefont{C.}~\bibnamefont{Bonvin}} \bibnamefont{and}
  \bibinfo{author}{\bibfnamefont{R.}~\bibnamefont{Durrer}},
  \bibinfo{journal}{Phys. Rev.} \textbf{\bibinfo{volume}{D84}},
  \bibinfo{pages}{063505} (\bibinfo{year}{2011}), \eprint{1105.5280}.

\bibitem[{\citenamefont{Challinor and Lewis}(2011)}]{Challinor:2011bk}
\bibinfo{author}{\bibfnamefont{A.}~\bibnamefont{Challinor}} \bibnamefont{and}
  \bibinfo{author}{\bibfnamefont{A.}~\bibnamefont{Lewis}},
  \bibinfo{journal}{Phys. Rev.} \textbf{\bibinfo{volume}{D84}},
  \bibinfo{pages}{043516} (\bibinfo{year}{2011}), \eprint{1105.5292}.

\bibitem[{\citenamefont{Duniya et~al.}(2015)\citenamefont{Duniya, Bertacca, and
  Maartens}}]{Duniya:2015nva}
\bibinfo{author}{\bibfnamefont{D.~G.~A.} \bibnamefont{Duniya}},
  \bibinfo{author}{\bibfnamefont{D.}~\bibnamefont{Bertacca}}, \bibnamefont{and}
  \bibinfo{author}{\bibfnamefont{R.}~\bibnamefont{Maartens}},
  \bibinfo{journal}{Phys. Rev.} \textbf{\bibinfo{volume}{D91}},
  \bibinfo{pages}{063530} (\bibinfo{year}{2015}), \eprint{1502.06424}.

\bibitem[{\citenamefont{Dinda and Sen}(2016{\natexlab{a}})}]{Dinda:2016ibo}
\bibinfo{author}{\bibfnamefont{B.~R.} \bibnamefont{Dinda}} \bibnamefont{and}
  \bibinfo{author}{\bibfnamefont{A.~A.} \bibnamefont{Sen}}
  (\bibinfo{year}{2016}{\natexlab{a}}), \eprint{1607.05123}.

\bibitem[{\citenamefont{{Unnikrishnan}
  et~al.}(2008)\citenamefont{{Unnikrishnan}, {Jassal}, and
  {Seshadri}}}]{2008PhRvD..78l3504U}
\bibinfo{author}{\bibfnamefont{S.}~\bibnamefont{{Unnikrishnan}}},
  \bibinfo{author}{\bibfnamefont{H.~K.} \bibnamefont{{Jassal}}},
  \bibnamefont{and} \bibinfo{author}{\bibfnamefont{T.~R.}
  \bibnamefont{{Seshadri}}}, \bibinfo{journal}{\prd}
  \textbf{\bibinfo{volume}{78}}, \bibinfo{eid}{123504} (\bibinfo{year}{2008}),
  \eprint{0801.2017}.

\bibitem[{\citenamefont{Dinda and Sen}(2016{\natexlab{b}})}]{Bikash:2016ica}
\bibinfo{author}{\bibfnamefont{B.~R.} \bibnamefont{Dinda}} \bibnamefont{and}
  \bibinfo{author}{\bibfnamefont{A.~A.} \bibnamefont{Sen}},
  \bibinfo{journal}{ArXiv e-prints}  (\bibinfo{year}{2016}{\natexlab{b}}),
  \eprint{1607.05123}.

\bibitem[{\citenamefont{Duniya}(2016{\natexlab{a}})}]{Duniya:2016ibg}
\bibinfo{author}{\bibfnamefont{D.}~\bibnamefont{Duniya}},
  \bibinfo{journal}{ArXiv e-prints}  (\bibinfo{year}{2016}{\natexlab{a}}),
  \eprint{1606.00712}.

\bibitem[{\citenamefont{Duniya et~al.}(2013)\citenamefont{Duniya, Bertacca, and
  Maartens}}]{Duniya:2013eta}
\bibinfo{author}{\bibfnamefont{D.}~\bibnamefont{Duniya}},
  \bibinfo{author}{\bibfnamefont{D.}~\bibnamefont{Bertacca}}, \bibnamefont{and}
  \bibinfo{author}{\bibfnamefont{R.}~\bibnamefont{Maartens}},
  \bibinfo{journal}{JCAP} \textbf{\bibinfo{volume}{1310}}, \bibinfo{pages}{015}
  (\bibinfo{year}{2013}), \eprint{1305.4509}.

\bibitem[{\citenamefont{Kaiser}(1987)}]{Kaiser:1987qv}
\bibinfo{author}{\bibfnamefont{N.}~\bibnamefont{Kaiser}},
  \bibinfo{journal}{Mon. Not. Roy. Astron. Soc.}
  \textbf{\bibinfo{volume}{227}}, \bibinfo{pages}{1} (\bibinfo{year}{1987}).

\bibitem[{\citenamefont{Moessner et~al.}(1998)\citenamefont{Moessner, Jain, and
  Villumsen}}]{Moessner:1997qs}
\bibinfo{author}{\bibfnamefont{R.}~\bibnamefont{Moessner}},
  \bibinfo{author}{\bibfnamefont{B.}~\bibnamefont{Jain}}, \bibnamefont{and}
  \bibinfo{author}{\bibfnamefont{J.~V.} \bibnamefont{Villumsen}},
  \bibinfo{journal}{Mon. Not. Roy. Astron. Soc.}
  \textbf{\bibinfo{volume}{294}}, \bibinfo{pages}{291} (\bibinfo{year}{1998}),
  \eprint{astro-ph/9708271}.

\bibitem[{\citenamefont{Bonvin}(2014)}]{Bonvin:2014owa}
\bibinfo{author}{\bibfnamefont{C.}~\bibnamefont{Bonvin}},
  \bibinfo{journal}{Class. Quant. Grav.} \textbf{\bibinfo{volume}{31}},
  \bibinfo{pages}{234002} (\bibinfo{year}{2014}), \eprint{1409.2224}.

\bibitem[{\citenamefont{Jeong et~al.}(2012)\citenamefont{Jeong, Schmidt, and
  Hirata}}]{Jeong:2011as}
\bibinfo{author}{\bibfnamefont{D.}~\bibnamefont{Jeong}},
  \bibinfo{author}{\bibfnamefont{F.}~\bibnamefont{Schmidt}}, \bibnamefont{and}
  \bibinfo{author}{\bibfnamefont{C.~M.} \bibnamefont{Hirata}},
  \bibinfo{journal}{Phys. Rev.} \textbf{\bibinfo{volume}{D85}},
  \bibinfo{pages}{023504} (\bibinfo{year}{2012}), \eprint{1107.5427}.

\bibitem[{\citenamefont{Yoo et~al.}(2012)\citenamefont{Yoo, Hamaus, Seljak, and
  Zaldarriaga}}]{Yoo:2012se}
\bibinfo{author}{\bibfnamefont{J.}~\bibnamefont{Yoo}},
  \bibinfo{author}{\bibfnamefont{N.}~\bibnamefont{Hamaus}},
  \bibinfo{author}{\bibfnamefont{U.}~\bibnamefont{Seljak}}, \bibnamefont{and}
  \bibinfo{author}{\bibfnamefont{M.}~\bibnamefont{Zaldarriaga}},
  \bibinfo{journal}{Phys. Rev.} \textbf{\bibinfo{volume}{D86}},
  \bibinfo{pages}{063514} (\bibinfo{year}{2012}), \eprint{1206.5809}.

\bibitem[{\citenamefont{Bertacca et~al.}(2012)\citenamefont{Bertacca, Maartens,
  Raccanelli, and Clarkson}}]{Bertacca:2012tp}
\bibinfo{author}{\bibfnamefont{D.}~\bibnamefont{Bertacca}},
  \bibinfo{author}{\bibfnamefont{R.}~\bibnamefont{Maartens}},
  \bibinfo{author}{\bibfnamefont{A.}~\bibnamefont{Raccanelli}},
  \bibnamefont{and} \bibinfo{author}{\bibfnamefont{C.}~\bibnamefont{Clarkson}},
  \bibinfo{journal}{JCAP} \textbf{\bibinfo{volume}{1210}}, \bibinfo{pages}{025}
  (\bibinfo{year}{2012}), \eprint{1205.5221}.

\bibitem[{\citenamefont{Duniya}(2016{\natexlab{b}})}]{Duniya:2015dpa}
\bibinfo{author}{\bibfnamefont{D.}~\bibnamefont{Duniya}},
  \bibinfo{journal}{Gen. Rel. Grav.} \textbf{\bibinfo{volume}{48}},
  \bibinfo{pages}{52} (\bibinfo{year}{2016}{\natexlab{b}}),
  \eprint{1505.03436}.

\bibitem[{\citenamefont{De~Felice et~al.}(2016)\citenamefont{De~Felice,
  Heisenberg, Kase, Mukohyama, Tsujikawa, and Zhang}}]{DeFelice:2016yws}
\bibinfo{author}{\bibfnamefont{A.}~\bibnamefont{De~Felice}},
  \bibinfo{author}{\bibfnamefont{L.}~\bibnamefont{Heisenberg}},
  \bibinfo{author}{\bibfnamefont{R.}~\bibnamefont{Kase}},
  \bibinfo{author}{\bibfnamefont{S.}~\bibnamefont{Mukohyama}},
  \bibinfo{author}{\bibfnamefont{S.}~\bibnamefont{Tsujikawa}},
  \bibnamefont{and} \bibinfo{author}{\bibfnamefont{Y.-l.} \bibnamefont{Zhang}},
  \bibinfo{journal}{JCAP} \textbf{\bibinfo{volume}{1606}}, \bibinfo{pages}{048}
  (\bibinfo{year}{2016}), \eprint{1603.05806}.

\bibitem[{\citenamefont{{Akeret} et~al.}(2015)\citenamefont{{Akeret}, {Gamper},
  {Amara}, and {Refregier}}}]{2015A&C....10....1A}
\bibinfo{author}{\bibfnamefont{J.}~\bibnamefont{{Akeret}}},
  \bibinfo{author}{\bibfnamefont{L.}~\bibnamefont{{Gamper}}},
  \bibinfo{author}{\bibfnamefont{A.}~\bibnamefont{{Amara}}}, \bibnamefont{and}
  \bibinfo{author}{\bibfnamefont{A.}~\bibnamefont{{Refregier}}},
  \bibinfo{journal}{Astronomy and Computing} \textbf{\bibinfo{volume}{10}},
  \bibinfo{pages}{1} (\bibinfo{year}{2015}), \eprint{1410.4345}.

\end{thebibliography}

\end{document}